\documentclass[journal]{IEEEtran}

\usepackage[cmex10]{amsmath}
\usepackage{amssymb}
\usepackage{graphicx}
\usepackage{caption}
\usepackage{cite}
\usepackage{subeqnarray}
\usepackage{diagbox}
\usepackage{booktabs}
\usepackage{multirow}
\usepackage{multicol}
\usepackage{algorithm}
\usepackage{algorithmic, setspace}
\usepackage{url}
\usepackage{xcolor}
\usepackage{array}
\usepackage{footnote}
\usepackage{enumerate}
\usepackage{framed}
\usepackage{lineno}
\usepackage{threeparttable}
\makesavenoteenv{tabular}

\newtheorem{example}{Example}

\newcommand{\mb}{\mathbf}
\newcommand{\mc}{\mathcal}

\begin{document}
	\title{Generalized Mutual Information-Maximizing Quantized Decoding of LDPC Codes with Layered Scheduling\\
	\text{\Large(Extended Version)}
}
	\author{
		\IEEEauthorblockN{Peng Kang$^\dag$,\! Kui Cai$^\dag$,\! Xuan He$^\dag$,\! Shuangyang Li$^\ast$, \! Jinhong Yuan$^\ast$\\}
		\IEEEauthorblockA{${}^\dag$Singapore University of Technology and Design (SUTD), Singapore, 487372\\}
		\IEEEauthorblockA{${}^\ast$The University of New South Wales, Sydney, Australia, 2032}
	\vspace{-1cm}}
\maketitle 
\begin{abstract}
	In this paper, we propose a framework of the mutual information-maximizing (MIM) quantized decoding for low-density parity-check (LDPC) codes by using simple mappings and fixed-point additions.
	Our decoding method is generic in the sense that it can be applied to LDPC codes with arbitrary degree distributions, and can be implemented based on either the belief propagation (BP) algorithm or the min-sum (MS) algorithm.
	In particular, we propose the MIM density evolution (MIM-DE) to construct the lookup tables (LUTs) for the node updates.
	The computational complexity and memory requirements are discussed and compared to the LUT decoder variants.
	For applications with low-latency requirement, we consider the layered schedule to accelerate the convergence speed of decoding quasi-cyclic LDPC codes.
	In particular, we develop the layered MIM-DE to design the LUTs based on the MS algorithm, leading to the MIM layered quantized MS (MIM-LQMS) decoder.
	An optimization method is further introduced to reduce the memory requirement for storing the LUTs.
	Simulation results show that the MIM quantized decoders outperform the state-of-the-art LUT decoders in the waterfall region with both 3-bit and 4-bit precision over the additive white Gaussian noise channels.
	For small decoding iterations, the MIM quantized decoders also achieve a faster convergence speed compared to the benchmarks.
	Moreover, the 4-bit MIM-LQMS decoder can approach the error performance of the floating-point layered BP decoder within $0.3$ dB in the moderate-to-high SNR regions, over both the AWGN channels and the fast fading channels.
\end{abstract}

\IEEEpeerreviewmaketitle


\section{Introduction}
Low-density parity-check (LDPC) codes \cite{Gallager62} and their variations \cite{Kou01,Mitchell2015scldpc,Xie2016EGSC,chen2018rate,Nasseri2021gcldpc} have been widely applied to wireless communication and data storage systems, for their capacity approaching performance under iterative message passing decoding \cite{Richardson01capacity}. 
Many researchers have put their efforts in designing the LDPC coding schemes to enhance the system performance for different channels, such as the partial-response channels and the fading channels e.g., \cite{Kurkoski2002partial,Fang2016qsfading,Shakiba}.
From the implementation perspective, the development of LDPC decoders also has drawn great attention in terms of the decoding algorithms \cite{MacKay1999SPA,Jinghu2005reduced,Valentin2008selfCor} and architectures \cite{Marc2001osd,Varnica2007ABP,Kang2018reliability,Kang2020eqml}, for achieving a good trade-off between the error performance and decoding complexity.
In practical systems, the LDPC decoders require inputs with finite precision and hence it is necessary to perform quantization on the messages exchanged within the decoders.
To support applications such as vehicle-to-everything and automated driving in recent fifth-generation (5G) systems and beyond, the design of LDPC decoders is devoted to achieving ultra-reliability and fast convergence speed with low decoding complexity.
In this context, the LDPC decoders use coarsely quantized messages attract much attention in recent literature \cite{Nguyen2018faid,Ghaffari2019pfaid,Romero2016mim, Lewandowsky18, Stark2018ib, Stark2020ib, Meidlinger2020irr}.
Due to the low-precision messages used for decoding, the complexity of the LDPC decoders can be significantly reduced.
One approach to implement the LDPC decoders with coarse quantization is to adopt lookup tables (LUTs) for the variable node (VN) and the check node (CN) updates.
For example, the non-surjective finite alphabet iterative decoder (NS-FAID) \cite{Nguyen2018faid} optimizes a single LUT for the VN update based on the density evolution (DE) \cite{Richardson01capacity} to achieve a high throughput. 
In \cite{Ghaffari2019pfaid}, the error performance of the NS-FAID was improved by using multiple LUTs with optimized usage probabilities.
Moreover, the mutual information-maximizing LUT (MIM-LUT) decoders were proposed in \cite{Romero2016mim, Lewandowsky18, Stark2018ib, Stark2020ib, Meidlinger2020irr}.
These decoders replace the arithmetic computation for the node updates with purely table lookup operations for low decoding complexity.
Note that the LUTs of these decoders are designed based on the DE by considering an optimal or a sub-optimal quantization scheme to maximize the mutual information between the quantized messages and the associated coded bits.
More specifically, the LUTs constructed in \cite{Romero2016mim} are designed based on the optimal quantization by using the dynamic programming (DP) \cite{Kurkoski14}.
In \cite{Lewandowsky18}, the information bottleneck method is utilized to generate the LUTs for decoding regular LDPC codes.  
Following \cite{Lewandowsky18}, Stark {\it et al.} proposed the message alignment method in \cite{Stark2018ib} and extended the framework to decode irregular LDPC codes. 
The message alignment method was further improved in \cite{Stark2020ib} by considering the degree distributions of the LDPC codes in the design of the LUTs.
In addition, the min-LUT decoder \cite{Meidlinger2020irr} replaces the belief propagation (BP) algorithm \cite{MacKay1999SPA} by the min-sum (MS) algorithm for the CN update to reduce the decoding complexity.
However, as the size of LUTs grows exponentially with the node degrees, the large LUTs in \cite{Romero2016mim, Lewandowsky18, Stark2018ib, Stark2020ib, Meidlinger2020irr} have to be decomposed into a series of cascaded small LUTs to achieve a manageable memory usage, where each LUT has two inputs and single output symbols.
This degrades the performance of these decoders due to the loss of mutual information \cite{cover2012elements}.

To improve the error performance, the mutual information-maximizing quantized BP (MIM-QBP) decoder was proposed in \cite{He2019qbp}, and elaborated in \cite{he2019mutual} for regular LDPC codes.
Instead of considering only two input symbols in the LUT design, the MIM-QBP decoder reconstructs each incoming message into a specific integer number and computes all possible combinations of the messages for the node updates. 
Then an optimal quantizer is determined by the DP method in \cite{he2021dynamic} to generate the threshold sets for the node updates, which maps the combined messages to the outgoing messages.
As shown in \cite{he2019mutual}, the MIM-QBP decoder with only 3-bit precision can outperform the floating-point BP decoder at the high signal-to-noise ratio (SNR) region for regular LDPC codes.
Meanwhile, Wang {\it et al.} proposed a decoding method in \cite{wang2020rcq} based on the reconstruction-computation-quantization (RCQ) approach to mitigate the performance degradation caused by LUT decomposition.
Similar to the MIM-QBP decoder, the RCQ decoder maps each input message to a corresponding log-likelihood ratio (LLR) and computes the combined messages based on either the BP or the MS algorithm \cite{Meidlinger2020irr}.
However, the quantization threshold sets for the RCQ decoder are designed in a sub-optimal way for less computational complexity.
The reconstructed LLR values are naturally with floating-point precision and require extra quantization for implementing the decoder with finite precision.

On the other hand, the LUTs of the LDPC decoders \cite{Nguyen2018faid, Ghaffari2019pfaid, Romero2016mim, Lewandowsky18, Stark2018ib, Stark2020ib, Meidlinger2020irr} and \cite{wang2020rcq} are commonly designed by considering the flooding schedule \cite{Kschischang1998fsd}, where all the CN and VN messages are simultaneously updated and propagated along the edges of the Tanner graph \cite{Tanner81}.
However, the flooding schedule requires a relatively large number of iterations for a satisfying decoding error performance.
This results in a slow convergence speed and high hardware complexity \cite{Mansour2003cnLayer}.
To accelerate the convergence speed of the decoder, the layered schedules have been investigated in \cite{Mansour2003cnLayer,Zhang2009cnlayer,Zhang2005vnLayer,Cui2011vnLayer,Aslam2017vnLayer} to generate the messages in a sequential order by using the latest available information.
In particular, the horizontal layered schedules in \cite{Mansour2003cnLayer,Zhang2009cnlayer} update the CN messages sequentially while the vertical layered schedules in \cite{Zhang2005vnLayer,Cui2011vnLayer,Aslam2017vnLayer} was based on a sequential update of the VNs' messages.	 
The analytical studies in \cite{Mansour2003cnLayer,Zhang2009cnlayer,Zhang2005vnLayer,Cui2011vnLayer,Aslam2017vnLayer} have proved that the layered schedules converge twice as fast as the flooding schedule at no cost in decoding complexity.
However, designing the MIM-LUT LDPC decoders for the layered schedule is still a challenging problem.
%
Although an overview of the layered LUT-based decoder using the information bottleneck method is presented in \cite{Mohr2021iblayer}, there exist few works to provide a systematic way to construct the layered LUT-based LDPC decoders in detail.

In this paper, we present a generalized framework of the mutual information-maximizing (MIM) quantized decoding for LDPC codes with arbitrary degree distributions.
The proposed decoding method only requires simple mappings and fixed-point additions, and hence facilitates easier hardware implementation.    
To further speed up the convergence of the decoder for quasi-cyclic LDPC (QC-LDPC) codes, we propose the framework of the MIM layered decoder, which can significantly benefit the applications with low-latency requirement such as the vehicular communications.
The main contributions of this work are summarized below:
\begin{itemize}
	\item We generalize the framework of the MIM quantized decoding for the LDPC codes with arbitrary degree distributions.
	In particular, we derive the density evolution (DE) by using the MIM quantization \cite{he2021dynamic} to construct the LUTs of the reconstruction and quantization mappings for the MIM-QBP decoder.
	To reduce the computational complexity of decoding, we approximate the CN update of the MIM-QBP decoder by the MS algorithm and obtain the MIM quantized MS (MIM-QMS) decoder. 
	We also compare the differences between the proposed MIM quantized decoders and that of the existing LUT decoder variants.
	\item We analyze the decoding complexity of the proposed MIM quantized decoders.
	We point out that the MIM quantized decoders have slightly increased computational complexity but significantly reduce the memory requirement compared to the MIM-LUT decoders. 
	\item We compare the error performance and the convergence behaviors of the MIM quantized decoders with 3-bit and 4-bit precision to those of the different MIM-LUT decoders over the additive white Gaussian noise (AWGN) channels.
	Simulation results show that the MIM quantized decoders achieve a faster convergence compared to the state-of-the-art MIM-LUT decoders for small decoding iterations.
	Moreover, the MIM-QBP decoder can outperform the benchmark decoders with the same message precision in the waterfall region.
	In addition, the 4-bit MIM-QMS decoder can even surpass the floating-point BP decoder in the error-floor region for the length-$1296$ IEEE 802.11n standard LDPC code with rate $1/2$.
	\item We further develop the layered MIM-DE (LMIM-DE) and use it to design the LUTs of the reconstruction and quantization mappings at each layer and each iteration for the MIM layered QMS (MIM-LQMS) decoder.
	To reduce the memory requirement for storing the LUTs by different layers and iterations, we further optimize the LUTs and make them only vary with different iterations with neglectable degradation of mutual information.
	\item We evaluate the error performance of the constructed MIM-LQMS decoders with 3-bit and 4-bit precision over the AWGN channels and the fast fading channels.
	Simulation results show that the 4-bit MIM-LQMS decoder achieves error performance less than $0.3$ dB away from the floating-point layered BP decoder in the moderate-to-high SNR regions.
	\vspace{-0.1cm}
	%
\end{itemize}

The rest of this paper is organized as follows.
In Section II, we review the basic framework of the MIM-QBP decoder for decoding regular LDPC codes.
In Section III, we present the generalized framework of the MIM quantized decoders for any ensemble of the LDPC codes that can be characterized by degree distributions.
We also highlight the differences in the design of the proposed decoder compared to the existing MIM-LUT decoders.
The decoding complexity of the MIM quantized decoders is discussed in Section IV.
We also investigate the error performance of the proposed MIM quantized decoders and compare it with that of different LUT decoder variants.
Section V introduces the design of the MIM-LQMS decoder, presents the LUT optimization method, and also illustrates the simulation results of the proposed MIM-LQMS decoders.
Section VII concludes the paper.
\begin{table*}[t!]
	\normalsize
	\renewcommand{\arraystretch}{1.1}
	\caption{List of Notations for the MIM-QBP Decoder}
	\label{notations}
	\centering
	\begin{tabular}{|l|l|}
		\hline
		$q_m$       & The precision (bit widths) of outer messages                              \\ \hline
		$q_c$       & The precision (bit widths) of inner messages at CNs                       \\ \hline
		$q_v$       & The precision (bit widths) of inner messages at VNs                      	\\ \hline
		$\mc{X}$    & The alphabet set of coded bits                                        	\\ \hline
		$X$         & A random variable of the coded bit                                    	\\ \hline
		$\mc{L}$    & The alphabet set of the channel quantization output                       \\ \hline
		$L$ and $l$ & A random variable and its realization of the channel quantization output  \\ \hline
		$P_{L|X}$   & The channel transition probability for output $L$ conditioned on input $X$                      	\\ \hline
		$\mc{R}$    & The alphabet set of variable-to-check (V2C) messages                  	\\ \hline
		$R$ and $r$ & A random variable and its realization of the V2C message              	\\ \hline
		$P_{R|X}$   & The probability mass function (pmf) of V2C message $R$ conditioned on $X$ \\ \hline
		$\mc{S}$    & The alphabet set of check-to-variable (C2V) messages                  	\\ \hline
		$S$ and $s$ & A random variable and its realization of the C2V message              	\\ \hline
		$P_{S|X}$   & The pmf of C2V message $S$ conditioned on $X$                            	\\ \hline
	\end{tabular}
\end{table*}
\section{Preliminaries}
\label{section: preliminaries}
\subsection{Notations}
In this paper, we use the calligraphy capitals to define an alphabet set.
The normal capitals denote the random variables.
The lower-case letters denote the realization of a random variable.
Boldface letters are used to define a vector.
For convenience, we list the notations of the MIM-QBP decoder in TABLE \ref{notations}.
\subsection{The MIM-QBP decoder for ($d_v$, $d_c$) LDPC codes}
\begin{figure}[h!]
	\centering
	\includegraphics[width=2.75in]{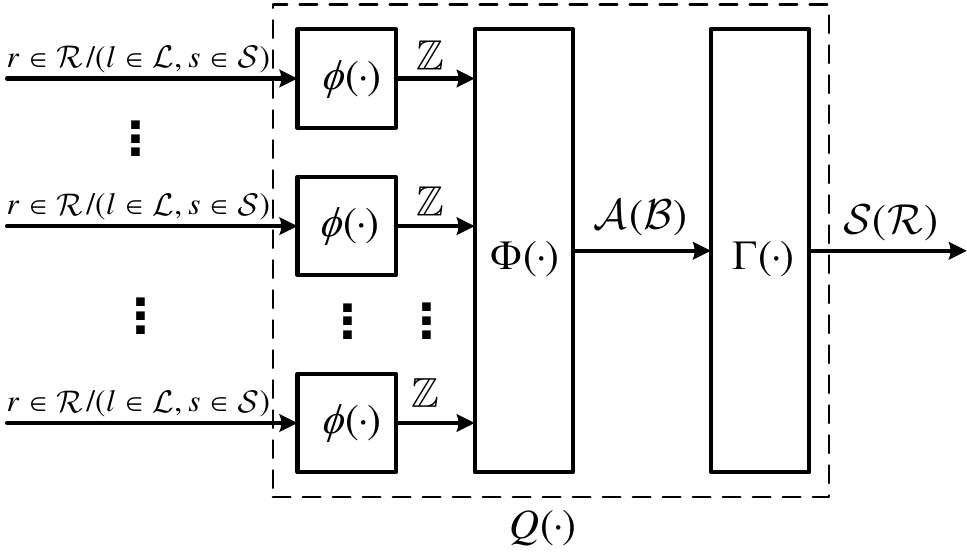}
	\caption{The framework of the MIM-QBP decoder for decoding the ($d_v$, $d_c$) LDPC codes.}
	\label{fig: general_struct_QBP}
\end{figure}
We briefly review the framework of the MIM-QBP decoder for ($d_v$, $d_c$) LDPC codes in \cite{he2019mutual}.
As shown in Fig. \ref{fig: general_struct_QBP}, the node (both CN and VN) updates of the MIM-QBP decoder is divided into three steps: reconstruction, calculation, and quantization.
We represent these three steps correspondingly by three functions, i.e., the reconstruction function $\phi(\cdot)$, the computing function $\Phi(\cdot)$, and the quantization function $\Gamma(\cdot)$.
We call the messages at the reconstruction function input and the quantization function output as outer messages, and those related to the computing function as inner messages.
The precision of the MIM-QBP decoder can be fully described by the bit widths $q_m$, $q_c$, and $q_v$.
With the preset value of $q_m$, $q_c$, and $q_v$, the MIM-QBP decoder is designed under the discrete memoryless channel (DMC).
Without loss of generality, we obtain the DMC from a binary-input AWGN channel, where the coded bit $X$ takes values from $\mc{X} = \{0,1\}$.
In the design procedure, the MIM-QBP decoder associates the LLRs of the messages exchanged between CNs and VNs to the symbols in the finite alphabet sets.
More specifically, we let $\mc{L} = \{0, 1, \ldots, |\mc{L}|-1\}$ be the alphabet set of the channel output, where $|\cdot|$ is the size of the alphabet set.
We also consider the alphabet sets of V2C and C2V messages as $\mc{R} = \{0, 1, \ldots, |\mc{R}|-1\}$ and $\mc{S} = \{0, 1, \ldots, |\mc{S}|-1\}$, respectively.
In general, we have $|\mc{R}|=|\mc{S}|=|\mc{L}|=2^{q_m}$ and assume that the symbols in the alphabet set are sorted in a way such that their associated LLRs are in descending order.
The design of the node updates for the MIM-QBP decoder is presented in detail as follows.
\subsubsection{Reconstruction $\phi(\cdot)$}
The reconstruction function $\phi(\cdot)$ maps each $q_m$-bit outer message to a specific number in the integer domain $\mathbb{Z}$.
We denote the reconstruction function for the channel output by $\phi_{ch}$, and those for the CN and VN updates by $\phi_c$ and $\phi_v$, respectively.
For the CN update, $\phi_c$ in \cite{he2019mutual} is designed to indicate the unreliability of the V2C message $R$ such that larger $|\phi_c|$ means smaller magnitude of the associated LLR value.
%
%
Let $g(r) = P_{R|X}(r|0) - P_{R|X}(r|1)$ for $r \in \mc{R}$.
Then we have \cite{he2019mutual}
\begin{equation}
	\label{phi_c}
	\setstretch{1.3}
	{\phi _c}(r) = \left\{ {\begin{array}{*{20}{l}}
			sgn(g(r)) \cdot \text{max} \{1, &\\  \, \left\lfloor {{\textstyle{{{2^{{q_c} - 1}} - 1} \over {{d_c}}}} \cdot \left| {\log (|g(r)|)} \right|/\alpha  + 0.5} \right\rfloor\},&{{\rm{|}}g(r){\rm{| \ne 0}},}\\
			{\left\lfloor {({2^{{q_c} - 1}} - 1)/{d_c}} \right\rfloor ,}&{{\rm{|}}g(r){\rm{| = 0}},}
	\end{array}} \right.
\end{equation}
where $\alpha  = \text{max} \{ {\left|\log (|g(r)|)\right|}:r \in {\mc{R}},\,g(r) \ne 0\}$ is the normalized factor to ensure $\sum\nolimits_{k = 1}^{{d_c}} {\left| {{\phi _c}({r_k})} \right|}  \le {2^{{q_c} - 1}} - 1$, and $sgn(\cdot)$ refers to the sign of the message such that
\begin{equation}
	\setstretch{1}
	sgn(\alpha) =
	\begin{cases}
		-1, & \alpha < 0,\\
		1, & \alpha \ge 0.
	\end{cases} \nonumber
\end{equation}
For the VN update, $\phi_{ch}$ and $\phi_{v}$ in \cite{he2019mutual} are designed to reflect the reliability of the C2V message $S$ and the channel output $L$, where the larger magnitude of the reconstruction functions means the larger magnitude of the associated LLR values.
%
%
Let $h(l) = \log ({P_{L|X}}(l|0)/{P_{L|X}}(l|1))$ for $l \in \mc{L}$, and $h(s) = \log ({P_{S|X}}(s|0)/{P_{S|X}}(s|1))$ for $s \in \mc{S}$, respectively.
The reconstruction functions $\phi_{ch}$ and $\phi_v$ is given by \cite{he2019mutual}
\begin{equation}
	\label{phi_v_ch}
	\setstretch{1.3}
	\left\{ {\begin{array}{*{20}{l}}
			{{\phi _{ch}}(l) = sgn(h(l)) \cdot \left\lfloor {{\textstyle{{{2^{{q_v} - 1}} - 1} \over {{d_v} + 1}}} \cdot |h(l)|/\beta  + 0.5} \right\rfloor ,}\\
			{{\phi _v}(s) = sgn(h(s)) \cdot \;\left\lfloor {{\textstyle{{{2^{{q_v} - 1}} - 1} \over {{d_v} + 1}}} \cdot |h(s)|/\beta  + 0.5} \right\rfloor ,}
	\end{array}} \right.
\end{equation}
where $\beta  = \text{max} (\{ |h(s)|:s \in \mc{S}\}  \cup \{ |h(l)|:l \in \mc{L}\} )$ is the normalized factor to refine $|{\phi _{ch}}(l)| + {d_v} \cdot |{\phi _v}(s)| \le {2^{{q_v} - 1}} - 1$.
\subsubsection{Calculation $\Phi(\cdot)$}
Following reconstruction, the computing function $\Phi(\cdot)$ is used to calculate the inner message with $q_c$-bit or $q_v$-bit precision.
We denote the computing functions for the CN update and the VN update by $\Phi_c$ and $\Phi_v$, respectively.
Let $\mb{R} \in \mc{R}^{d_c - 1}$ be the vector of V2C messages at a CN of degree $d_c$.
For a realization $\mb{r} = (r_1,r_2,\ldots,r_{d_c-1}) \in \mc{R}^{d_c - 1}$, $\Phi_c$ in \cite{he2019mutual} is designed based on the BP algorithm as
\begin{equation}
	\label{eqn: def of Phi_c}
	{\Phi _c}(\mb{r}) = \left( {\prod\limits_{k = 1}^{{d_c} - 1} s gn({\phi _c}({r_k}))} \right)\sum\limits_{k = 1}^{{d_c} - 1} | {\phi _c}({r_k})|.
\end{equation}
Define $\mb{S} \in \mc{S}^{d_v - 1}$ as the vector of C2V messages at a VN of degree $d_v$.
For a realization $\mb{s} = (s_1,s_2,\ldots,s_{d_v-1}) \in \mc{S}^{d_v - 1}$, $\Phi_v(\cdot)$ in \cite{he2019mutual} operates as
\begin{equation}
	\label{eqn: def of Phi_v}
	\Phi_v(l, \mb{s}) = \phi_{ch}(l) + \sum_{k = 1}^{{d_v} - 1} \phi_v(s_k).
\end{equation}
We denote the alphabet sets of the inner message for the CN update and the VN update by integer sets $\mc{A} = \{a_1, a_2, \ldots, a_{|\mc{A}|}\} = \{\Phi_c(\mb{r}): \mb{r} \in \mc{R}^{d_c - 1}\}$, and $\mc{B} = \{b_1, b_2, \ldots, b_{|\mc{B}|}\} = \{\Phi_v(l, \mb{s}): (l, \mb{s}) \in \mc{L} \times \mc{S}^{d_v - 1}\}$, respectively.
The elements in $\mc{A}$ and $\mc{B}$ are listed in a certain order such that their associated LLRs are (or almost) decreasing.
Specifically, the elements in $\mc{A}$ are labeled to satisfy $a_1 \succ a_2 \succ \cdots \succ a_{|\mc{A}|}$, where $\succ$ refers to a binary relation on $\mathbb{Z}$ such that $\alpha \succ \beta \iff sgn(\alpha) > sgn(\beta) \text{~or~} (sgn(\alpha) = sgn(\beta) \text{~and~} \alpha < \beta)$ for $\alpha, \, \beta \in \mathbb{Z}$.
Here we give an example of the alphabet set $\mc{A}$ of size $|\mc{A}|=6$ to clarify the binary relation $\succ$ such that $\mc{A} = \{a_1=4, a_2=10, a_3 = 30, a_4 = -30, a_5=-10, a_6=-4\}$.
For the alphabet set $\mc{B}$, we have $b_{1} > b_2 > \cdots > b_{|\mc{B}|}$.
\subsubsection{Quantization $\Gamma(\cdot)$}
Based on $\mc{A}$ and $\mc{B}$, we quantize the $q_c$ and $q_v$ bits inner message to $q_m$ bits outer message by an optimal $q_m$-bit sequential deterministic quantizer (SDQ) \cite{he2021dynamic}. 
Assume that the input message $Y$ takes values from $\mc{Y} = \{y_1, y_2, \ldots, y_N\}$, where we have $\mc{Y} = \mc{A}$ for the CN update and $\mc{Y} = \mc{B}$ for the VN update.
A $q_m$-bit SDQ for $Y$ can be expressed as
\begin{equation}
	\setstretch{1}
	Q(x) = \left\{ {\begin{array}{*{20}{l}}
			{0,}&{x \in {\mkern 1mu} \{ {y_{1}},{y_{2}}, \ldots ,{y_{{\lambda _1}}}\} ,}\\
			{1,}&{x \in \{ {y_{{\lambda _1} + 1}},{y_{{\lambda _1} + 2}}, \ldots ,{y_{{\lambda _2}}}\} ,}\\
			\vdots & \vdots \\
			{{2^{{q_m}}} - 1,}&{x \in \{ {y_{{\lambda _{{2^{q_m}} - 1}} + 1}},{y_{{\lambda _{{2^{q_m}} - 1}} + 2}}, \ldots ,{y_{N}}\}, }
	\end{array}} \right.
\end{equation}
where $0 < \lambda_1 < \cdots < \lambda_{2^{q_m}-1} < N$ are the $2^{q_m}-1$ element indices to form the threshold set $\Gamma = \{\gamma_k: \, \gamma_k=y_{\lambda_k}, \, k=1,2,\ldots,2^{q_m}-1\}$.
The threshold set of the optimal $q_m$-bit SDQ is determined by the DP method \cite{he2021dynamic}, which considers the MIM criterion such that the mutual information between the coded bit $X$ and the C2V (resp. V2C) message $S$ (resp. $R$) can be maximized.
We denote the MIM quantization for the CN and VN updates by
\begin{equation}
	\mathop {\arg \max }\limits_Q I(X;S), \,\text{and}\, \mathop {\arg \max }\limits_Q I(X;R),
\end{equation}
respectively.
We denote the threshold set for the CN update by $\Gamma_c$, and denote the threshold set for the VN update by $\Gamma_v$, which operate as
\begin{equation}
	\label{gamma}
	\setstretch{1}
		\Gamma _c(x) \!=\! \left\{ {\begin{array}{*{20}{l}}
				{\!\!0,}&{\!\!\!\!x \!\succeq\! {\gamma _1},}\\
				{\!\!{2^{{q_m}}}\!\! -\!\! 1,}&{\!\!\!\!{\gamma _{{2^{{q_m}}} - 1}} \!\succ\! x,}\\
				{\!\!k,}&{\!\!\!\!{\gamma _k} \!\succ\! x \!\succeq\! {\gamma _{k + 1}},}
		\end{array}} \right.
		\hspace{-0.3cm}
		{\Gamma _v}(x) \!=\! \left\{ {\begin{array}{*{20}{l}}
				{\!\!0,x \!\ge\! {\gamma _1},}&{\!\!\!}\\
				{\!\!{2^{{q_m}}} \!\!-\!\! 1,x \!<\! {\gamma _{{2^{{q_m}}} - 1}},}&{\!\!\!}\\
				{\!\!k,{\gamma _k} \!>\! x \!\ge\! {\gamma _{k + 1}}.}&{\!\!\!}
		\end{array}} \right. 
	\end{equation}
	Here $\succeq$	is defined as a binary relation on $\mathbb{Z}$ such that $\alpha \succeq \beta \iff \alpha \succ \beta \text{~or~} \alpha = \beta$ for $\alpha, \beta \in \mathbb{Z}$.
	To design the MIM-QBP decoder, we first discretize a binary-input AWGN channel by considering the MIM quantization.
	The threshold set for quantizing the channel output is denoted by $\Gamma_{ch}$, which has the same operations as $\Gamma_{v}$.
	The channel parameter of the AWGN channel, denoted by $\sigma_d$, is the design noise standard deviation assumed to be known at the receiver.
	Based on the constructed DMC, we perform reconstruction, calculation, and quantization for the node updates and track the evolution of $P_{R|X}$ and $P_{S|X}$ to construct the reconstruction functions and threshold sets at each iteration accordingly.
	\footnote{In this paper, we consider reconstruction functions and threshold sets varying with iterations.
		Here we do not specify these notations for the associated iterations because once the decoder design for one iteration is determined, it is straightforward to generalize it for the other iterations.
	}
	The associated reconstruction functions and threshold sets are stored as the LUTs for implementing the decoding.
	Define the mapping functions $Q_{c}: \mc{R}^{d_c - 1} \to \mc{S}$ and $Q_v: \mc{L} \times \mc{S}^{d_v - 1} \to \mc{R}$ for the CN and the VN updates, respectively, which consist of the above three steps.
	In addition, the mapping function for bit decision $Q_{e}: \mc{L} \times \mc{S}^{d_v} \to \mc{X}$ is designed similarly to $Q_v$ based on the same reconstruction functions $\phi_v$ and $\phi_{ch}$.
	However, the alphabet set of the inner messages becomes $\{\Phi_v(l, \mb{s}) : (l, \mb{s}) \in \mc{L} \times \mc{S}^{d_v}\}$ and we have $|\Gamma_v|=1$ for hard decision.
	We ignore the details here due to space limitations.
	\section{Generalized MIM Quantized Decoding for Irregular LDPC Codes}
	\label{section: irregular design}
	\begin{figure*}[h!]
		\centering
		\includegraphics[width=3.9in]{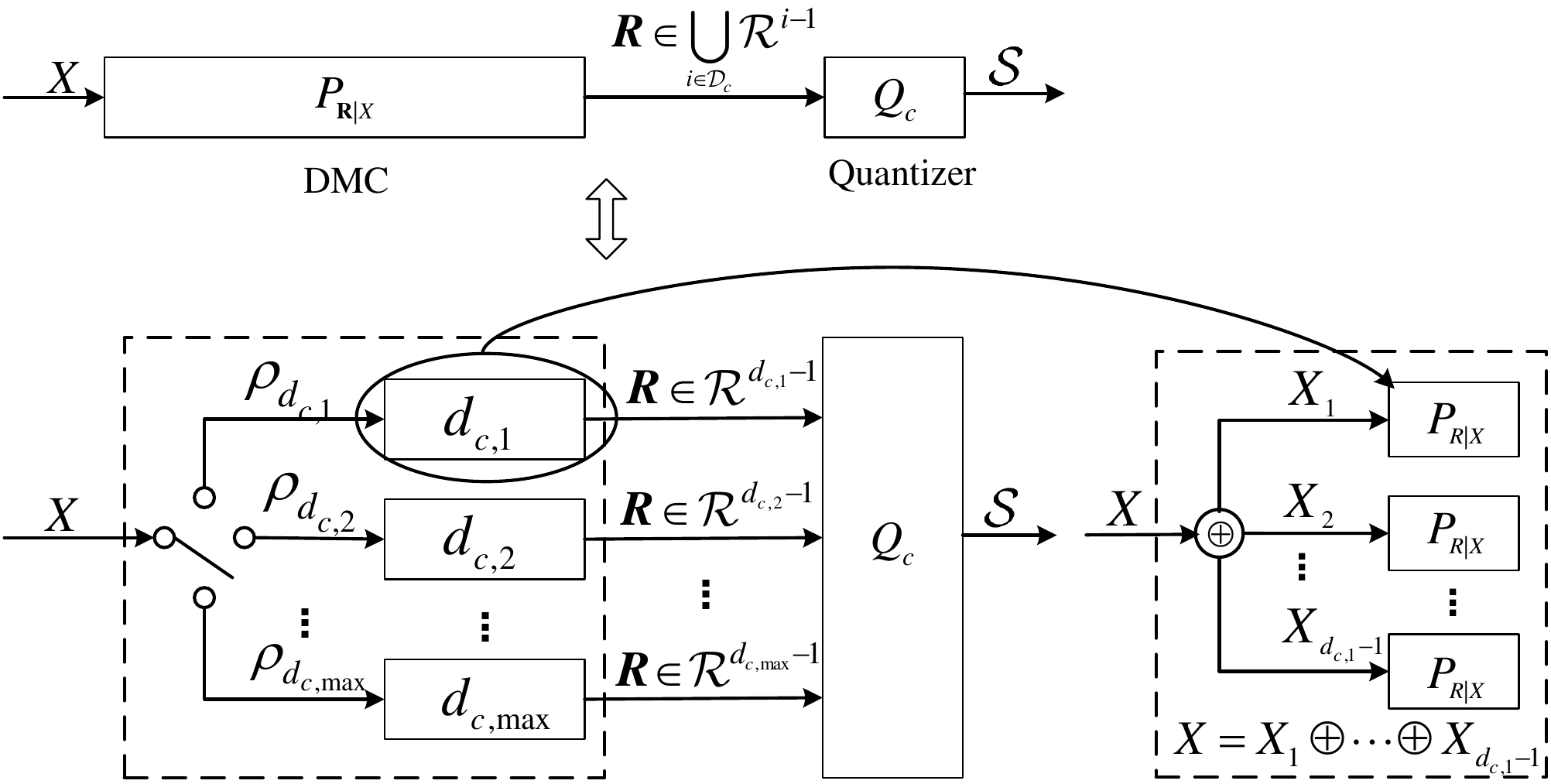}
		\caption{The MIM-QBP decoder design at CN and its equivalence to quantization for DMC.} 
		\label{fig: CN_update_channel_irregular}
	\end{figure*}
	In this section, we extend the design principle of the MIM-QBP decoder to irregular LDPC codes and propose the generalized framework of the MIM quantized decoding for the LDPC codes with arbitrary degree distributions.
	As shown in Section \ref{section: preliminaries}, the reconstruction functions and threshold sets of the MIM-QBP decoder are constructed by quantizing the pmf between the coded bits and the messages sent from the CNs or VNs.
	Therefore, we first derive the MIM density evolution (MIM-DE) to trace the evolution of the pmf when decoding the irregular LDPC codes.
	The main difference between our MIM-DE and the conventional DE \cite{Richardson01capacity} is that we consider the node update with reconstruction-calculation-quantization architecture, and aim to maximize the mutual information between the coded bits and the message from the CN or VN, while the conventional DE uses uniform quantization and considers the asymptotic error probability.
	To reduce the decoding complexity, we further simplify the CN update of the MIM-QBP decoder by the MS algorithm \cite{Meidlinger2020irr}, which leads to the MIM-QMS decoder.
	We refer to either the MIM-QBP decoder or the MIM-QMS decoder as the MIM quantized decoder.
	In addition, we also compare the proposed design method of the MIM quantized decoders with those of the prior state-of-art LUT decoder variants.

	Denote the sets of the CN and VN degrees by $\mc{D}_c = \{d_{c, 1}, d_{c, 2}, \ldots, d_{c, \text{max}}\}$ with $d_{c, 1} < d_{c, 2} < \cdots < d_{c, \text{max}}$ and $\mc{D}_v = \{d_{v, 1}, d_{v, 2}, \ldots, d_{v, \text{max}}\}$ with $d_{v, 1} < d_{v, 2} < \cdots < d_{v, \text{max}}$, respectively.
	An ensemble of binary LDPC codes can be characterized by the degree distributions as
	\begin{equation}
		\label{degreeDis}
		\rho(x) = \sum_{i \in \mc{D}_c} \rho_i x^{i-1}, \, \theta(x) = \sum_{j \in \mc{D}_v} \theta_j x^{j-1},
	\end{equation}
	where $\rho_i$ and $\theta_j$ are the fractions of edges incident to the CNs with degree-$i$ and the VNs with degree-$j$, respectively.
	In the following, we illustrate the design of the MIM-QBP decoder for irregular LDPC codes.
	\subsection{MIM-DE at CN Update}
	\label{subsection: cn}
	As depicted in Fig. \ref{fig: CN_update_channel_irregular}, the design of the mapping function $Q_c$ for the CN update of the irregular LDPC codes is equivalent to solving the DMC quantization problem as
	\begin{equation}
		Q_{c}: \bigcup\limits_{i \in {\mc{D}_c}} {{\mc{R}^{i - 1}}}   \to \mc{S} \nonumber
	\end{equation}
	such that $I(X;S), \,S \in \mc{S}$ is maximized.
	Note that the DMC output can be characterized by $|\mc{D}_c|$ independent alphabets corresponding to different CN degrees, where we have $|\mc{D}_c| = 1$ for regular LDPC codes.
	From Section \ref{section: preliminaries}, we know that updating the degree-$i$ ($i \in \mc{D}_c$) CNs requires $i-1$ V2C messages from its connected VNs.  
	Define $\mb{R} \in \mc{R}^{i-1}$ as the vector of the V2C messages sent to the degree-$i$ CN.
	Based on the independence and identical distribution (i.i.d) assumption in the DE \cite{Richardson01capacity}, each V2C message has the same pmf $P_{R|X}$.
	Therefore, the joint pmf $P_{\mb{R}|X}$ of $\mb{R}$ conditioned on the coded bit $X$ is given by
	\begin{equation}
		\label{eqn: joint P_R|X}
		P_{\mb{R}|X}(\mb{r}|x) = \left(\frac{1}{2}\right)^{i - 2} \sum_{\mb{x}: \oplus \mb{x} = x} \prod_{k = 1}^{i-1} P_{R|X}(r_k|x_k),
	\end{equation}
	where $x \in \mc{X}$ is a realization of coded bit $X$, and $\mb{x} = {\rm{ }}({x_1},{x_2}, \ldots ,{x_{i - 1}})$ is the vector of input coded bits associated with the CN's connecting VNs.
	Note that $\oplus \mb{x} = x$ refers to the case where the checksum of the CN is satisfied. 

	Let $\mc{A}_i = \{a_{i, 1}, a_{i, 2}, \ldots, a_{i, |\mc{A}_i|}\} = \{\Phi_c(\mb{r}) : \mb{r} \in \mc{R}^{i-1}, i \in {\mc{D}_c}\}$ be the alphabet set of the inner message calculated for the degree-$i$ CNs, where $\Phi_c(\cdot)$ is given by (\ref{eqn: def of Phi_c}).
	Denoted by $A_i$ the random variable for the inner message taking values from $\mc{A}_i$, we obtain the conditional probability
	\begin{equation}
		\label{Prx_each}
		P_{A_i|X}(a_{i,k}|x) = \sum_{\mb{r} \in \mc{R}^{i-1}, \hfill\atop \Phi_c(\mb{r}) = a_{i,k}} P_{\mb{R} | X}(\mb{r} | x), \, k=1,2,\ldots,|\mc{A}_i|.
	\end{equation}
	According to (\ref{degreeDis}), we obtain $\mc{A} = \cup_{i \in \mc{D}_c} \mc{A}_i$ by considering all $|\mc{D}_c|$ independent alphabets for the CN update.
	Define $A \in \mc{A}$ as the inner message computed for all CNs of different degrees. 
	Based on the fraction $\rho_i$ of the degree-$i$ CNs, we have $A \in \mc{A}_i$ with probability $\rho_{i}$. 
	Let $P_{A|X}$ denote the pmf of $A$ conditioned on the coded bit $X$, which is given by
	\begin{equation}
		\label{Prx_all}
		P_{A|X}(a|x) = \sum\limits_{i \in {\mc{D}_c}} {{\rho _{i}}\cdot{P_{{A_i}|X}}} (a|x),
	\end{equation}
	where $a \in \mc{A}$ is a realization of $A$.
	Based on $\mc{A}$ and $P_{A|X}$, we perform the MIM quantization by using the DP method \cite{he2021dynamic} to determine the threshold set $\Gamma_c$ and the associated pmf ${P_{S|X}}$, which is denoted by
	\begin{equation}
		\label{eqn: P_(S|X) lambda}
		[{P_{S|X}},{\Gamma _c}] = \textbf{DP}(\mc{A}, \, {P_{A|X}}).
	\end{equation}
	Consequently, the reconstruction functions $\phi_{ch}$ and $\phi_v$ for current iteration can be constructed according to (\ref{phi_v_ch}) for the VN update.
	\begin{figure*}[h!]
		\centering
		\includegraphics[width=3.9in]{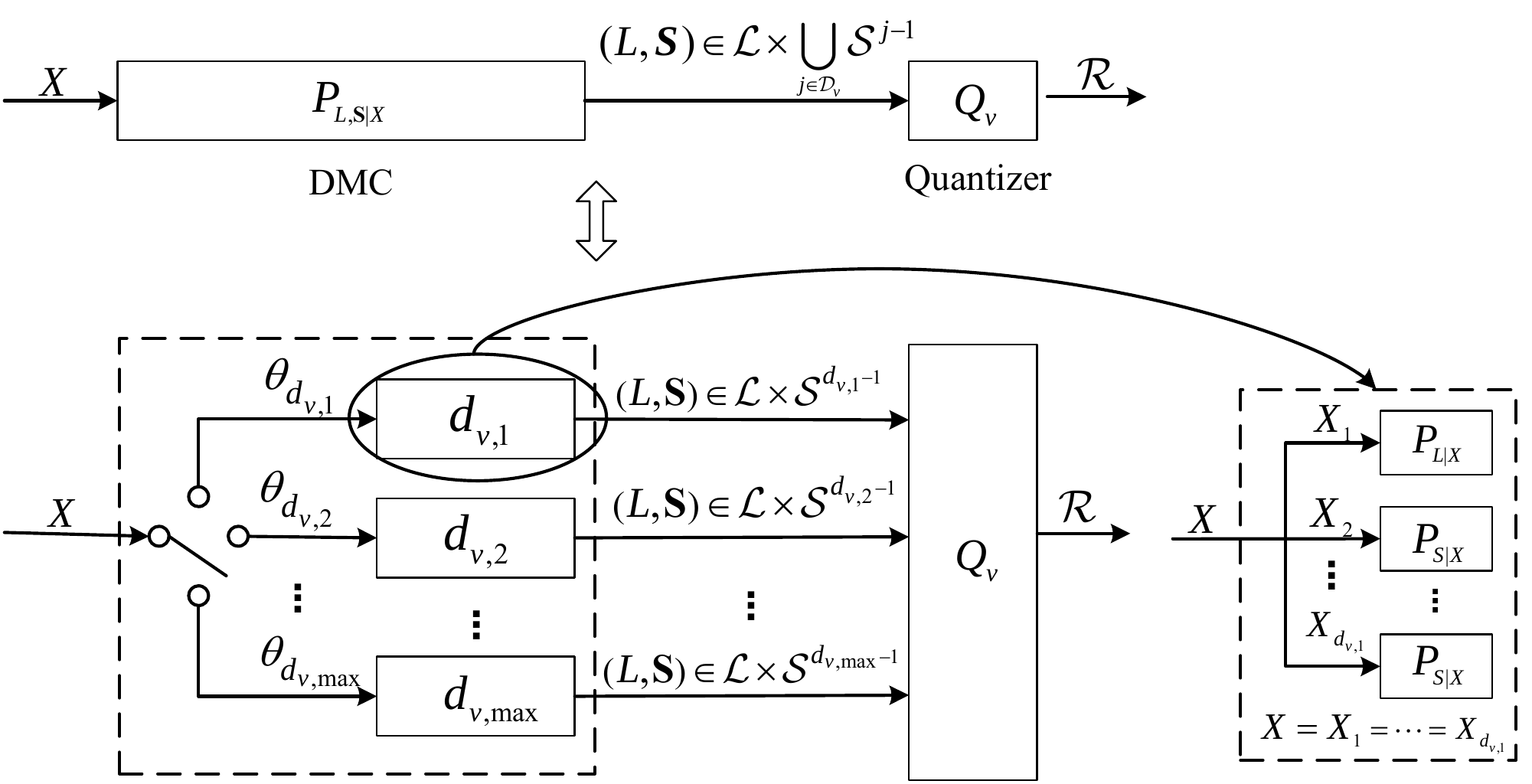}
		\caption{The MIM-QBP decoder design at VN and its equivalence to quantization for DMC.}
		\label{fig: VN_update_channel_irregular}
	\end{figure*}
	\subsection{MIM-DE at VN Update}
	As shown by Fig. \ref{fig: VN_update_channel_irregular}, the design of the mapping function $Q_v$ for the VN update of irregular LDPC codes is equivalent to solving the quantization problem over DMC such that
	\begin{equation}
		Q_v:\, \mc{L} \times \bigcup\limits_{j \in {\mc{D}_v}} {{\mc{S}^{j - 1}}}  \to \mc{R}, \nonumber
	\end{equation}
	which maximize $I(X;R), \,R \in \mc{R}$.
	Note that the DMC output is classified into $|\mc{D}_v|$ independent alphabets corresponding to different VN degrees, where $|\mc{D}_v| = 1$ is a special subcase corresponding to regular LDPC codes.
	To update a degree-$j$ ($j \in \mc{D}_v$) VN, we need the message from the channel output and $j-1$ C2V messages from its connected CNs.  
	Define $\mb{S} \in \mc{S}^{j-1}$ as the vector of the C2V messages sent to the degree-$j$ VN.
	We denote the vector of the C2V messages and the channel output by $(L,\mb{S}) \in \mc{L} \times \mc{S}^{j-1}$.
	Similar to the case of the CN update, we have the pmf $P_{S|X}$ for each C2V message and $P_{L|X}$ for the channel output.
	As a result, the joint pmf $P_{L,\mb{S}|X}$ can be computed by
	\begin{equation}
		\label{Plsx}
		P_{L,\mb{S}|X}(l, \mb{s}|x) =  P_{L|X}(l|x) \prod_{k = 1}^{j-1} P_{S|X}(s_k|x), \, j \in \mc{D}_v.
	\end{equation}

	Let $\mc{B}_j = \{b_{j, 1}, b_{j, 2}, \ldots, b_{j, |\mc{B}_j|}\} = \{\Phi_v(l, \mb{s}) : l \in \mc{L}, \mb{s} \in \mc{S}^{j-1}\}$ be the alphabet set of the inner message computed for the VNs of degree-$j$, where $\Phi_v(\cdot)$ is given by (\ref{eqn: def of Phi_v}).
	Denoted by $B_j$ the random variable for the inner message taking values from $\mc{B}_j$, we obtain
	\begin{equation}
		\label{Pbx_each}
		P_{B_j|X}(b_{j,k}|x) = \sum_{(l, \mb{s}) \in \mc{L} \times \mc{S}^{j-1}, \hfill\atop \Phi_v(l, \mb{s}) = b_{j,k}} P_{L, \mb{S} | X}(l, \mb{s} | x), \, k=1,2,\ldots,|\mc{B}_j|. 	\end{equation}
	\\Based on (\ref{degreeDis}), we obtain $\mc{B} = \cup_{j \in \mc{D}_v} \mc{B}_j$ by considering all $|\mc{D}_v|$ independent alphabet sets for the VN update.
	Define $B \in \mc{B}$ as the random variable for the inner message calculated by considering all VNs of different degrees.
	With the fraction $\theta_{j}$ of the degree-$j$ VNs, we have $B \in \mc{B}_j$ with probability $\theta_{j}$.
	Let $P_{B|X}$ denote the pmf of $B$ conditioned on the coded bit $X$, which is given by
	\begin{equation}
		\label{Pbx_all}
		P_{B|X}(b|x) = \sum_{j \in \mc{D}_v} \theta_{j} \cdot P_{B_j|X}(b|x),
	\end{equation}
	where $b \in \mc{B}$ is a realization of $B$.
	In the end, the MIM quantization \cite{he2021dynamic} is conducted based on $\mc{B}$ and $P_{B|X}$, which is denoted by
	\begin{equation}
		\label{eqn: P_(R|X) lambda}
		[{P_{R|X}},{\Gamma _v}] = \textbf{DP}(\mc{B},{P_{B|X}}).
	\end{equation}
	As a result, the threshold set $\Gamma_v$ for the current iteration is determined and the reconstruction function $\phi _c$ for the next iteration can be built based on (\ref{phi_c}) and ${P_{R|X}}$.
	In addition, the mapping function for the bit decision is $Q_{e}: \, \mc{L} \times \bigcup\limits_{j \in {\mc{D}_v}} {{\mc{S}^{j}}}  \to \mc{X}$ for irregular LDPC codes.
	Since the design of $Q_e$ is similar to that of $Q_v$, we ignore the details to save space.
	Note that the proposed MIM-DE does not consider the all-zero codeword assumption because it may result in asymmetrical pmf when using the DP method to realize the quantization \cite{Romero2016mim}.\vspace{-0.3cm}
	\begin{table}[h!]
		\begin{algorithm}[H]
			\normalsize
			\setstretch{1.15}
			\caption{Deisgn Flow of the MIM-QBP Decoder}
			\label{alg:designFlow}
			\begin{algorithmic}[1]
				\REQUIRE ${\sigma _d}, \, \rho(x), \, \theta(x), \, q_m, \, q_c, \, q_v, \, I_\text{max}$.
				\ENSURE $\Gamma _{ch}, \, \Gamma _{c}^{(t)}, \, \Gamma _{v}^{(t)}, \, \phi _{c}^{(t)}, \, \phi _{v}^{(t)}, \, \phi _{ch}^{(t)}$.
				\STATE {Compute $P_{L|X}$ and $\Gamma _{ch}$ with MIM quantization}
				\STATE{${P_{R|X}^{(0)}} = {P_{L|X}}$}
				\FOR {$t=1:I_\text{max}$}
				\STATE{Construct $\phi_{c}^{(t)}$ with ${P_{R|X}^{(t-1)}}$ based on (\ref{phi_c})}
				\STATE{Calculate $\mc{A}^{(t)}$ and $P_{A|X}^{(t)}$ using (\ref{eqn: def of Phi_c}) and (\ref{Prx_all}), respectively}
				\STATE{$[{P_{S|X}^{(t)}},{\Gamma _{c}^{(t)}}] = \textbf{DP}(\mc{A}^{(t)},{P_{A|X}}^{(t)})$}
				\STATE{Construct $(\phi_{v}^{(t)}, \, \phi_{ch}^{(t)})$ with ${P_{S|X}^{(t)}}$ and ${P_{L|X}}$ based on (\ref{phi_v_ch})}\label{aaa1}
				\STATE{Calculate $\mc{B}^{(t)}$ and $P_{B|X}^{(t)}$ using (\ref{eqn: def of Phi_v}) and (\ref{Pbx_all}), respectively}
				\STATE{$[{P_{R|X}^{(t)}},{\Gamma _{v}^{(t)}}] = \textbf{DP}(\mc{B}^{(t)},{P_{B|X}^{(t)}})$}\label{aaa2}
				\STATE{Repeat \ref{aaa1}-\ref{aaa2} with $\mb{s} \in \mc{S}^{j}, j \in \mc{D}_v$ and $q_m=1$ to construct ${\Gamma _{e}^{(t)}}$}
				\ENDFOR
			\end{algorithmic}
		\end{algorithm}\vspace{-0.4cm}
	\end{table}

	Inspired by the observations investigated empirically in \cite{Romero2016mim, Lewandowsky18, Meidlinger2020irr, Stark2018ib, he2019mutual, Stark2020ib}, the design noise standard deviation ${\sigma _d}$ is selected based on the mutual information-maximizing criterion.
	More precisely, we perform the MIM-DE and choose ${\sigma _d}$ to make the decoder converge at the maximum decoding iteration $I_\text{max}$.
	In this paper, ${\sigma _d}$ is determined via the bisection search.
	After selecting ${\sigma _d}$, the pmf $P_{L|X}$ and the threshold set $\Gamma_{ch}$ of the associated DMC are obtained by performing the MIM quantization \cite{he2021dynamic}.
	Then the pmfs $P_{S|X}$ and $P_{R|X}$ are iteratively updated by using the MIM-DE to construct $\phi_c$, $\phi_v$, $\phi_{ch}$, $\Gamma_c$, and $\Gamma_v$, respectively.
	We summarize the design flow of the MIM-QBP decoder in \textbf{Algorithm \ref{alg:designFlow}}.
	The superscript $t$ indicates the iteration number.
	Note that our proposed method can design the parameters of the decoders systematically based on the MIM-DE, given the degree distributions of the target LDPC codes and the design noise standard deviation $\sigma_d$. 
	This means that the LUTs of the reconstruction functions and the threshold sets are constructed depending on the selected $\sigma_d$, and they do not change with the received SNRs when implementing decoding.
	Moreover, our proposed method has the potential to be extended to other channels whose channel transition probabilities can be expressed explicitly.
	In addition, we would like to point out that the computational complexity of the design process mainly originates from the calculation and quantization steps.
	As motivated by \cite{he2019mutual}, the computational complexity is $O(2^{q_c+q_m} \cdot (d_{c, \text{max}} + 2^{q_c}) )$ for the CN update and $O(2^{q_v+q_m} \cdot (d_{v, \text{max}} + 2^{q_v}) )$ for the VN update at each iteration.
	This is not a big issue for the offline process.

	\subsection{The MIM-QMS Decoder with Reduced-Complexity}
	\label{section: qms design}
	Compared to the outer messages with $q_m$-bit precision, the computing function given by (\ref{eqn: def of Phi_c}) requires a large $q_c$-bit precision and a large amount of fixed-point additions especially for the LDPC codes with high CN degrees.
	To reduce the computational complexity, we approximate (\ref{eqn: def of Phi_c}) by the MS algorithm \cite{Meidlinger2020irr}.
	In this case, we define a function $f(\cdot)$ to reflect the relationship of the relative LLRs of the V2C messages for MS operation.
	More specifically, $f(\cdot)$ maps the V2C messages in $\mc{R} = \{0, 1, \ldots, {{2^{{q_m}}} - 1}\}$ to an integer set $\{ {2^{{q_m} - 1}}, \ldots ,1, - 1, \ldots  - {2^{{q_m} - 1}}\}$.
	To differentiate the updating rule for the CN, we denote the computing function based on the BP algorithm and the MS algorithm by $\Phi _c^{\text{BP}}$ and $\Phi _c^{\text{MS}}$, respectively.
	For a degree-$i \, (i \in \mc{D}_c)$ CN, $\Phi _c^{\text{MS}}$ is given by\vspace{-0.1cm}
	\begin{equation}
		\label{eqn: def of Phi_c ms}
		{\Phi _c^{\text{MS}}}(\mb{r}) = f^{-1}\left( {\prod\limits_{k = 1}^{{i} - 1} s gn(f({r_k}))} \cdot \mathop {\min }\limits_{k \in 1,2, \ldots, {i} - 1} (\left| {f({r_k})} \right|)\right),\vspace{-0.1cm}
	\end{equation}  
	where $f^{-1}(\cdot)$ is the inverse function of $f(\cdot)$.
	We call the decoder designed based on $\Phi _c^{\text{MS}}$ as the MIM quantized MS (MIM-QMS) decoder.
	In this paper, we compute $\Phi _c^{\text{MS}}$ recursively on $k = 1,2,\ldots,i-1$ to achieve low complexity.
	A similar recursive method is adopted by \cite{Kameni2015dems} for the DE to select a proper channel scaling factor of the MS decoder.
	Note that the inner messages computed by the MS operation have the same cardinality as that of the outer messages, i.e., $q_c=q_m$, for all iterations.  
	Therefore, $\Gamma_c$ is not required and the pmf $P_{S|X}$ in (\ref{eqn: P_(S|X) lambda}) can be directly derived from $P_{\mb{R}|X}$ as  
	\begin{equation}
		\label{psx_ms}
		{P_{S|X}}(s|x) = \sum\limits_{i \in {\mc{D}_c}} {{\rho _i} \cdot \!\! \sum\limits_{\mb{r} \in {\mc{R}^{i - 1}}, \hfill\atop \Phi _c^{{\text{MS}}}(\mb{r}) = s} {{P_{\mb{R}|X}}} (\mb{r}|x)},
	\end{equation}   
	where $P_{\mb{R}|X}$ is the joint pmf given by (\ref{eqn: joint P_R|X}).
	Compared to the MIM-QBP decoder, the MIM-QMS decoder requires less bit widths used for the CN update.
	Due to $q_c=q_m$, the precision of the MIM-QMS decoder can be simply determined by $q_m$ and $q_v$.
	Moreover, the MIM-QMS decoder also reduces the computational complexity caused by the fixed-point additions in $\Phi _c^{\text{BP}}$. 
	\subsection{Remarks}
	We now discuss the difference between the design of the MIM quantized decoders and the LUT decoder variants \cite{Nguyen2018faid, Stark2018ib, Ghaffari2019pfaid, Stark2020ib, Meidlinger2020irr, wang2020rcq} that utilize the DE techniques to design the LUTs for irregular LDPC codes.
	In \cite{Nguyen2018faid} and \cite{Ghaffari2019pfaid}, the DE is employed to optimize one or multiple iteration-invariant LUTs for the VN update, which aims to minimize the asymptotic error probability with a preset memory constraint of the decoder.
	Instead of considering the error probability, the MIM quantized decoders are designed from the perspective of maximizing the mutual information between the coded bits and the messages from the CN or VN.

	Furthermore, the MIM-LUT decoders in \cite{Stark2018ib, Stark2020ib, Meidlinger2020irr} adopt DE to construct the LUTs for node updates based on the tree-like decomposition structure with $d_{c, \text{max}}$ or $d_{v, \text{max}}$ levels.
	Note that each level of the tree-like structure is corresponding to the nodes with a specific degree.
	More precisely, the message alignment method is adopted in \cite{Stark2018ib} as a post-processing step by considering the degree distributions after designing the LUTs for all levels.
	The implicit message alignment \cite{Stark2020ib} and the min-LUT decoder \cite{Meidlinger2020irr} take the degree distributions into account to design the LUTs for the VN update at every $j$-th level, where $j \in \mc{D}_v$.
	Obviously, the design methods in \cite{Stark2018ib, Stark2020ib, Meidlinger2020irr} deteriorates the performance of the MIM-LUT decoders since table decomposition causes the loss of mutual information \cite{cover2012elements}.
	It is worth noting that our proposed MIM quantized decoders utilize fixed-point additions to handle all inner messages simultaneously without using the decomposition technique. 
	In this way, the MIM quantized decoders can minimize the degradation of the error performance. 

	Apart from the aforementioned decoders, Wang et al. proposed the RCQ decoder in \cite{wang2020rcq} based on the DE, which has a similar framework of the MIM quantized decoders in terms of reconstruction, calculation, and quantization.
	However, the reconstruction functions of the RCQ decoder are designed in the floating-point domain, and the quantization scheme considered in the RCQ decoder is sub-optimal compared to the optimal DP method \cite{he2021dynamic}.
	\section{Complexity and Performance Analysis of MIM Quantized Decoders}
	\label{section: practical design}
	In this section, we investigate the decoding complexity of the proposed MIM quantized decoders, which is a critical issue that needs to be considered due to system constraints.
	More specifically, we discuss the computational complexity of the MIM quantized decoders, and compare the implementation complexity in terms of the memory requirement with the MIM-LUT decoders.
	Moreover, the error performance of the MIM quantized decoders and the MIM-LUT decoders with low precision is evaluated and compared via Monte-Carlo simulations.
	\subsection{Computational Complexity}
	As known from Section \ref{section: preliminaries}, the computational complexity of the MIM quantized decoders mainly originates from the calculation and quantization steps since the reconstruction step is a one-to-one mapping by the corresponding reconstruction functions.
	Denote the average CN and VN degrees by
	\begin{equation}
		\bar{d_c} = 1 \Big/ \sum_{i \in \mc{D}_c} \frac{\rho_i}{i} \text{~and~} \bar{d_v} = 1 \Big/ \sum_{j \in \mc{D}_v} \frac{\theta_j}{j}, \nonumber
	\end{equation}
	respectively.
	According to \cite{Lee05}, the calculation of $\Phi _c^{\text{BP}}$ for the CN update of the MIM-QBP decoder requires $\bar{d_c} - 1$ addition operations to compute the summation of $\bar{d_c}$ inner messages, and $\bar{d_c}$ subtraction operations to compute the $\bar{d_c}$ outer messages.  
	Moreover, the table lookup operations for $q_m$-bit outer messages based on $\Gamma_c$ require $\bar{d_c} \cdot q_m$ comparisons.
	Therefore, we have the average computational complexity of $\bar{d_c} \cdot (q_m + 2) - 1$ for one CN per iteration.
	For the MIM-QMS decoder, the MS operation can be implemented by an efficient method such as \cite{Lee2015ms}, which leads to a computational complexity $2\bar{d_c} + \left\lceil {{{\log }_2}({\bar{d_c}}} ) \right\rceil - 2$ for one CN per iteration.
	Furthermore, the computation of $\Phi _v$ for the VN update requires $\bar{d_v}$ addition and subtraction operations to compute the summation of $\bar{d_v}+1$ inner messages and the $\bar{d_v}$ outer messages, respectively.  
	The table lookup operations based on $\Gamma_v$ result in $\bar{d_v} \cdot q_m$ comparisons.
	Consequently, the average computational complexity is $\bar{d_v} \cdot (q_m + 2)$ for one VN per iteration.
	In addition, there is only one extra comparison operation required for the mapping function $Q_e$ because the summation computed by $\Phi _v$ can be utilized for the hard decision. 
	We summarize the computational complexity of the MIM quantized decoders in TABLE \ref{table: comlexity of MIM decoder}.
	Although extra bit widths ($q_c$ and $q_v$) are required for the computing functions, it is notable to point out that the choice of $q_c$ and $q_v$ merely impacts on the computational complexity in the design procedure, but not in the decoding process.
	\begin{table}[t!]
		\normalsize
		\renewcommand{\arraystretch}{1.2}
		\caption{Computational Complexity of One Node per Iteration for the MIM Quantized Decoders}
		\label{table: comlexity of MIM decoder}
		\centering
		\resizebox{\columnwidth}{!}{
			\begin{tabular}{|c|c|c|}
				\hline
				Mapping functions & 		QBP        												& QMS       																\\ \hline
				$Q_c$  		& $\bar{d_c} \cdot (q_m + 2) - 1$         	 			& $2\bar{d_c} + \left\lceil {{{\log }_2}({\bar{d_c}}} ) \right\rceil - 2$    \\ \hline
				$Q_v$  		& \multicolumn{2}{c|}{$\bar{d_v} \cdot (q_m + 2)$ } 																		 \\ \hline
				$Q_e$  		& \multicolumn{2}{c|}{$1$} 																								 \\ \hline
			\end{tabular}
		}
	\end{table}
	\subsection{Memory Requirements}
	Apart from the computational complexity, the implementation complexity is also of significant importance for various applications.
	Due to table lookup operations adopted in the node updates, the memory requirements complexity of the MIM-LUT decoders and the MIM quantized decoders is mainly affected by the memory requirement for storing the LUTs.
	This depends on the number of LUTs used for decoding and the number of entries in each LUT.
	In \cite{Stark2018ib, Stark2020ib, Meidlinger2020irr}, the tree-like structure of table lookup operations leads to the memory requirement for storing various LUTs for updating the nodes with different degrees.
	Each LUT has the same number of entries.
	Given $q_m$-bit precision of the messages, the number of entries in one LUT is ${2^{2{q_m}}}$.
	According to \cite{Stark2018ib}, there are $2 \cdot \left\lfloor {{{\log }_2}({d_{c, \text{max}}}} ) \right\rfloor$ and $2 \cdot \left\lfloor {{{\log }_2}({d_{v, \text{max}}}} ) \right\rfloor$ LUTs required for the CN and the VN update, respectively.
	As a result, the memory requirement at each iteration for the tree-like structure is $2^{2{q_m}+1} \cdot \left\lfloor {{{\log }_2}({d_{c, \text{max}}}} ) \right\rfloor$ for the CN update, and $2^{2{q_m}+1} \cdot \left\lfloor {{{\log }_2}({d_{v, \text{max}}}} ) \right\rfloor$ for the VN update.  

	For our proposed MIM quantized decoders, we use the same LUTs for the reconstruction functions and threshold sets to update the nodes of different degrees.
	By considering the $q_m$-bit outer messages, the LUTs of one reconstruction function and one threshold set have $2^{q_m}$ and $2^{q_m}-1$ entries at each iteration, respectively.  
	Therefore, the total memory requirement for the CN update of the MIM-QBP decoder with $\phi_c$ and $\Gamma_c$ is $2^{q_m+1}-1$ for each iteration.
	For the CN update of the MIM-QMS decoder, the function $f(\cdot)$ is totally determined by $q_m$.
	Thus we do not need to store the LUT for $f(\cdot)$.
	Moreover, the LUTs of the reconstruction functions ($\phi_{v}$, $\phi_{ch}$) and the threshold sets ($\Gamma_{v}$, $\Gamma_{ch}$) for the VN update have $2^{q_m+2}-2$ entries in total for each iteration.
	Since the same reconstruction functions $\phi_{v}$ and $\phi_{ch}$ are used for the mapping function $Q_e$, there is only one entry in the threshold set $\Gamma_{e}$ for hard decision at each iteration.
	We summarize the memory requirements of various decoders in TABLE \ref{table: implementation of MIM decoder}.
	Compared to the MIM-LUT decoders \cite{Stark2018ib, Stark2020ib, Meidlinger2020irr}, our proposed MIM quantized decoders significantly reduce the memory requirement for the decoding process.
	More importantly, the memory demand for the MIM quantized decoders is only affected by the precision of the outer message, which is independent of the degree distributions of the LDPC codes.
	This benefits the decoding of the LDPC codes especially for those with high code rates.
	This is because the LDPC codes with high code rates incorporate very high node degrees, resulting in an undesirable growth in memory demand for the MIM-LUT decoders.
	\begin{table}[t!]
		\renewcommand{\arraystretch}{1.2}
		\Large
		\caption{Memory Demand per Iteration for Various Decoders}
		\label{table: implementation of MIM decoder}
		\centering
		\resizebox{\columnwidth}{!}{
			\begin{tabular}{|c|c|c|c|}
				\hline
				Decoders                   & MIM-QBP                				& MIM-QMS             & MIM-LUT \cite{Stark2018ib, Stark2020ib, Meidlinger2020irr} \\ \hline
				\multicolumn{1}{|c|}{$Q_c$} & \multicolumn{1}{c|}{$2^{q_m+1}-1$} 	& \multicolumn{1}{c|}{$-$}  & $2 \cdot \left\lfloor {{{\log }_2}({d_{c, \text{max}}}} ) \right\rfloor  \cdot {2^{2{q_m}}}$                    \\ \hline
				\multicolumn{1}{|c|}{$Q_v$} & \multicolumn{2}{c|}{$2^{q_m+2}-2$}  & $2 \cdot \left\lfloor {{{\log }_2}({d_{v, \text{max}}}} ) \right\rfloor  \cdot {2^{2{q_m}}}$                    \\ \hline
				\multicolumn{1}{|c|}{$Q_e$} & \multicolumn{2}{c|}{$1$}  & $2 \cdot \left\lfloor {{{\log }_2}({d_{v, \text{max}}+1}} ) \right\rfloor  \cdot {2^{2{q_m}}}$  \\ \hline
			\end{tabular}
		}\vspace{-0.1cm}
	\end{table}
	\subsection{Performance Comparison}\label{MIM quantized fer}
\begin{figure}[t!]
		\centering
		\includegraphics[width=3in]{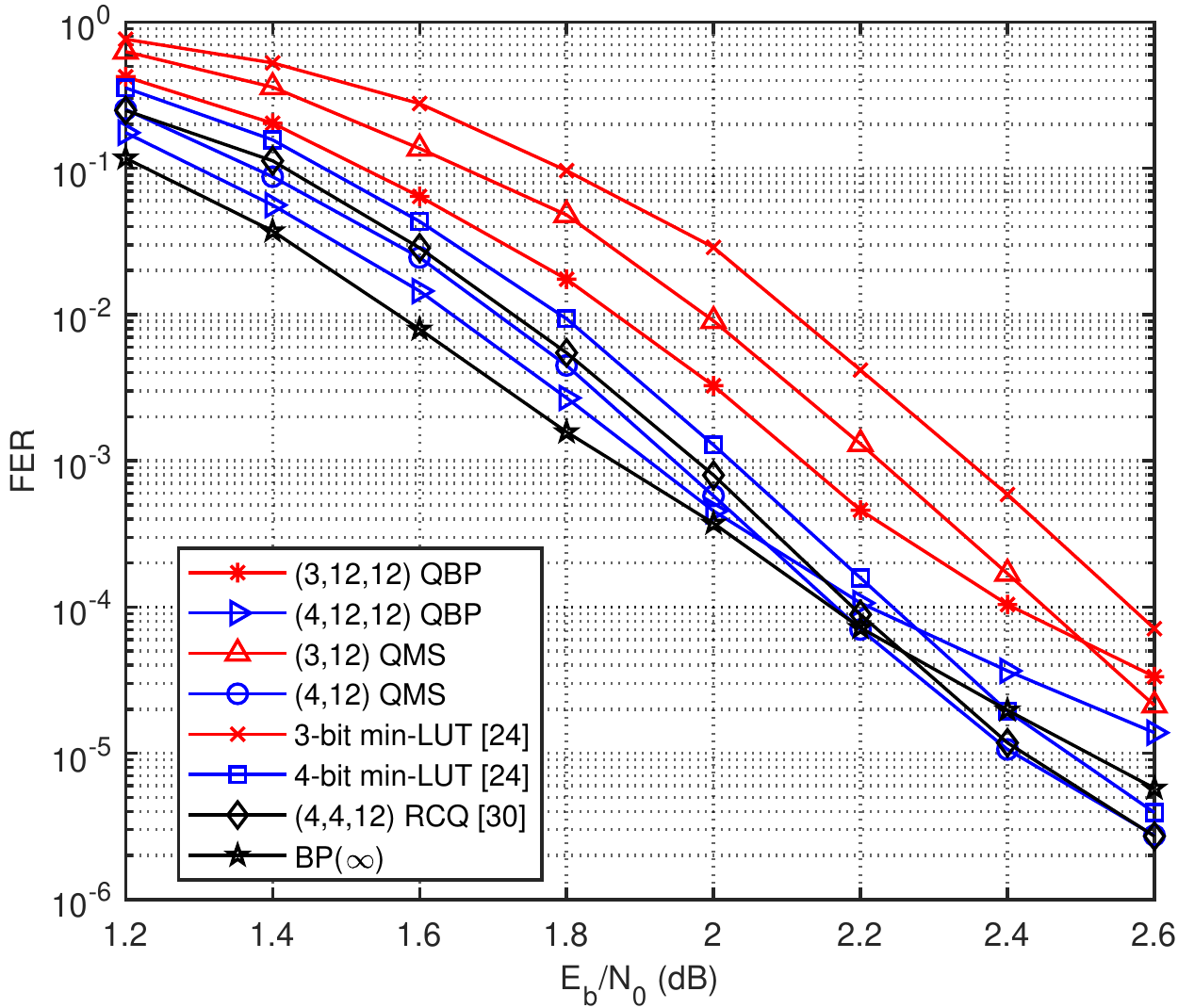}
		\caption{FER of the length-$1296$ IEEE 802.11n LDPC code \cite{IEEESTD802_11n} with code rate $1/2$ over AWGN channels, $I_\text{max} = 50$.} 
		\label{fig: 802_11n}
	\end{figure}

	We further investigate the frame error rate (FER) performance of the proposed MIM quantized decoders via Monte-Carlo simulations and compared to that of the MIM-LUT decoders. 
	Assume that binary LDPC codewords are modulated by binary phase-shift keying (BPSK) and transmitted over the AWGN channels.
	For convenience, we use the tuples ($q_m, q_c, q_v$) and ($q_m, q_v$) to represent the precision of the MIM-QBP decoder and the MIM-QMS decoder, respectively.
	Note that we consider the low precision quantization, i.e., $q_m = 3$ and $4$ for practical implementations. 
	The floating-point precision is denoted by ``$\infty$''.
	The irregular LDPC code is selected from the IEEE 802.11n standard \cite{IEEESTD802_11n} with length $1296$ and code rate $1/2$.
	The degree distributions of the simulated code are
	\begin{equation}\notag
		\begin{aligned}
			\rho (x) &= {\rm{0}}{\rm{.8140}}{x^6} + {\rm{0}}{\rm{.1860}}{x^7},\\
			\theta (x) &= {\rm{0}}{\rm{.2558}}x + {\rm{0}}{\rm{.3140}}{x^2} + {\rm{0}}{\rm{.0465}}{x^3} + {\rm{0}}{\rm{.3837}}{x^{10}}.
		\end{aligned}
	\end{equation}
	We summarize the design noise standard deviation ${\sigma _d}$ for the associated MIM quantized decoders in TABLE \ref{table: simulation parameters1}.
	At least $300$ error frames are collected for each simulated SNR.
	(See Appendix A in \cite{kang2020generalized} for details of the constructed LUTs.)
	\begin{table}[h!]
		\renewcommand{\arraystretch}{1.2}
		\caption{Noise Standard Deviation ${\sigma _d}$ for MIM Quantized Decoders}
		\label{table: simulation parameters1}
		\centering
		\footnotesize
		\resizebox{\columnwidth}{!}{
			\begin{tabular}{|cc|cc|}
				\hline
				\multicolumn{2}{|c|}{($q_m, q_c, q_v$) MIM-QBP, ${\sigma _d}$} & \multicolumn{2}{c|}{($q_m, q_v$) MIM-QMS, ${\sigma _d}$}  \\ \hline
				\multicolumn{1}{|c|}{($3,12,12$)}      & ($4,12,12$)      & \multicolumn{1}{c|}{($3,12$)}      & ($4,12$)      \\ \hline
				\multicolumn{1}{|c|}{$0.8778$} & $0.9003$ & \multicolumn{1}{c|}{$0.8501$} & $0.8998$ \\ \hline
			\end{tabular}
		}
	\end{table}
	\begin{example}\label{802.11n}
		Fig. \ref{fig: 802_11n} depicts the FER performance of the proposed MIM quantized decoders with the maximum number of iterations $I_{\text{max}} = 50$.
		For comparison, we also present the FER performance of the 3-bit/4-bit min-LUT decoders \cite{Meidlinger2020irr}, the ($4,4,12$) RCQ decoder in \cite{wang2020rcq}, and the BP($\infty$) decoder.
		It is shown that the FER performance of our proposed ($3,12,12$) and ($4,12,12$) MIM-QBP decoder can approach that of the BP($\infty$) decoder by about $0.1$ dB and $0.3$ dB, respectively.
		For $E_b/N_0 < 2$ dB, the proposed $(3,12,12)$ MIM-QBP decoder has a performance gain of about $0.2$ dB compared to the 3-bit min-LUT decoder.  
		Furthermore, the $(4,12,12)$ MIM-QBP decoder performs better than both the 4-bit min-LUT decoder and the ($4,4,12$) RCQ decoder when $E_b/N_0 < 2$ dB.
		In addition, the $(3,12)$ MIM-QMS decoder achieves about $0.1$ dB gain compared to the 3-bit min-LUT decoder and its FER performance is around $0.4$ dB away from the BP($\infty$) decoder for $E_b/N_0 < 2$ dB.
		When $E_b/N_0 \geq 2$ dB, the FER performance of the $(3,12)$ MIM-QMS decoder approaches that of the $(3,12,12)$ MIM-QBP decoder and the BP($\infty$) decoder with a difference of less than $0.1$ dB and $0.3$ dB, respectively.
		It is notable that the $(4,12)$ MIM-QMS decoder can surpass the ($4,4,12$) RCQ decoder and achieves the FER performance of less than $0.1$ dB away from the BP($\infty$) decoder when $E_b/N_0 < 2.2$ dB.
		When $E_b/N_0 \geq 2.2$ dB, the $(4,12)$ MIM-QMS decoder even outperforms the BP($\infty$) decoder by about $0.1$ dB.   

		Due to the existing of the degree-2 VNs in the code graph, the BP($\infty$) decoder and the MIM-QBP decoders have higher error floor than the MIM-QMS decoders for the simulated LDPC code.
		This is because the cycles confined among the degree-2 VNs result in low-weight codewords, which become the most detrimental objects for the BP updating rule \cite{Richardson03}.
		%
		%
		Similar phenomena are also observed in \cite{Romero2016mim} and \cite{wang2020rcq}, which show that using the MIM quantization is capable of mitigating the effect of certain low-weight codewords or trapping sets.
	\end{example}

\begin{figure}[t!]
		\centering
		\includegraphics[width=3in]{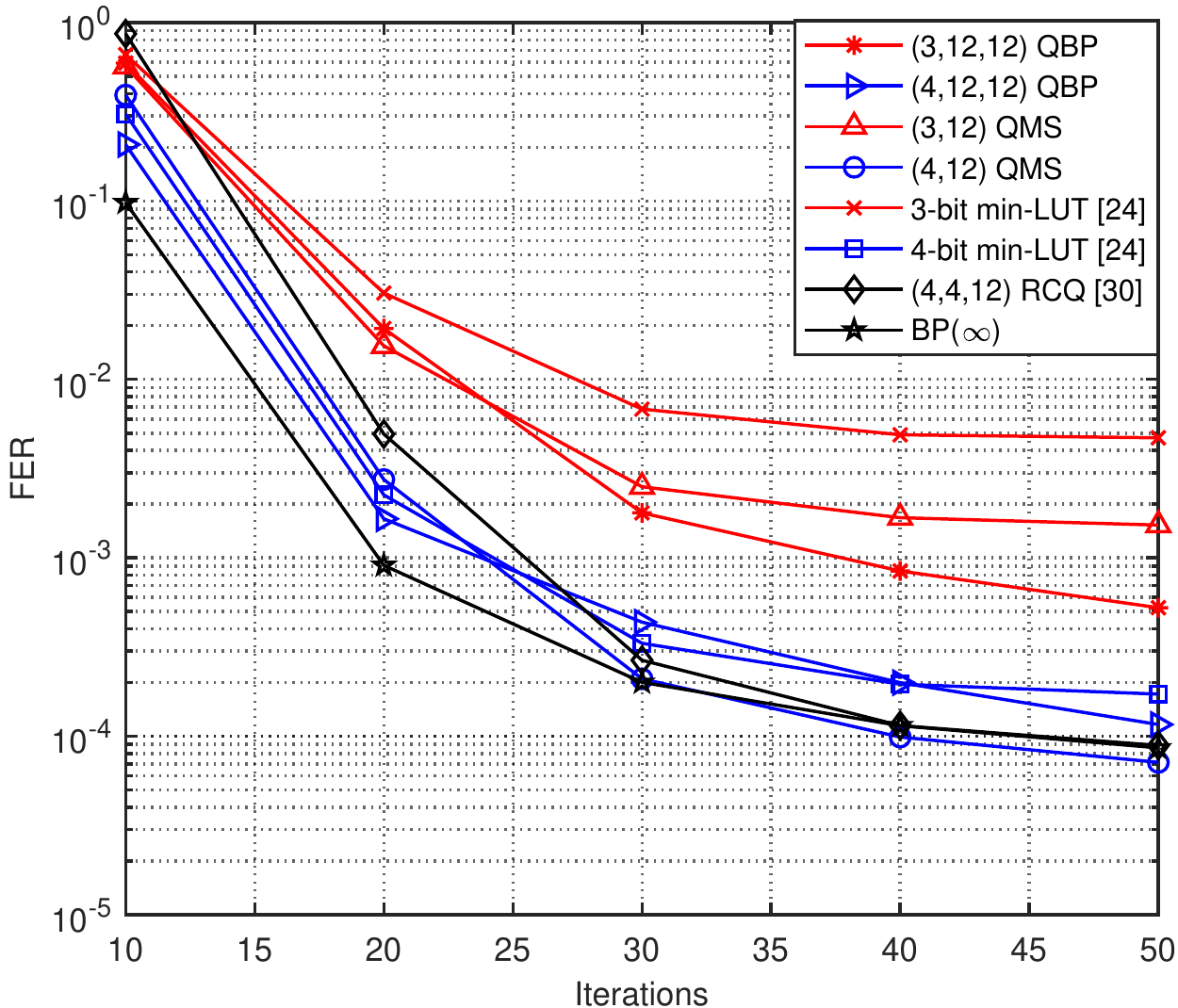}
		\caption{FER-versus-iteration performance of the length-$1296$ IEEE 802.11n LDPC code \cite{IEEESTD802_11n} with code rate $1/2$ over AWGN channels, $E_b/N_0 = 2.2$ dB.}
		\vspace{-0.5cm}
		\label{fig: ferVSiter}
	\end{figure}

	We also investigate the convergence behaviors of the proposed MIM quantized decoders in decoding the same LDPC code at $E_b/N_0 = 2.2$ dB over AWGN channels.
	Fig. \ref{fig: ferVSiter} shows the FER-versus-iteration performance for the proposed MIM quantized decoders.
	It can be seen that the $(3,12,12)$ MIM-QBP decoder and the $(3,12)$ MIM-QMS decoder outperform the $3$-bit min-LUT decoder for all decoding iterations.
	For decoding iteration $\le 20$, the ($4,12,12$) MIM-QBP decoder has a convergence speed close to that of the BP($\infty$) decoder, and it achieves a faster convergence speed compared to both the min-LUT decoder and the RCQ decoder with the 4-bit precision.  
	Moreover, the $(4,12)$ MIM-QMS decoder outperforms the ($4,4,12$) RCQ decoder and converges almost as fast as that of the 4-bit min-LUT decoder for decoding iteration $\le 30$.
	When decoding iteration $\le 30$, the $(4,12)$ MIM-QMS decoder shows almost the same FER performance compared to the BP($\infty$) decoder for the same decoding iterations.
	\section{The Layered Schedule for MIM-QMS Decoder}
	Compared to the MIM-LUT decoders, the proposed MIM-QMS decoder achieves a better trade-off between the error performance and the decoding complexity.
	However, there are still a relative large number of iterations required to obtain a desirable error performance due to the flooding schedule \cite{Kschischang1998fsd} used for decoding.
	To reduce the number of decoding iterations, the layered schedule has been adopted in many designs, e.g., \cite{Mansour2003cnLayer,Zhang2009cnlayer,Zhang2005vnLayer,Cui2011vnLayer,Aslam2017vnLayer}, to achieve a faster convergence speed of the decoder.
	However, we observe that there is a mismatch between the reconstruction functions and the threshold sets designed by the MIM-DE with those needed for layered decoding.
	More specifically, the MIM-DE is derived originally based on the flooding schedule at each iteration, and it constructs the LUTs of the reconstruction functions and threshold sets by assuming that all outer messages sent to the CN or VN have the same pmfs.
	However, the pmfs of the outer messages need to be updated sequentially at each iteration when the layered schedule is adopted.

	Motivated by this observation, in this section, we propose the design method of the MIM layered QMS (MIM-LQMS) decoder for the QC-LDPC codes.
	Note that we consider QC-LDPC codes because they are favorable for the implementation with layered scheduling due to the regularity in the parity check matrices.
	Particularly, we derive the layered MIM-DE (LMIM-DE) to construct the LUTs of the reconstruction functions and threshold sets for each layer and each iteration without mismatch.
	We call those reconstruction functions and threshold sets the layer-specific parameters.
	However, this design method increases the memory requirement of the MIM-LQMS decoder because the number of LUTs grows with the number of layers.
	To reduce the memory requirement, we further introduce an optimization method for the MIM-LQMS decoder by considering the iteration-specific reconstruction functions and threshold sets instead of the layer-specific ones.
	\subsection{The Design of MIM-LQMS Decoder}
	Let $H$ be the parity-check matrix of size $M \times N$ for an LDPC code, which can be represented by a Tanner graph \cite{Tanner81} containing $N$ VNs and $M$ CNs.  
	We define $\mc{N}(c_m) \backslash v_n$ as the set of VNs connected to CN $c_m$ ($1 \le m \le M$) with the node $v_n$ excluded and define $\mc{M}(v_n) \backslash c_m$ as the set of CNs connected to VN $v_n$ ($1 \le n \le N$) with the node $c_m$ excluded.
	The QC-LDPC codes are a class of the structured LDPC codes which can be constructed from an $M_b \times N_b$ compact matrix $H_b$.
	To obtain $H$, each element in $H_b$ is expanded by either a circulant permutation matrix \cite{ryan2009channel} or an all-zero matrix of size $Z \times Z$, where $Z$ is known as the lifting factor \cite{Mitchell2015scldpc}.
	Therefore, we have $Z = M/{M_b} = N/{N_b}$.

	Compared to the flooding schedule, the layered schedule updates the message in a sequential order to achieve a faster convergence speed.
	In this paper, we focus on the vertical layered schedule \cite{Zhang2005vnLayer} because it inherently has short critical paths and can minimize the loading latency of the messages \cite{Cui2011vnLayer}.	
	Denote the C2V message and the V2C message between a node $c_m \in \mc{M}(v_n)$ and $v_n$ by $S_{mn}$ and $R_{nm}$, respectively.
	At each layer and each iteration, the vertical layered decoder based on the MS algorithm performs \cite{Cui2011vnLayer}
	\begin{equation}
		\label{LMS_cn}
		S_{mn} = \prod\limits_{{v_{n'}} \in \mc{N}({c_m})\backslash {v_n}} {sgn(R_{n'm})} \cdot \mathop {\min }\limits_{{v_{n'}} \in \mc{N}({c_m})\backslash {v_n}} \left| {R_{n'm}} \right|,
	\end{equation}
	\begin{equation}
		\label{LMS_vn}
		R_{nm} = L({v_n}) + \sum\limits_{{c_{m'}} \in \mc{M}({v_n})\backslash {c_m}} {S_{m'n}},
	\end{equation}
	where $L({v_n})$ is the channel output message for the node $v_n$.
	At the end of each iteration, a posterior information of the node $v_n$, denoted by $L_n$, is given by
	\begin{equation}
		\label{LMS_app}
		L_n = R_{nm} + S_{mn}.
	\end{equation}

	The QC-LDPC codes are well suited to the vertical layered schedule by partitioning the VNs into $N_b$ groups, so-called the layers.
	Each layer consists of $Z$ VNs and the VNs of layer $h \, (h=1, 2, \ldots, N_b)$ are corresponding to the $(h \cdot Z-Z+1)$-th to $(h \cdot Z)$-th consecutive columns in $H$.
	Note that any two columns in each layer have at most one nonzero element in the same row.
	In this way, implementing the vertical layered schedule over $H$ is equivalent to performing the decoding on the base matrix $H_b$, where layer $h$ is corresponding to the $h$-th column in $H_b$.
	Moreover, we consider the decoding proceeds layer after layer in ascending order of layer index $h$.

\begin{figure}[!t]
		\centering
		\includegraphics[width=3.2in]{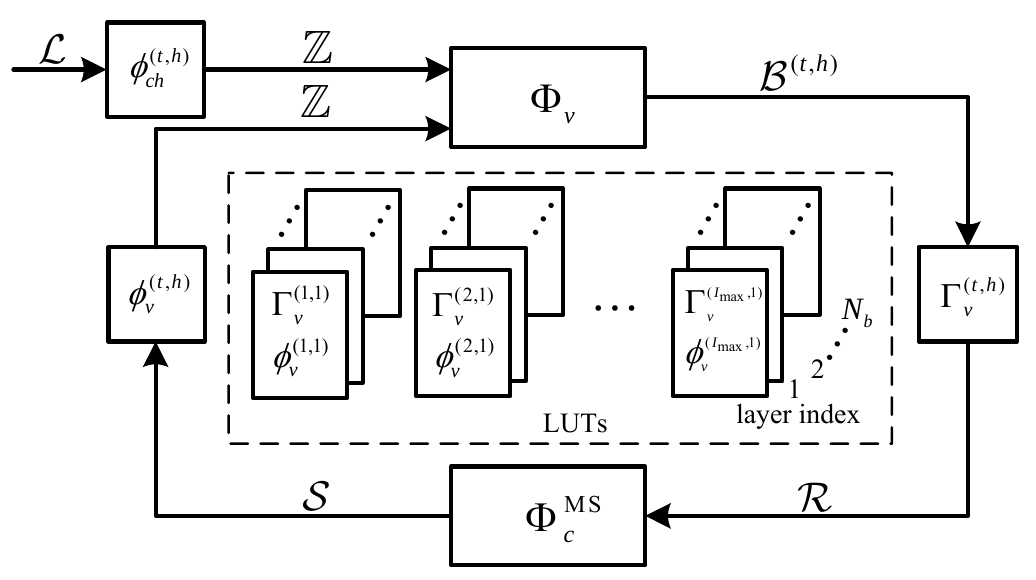}
		\caption{The architecture of MIM-LQMS decoder.}
		\label{fig: MIM-LQMS}
		\vspace{-0.6cm}
	\end{figure}

	In Fig. \ref{fig: MIM-LQMS}, we present the architecture of the MIM-LQMS decoder designed based on $H_b$.
	For convenience, we use the superscript $(t, h)$ to represent the parameter at layer $h$ and iteration $t$.
	Compared to the MIM-QMS decoder, the MIM-LQMS decoder has a similar framework to the MIM-QMS decoder but differs in the sense that the reconstruction functions and the threshold sets vary with both the associated layer and iteration.
	This is because the pmfs of the messages are different at each layer and each iteration due to the sequential updating order adopted by the layered scheduling.
	We denote the reconstruction function for the channel output and the VN update by ${\phi_{ch}^{(t,h)}}$ and $\phi_{v}^{(t,h)}$, respectively.
	We also denote the threshold set for the VN update at layer $h$ and iteration $t$ by $\Gamma_{v}^{(t, \, h)}$.

	Similar to the case of MIM-QMS decoder, designing ${\phi_{ch}^{(t,h)}}$, $\phi_{v}^{(t, \, h)}$ and $\Gamma_{v}^{(t, \, h)}$ for the MIM-LQMS decoder requires the pmfs between the coded bits and the outer messages at each layer for different iterations.
	Since the MIM-DE is only capable of tracking the evolution of the pmfs under the flooding schedule, it is necessary to develop the LMIM-DE to obtain the pmfs at each layer and each iteration.  
	Define $P_{R|X}^{(t, \, h)}$ and $P_{S|X}^{(t, \, h)}$ as the pmf of the V2C message $R$ and C2V message $S$ conditioned on the coded bit $X$ at layer $h$ and iteration $t$, respectively.
	In the following, we illustrate the design of the MIM-LQMS decoder based on the LMIM-DE in detail.
	\subsubsection{LMIM-DE at CN Update}
	At iteration $t$, a degree-$i$ ($i \in \mc{D}_c$) CN connected to a VN in layer $h$ receives the V2C messages from the layers that have already been updated (layer $h' < h$) at iteration $t$, and from the layers that were updated at iteration $t-1$ (layer $h' > h$).
	Therefore, we consider the expected pmf of the V2C message received at layer $h$ as
	\begin{equation}
		\label{Prx_layer}
		\begin{aligned}
			{\tilde P_{R|X}}^{(t, \, h)}(r|x) = \frac{1}{{{N_b} - 1}}\left( {\sum\limits_{h' = 1}^{h - 1}} {P_{R|X}^{(t, \, h')}(r|x)} \right. \\ \vspace{-0.1cm}
			\left. {+ \sum\limits_{h' = h + 1}^{{N_b}} {P_{R|X}^{(t - 1, \, h')}(r|x)}  } \right).
		\end{aligned}
	\end{equation}
	Note that we have $P_{R|X}^{(0, h)} = P_{L|X}$ for $h = 1,2,\ldots,N_b$.
	By replacing $P_{R|X}$ with $\tilde P_{R|X}^{(t, \, h)}$ in (\ref{eqn: joint P_R|X}), the joint pmf at layer $h$ and iteration $t$ is
	\begin{equation}
		\label{eqn: joint layered P_R|X}
		\tilde P_{\mb{R}|X}^{(t, \, h)}(\mb{r}|x) = \left(\frac{1}{2}\right)^{i - 2} \sum_{\mb{x}: \oplus \mb{x} = x} \prod_{k = 1}^{i-1} \tilde P_{R|X}^{(t, \, h)}(r_k|x_k).
	\end{equation}
	Inspired by (\ref{psx_ms}), the pmf of the C2V message at layer $h$ and iteration $t$ can be computed by 
	\begin{equation}
		\label{eqn: P_(S|X) layer}
		P_{S|X}^{(t,\,h)}(s|x) = \sum\limits_{i \in {D_c}} {{\rho _i} \cdot \sum\limits_{\mb{r} \in {\mc{R}^{i - 1}},\hfill\atop
				\Phi _c^{\text{MS}}(\mb{r}) = s\hfill} {{\tilde P_{\mb{R}|X}^{(t, \, h)}} } ({\mb{r}}|x)}.
	\end{equation}
	\subsubsection{LMIM-DE at VN Update}	
	The VN update for the MIM-LQMS decoder consists of the reconstruction by the reconstruction functions $\phi_{ch}^{(t, \, h)}$ and $\phi_{v}^{(t, \, h)}$, the calculation with $\Phi_v$ according to (\ref{eqn: def of Phi_v}), and the MIM quantization using the DP method.
	Based on $P_{S|X}^{(t,\,h)}$ and $P_{L|X}$, the reconstruction functions $\phi_{ch}^{(t, \, h)}$ and $\phi_{v}^{(t, \, h)}$ can be constructed accordingly by (\ref{phi_v_ch}).
	Due to the DE assumptions in \cite{Richardson01capacity}, we assume that the C2V messages entering a degree-$j$ ($j \in \mc{D}_v$) VN in the same layer are independent and identically distributed.
	Thus, the joint pmf at layer $h$ and iteration $t$ can be calculated by
	\begin{equation}
		\label{Plsx_layer}
		P_{L,\mb{S}|X}^{(t,\,h)}(l,\mb{s}|x) = {P_{L|X}}(l|x)\prod\limits_{k = 1}^{j - 1} {P_{S|X}^{(t,\,h)}({s_k}|x)}.
	\end{equation}
	After reconstruction, the computing function $\Phi_v$ calculates the alphabet set $\mc{B}^{(t, \, h)}$ of the inner message for each realization $(l, \mb{s}) \in \mc{L} \times \mc{S}^{j-1}$.
	Define $P_{B|X}^{(t,\,h)}$ as the pmf of the inner message $B$ at layer $h$ and iteration $t$.
	For a realization $b \in \mc{B}^{(t, \, h)}$, we have
	\begin{equation}
		\label{Pbx_layer}
		P_{B|X}^{(t,\,h)}(b|x) = \sum\limits_{j \in {D_v}} {{\theta _j} \cdot \sum\limits_{\scriptstyle(l,\,\mb{s}) \in \mc{L} \times {\mc{S}^{j - 1}},\hfill\atop
				\scriptstyle{\Phi _v}(l,\mb{s}) = b\hfill} {P_{L,\mb{S}|X}^{(t,\,h)}(l,\mb{s}|x)} }.
	\end{equation}
	With $\mc{B}^{(t, \, h)}$ and $P_{B|X}^{(t,\,h)}$, we perform the MIM quantization and obtain $P_{R|X}^{(t,\,h)}$ and $\Gamma_{v}^{(t,\,h)}$:
	\begin{equation}
		\label{eqn: P_(R|X) layer}
		[{P_{R|X}^{(t,\,h)}},{\Gamma_{v}^{(t,\,h)}}] = \textbf{DP}(\mc{B}^{(t, \, h)},{P_{B|X}^{(t,\,h)}}).
	\end{equation} 
	By implementing LMIM-DE, the pmfs $P_{R|X}^{(t, \, h)}$ and $P_{S|X}^{(t, \, h)}$ are updated at each layer and each iteration to construct $\phi_v^{(t, \, h)}$, $\phi_{ch}^{(t, \, h)}$, and $\Gamma_v^{(t, \, h)}$, respectively.
	These layer-specific parameters need to be stored as the LUTs for decoding.\vspace{-0.5cm}
	\begin{table}[h!]
		\begin{algorithm}[H]
			\normalsize
			\setstretch{1.2}
			\caption{The Iteration-Specific Design Flow of MIM-LQMS Decoder}
			\label{alg:designFlow_layer}
			\begin{algorithmic}[1]
				\REQUIRE ${\sigma _d}, \, \rho(x), \, \theta(x), \, q_m, \, q_v, \, I_\text{max}$.
				\ENSURE $\Gamma _{ch}, \, \Gamma _{e}^{(t)}, \Gamma _{v}^{(t)}, \, \phi _{v}^{(t)}, \, \phi _{ch}^{(t)}$.
				%
				%
				%
				\STATE {Compute $P_{L|X}$ and $\Gamma _{ch}$ with MIM quantization in \cite{he2021dynamic}}
				\STATE{$P_{R|X}^{(0, h)} = P_{L|X}$, $h = 1,2,\ldots,N_b$}
				\FOR {$t=1:I_\text{max}$}
				\FOR {$h=1:N_b$}
				\STATE{Compute $\tilde P_{R|X}^{(t, \, h)}$ and ${P_{S|X}^{(t, \, h)}}$ using (\ref{Prx_layer}) and (\ref{eqn: P_(S|X) layer}), respectively}\label{bbb1}
				\STATE{Construct $\phi_{ch}^{(t, \, h)}$ and $\phi_{v}^{(t, \, h)}$ based on (\ref{phi_v_ch})}	
				\STATE{Calculate $\mc{B}^{(t, \, h)}$ and $P_{B|X}^{(t, \, h)}$ using (\ref{eqn: def of Phi_v}) and (\ref{Pbx_layer}), respectively}\label{bbb2}
				\STATE{$[{P_{R|X}^{(t,\,h)}},{\Gamma_{v}^{(t,\,h)}}] = \textbf{DP}(\mc{B}^{(t, \, h)},{P_{B|X}^{(t,\,h)}})$}\label{bbb3}
				\ENDFOR
				\STATE{Compute $\tilde P_{S|X}^{(t)}$ using (\ref{Psx_iter})}
				\STATE{Construct $\phi_{ch}^{(t)}$ and $\phi_{v}^{(t)}$ based on (\ref{phi_v_ch}) with $\tilde P_{S|X}^{(t)}$}
				\STATE{Construct ${\Gamma _{v}^{(t)}}$ and ${\Gamma _{e}^{(t)}}$ using (\ref{eqn: def of Phi_v}), (\ref{Pbx_all}), and (\ref{eqn: P_(R|X) lambda}) with $\tilde P_{S|X}^{(t)}$}
				\FOR {$h=1:N_b$}
				\STATE{Perform steps 5 and 7 with $\phi_{ch}^{(t)}$ and $\phi_{v}^{(t)}$}
				%
				%
				%
				\STATE{Update ${P_{R|X}^{(t,\,h)}}$ based on $\mc{B}^{(t, \, h)}$, $P_{B|X}^{(t, \, h)}$, and ${\Gamma_{v}^{(t)}}$}
				\ENDFOR

				%
				%
				%
				%
				\ENDFOR
			\end{algorithmic}
		\end{algorithm}\vspace{-0.7cm}
	\end{table}
	\subsection{Iteration-Specific Optimization}
	As shown in the previous sub-section, the main disadvantage of the MIM-LQMS decoder is that storing the LUTs of layer-specific parameters significantly increases the memory requirement especially for a large number of layers.
	To reduce the memory demand for storing the LUTs, we optimize the proposed LMIM-DE and design the iteration-specific reconstruction functions and threshold sets for the MIM-LQMS decoder.
	More specifically, we first conduct the layer-specific design to obtain the pmf $P_{S|X}^{(t, \, h)}$.
	After that, a post-processing is performed, where we consider the expected pmf of the C2V message from all $N_b$ layers at iteration $t$ as
	\begin{equation}
		\label{Psx_iter}
		\tilde P_{S|X}^{(t)}(s|x) = \frac{1}{{{N_b}}}\sum\limits_{h = 1}^{{N_b}} {P_{S|X}^{(t, \, h)}(s|x)}.
	\end{equation}
	By replacing $P_{S|X}^{(t, \, h)}$ with $\tilde P_{S|X}^{(t)}$ in (\ref{phi_v_ch}), we can construct the iteration-specific reconstruction functions $\phi_{ch}^{(t)}$ and $\phi_v^{(t)}$, and the threshold set $\Gamma_v^{(t)}$ following the steps \ref{aaa1}-\ref{aaa2} in \textbf{Algorithm \ref{alg:designFlow}}.
	Then the layer-specific design need to be conducted again to update the pmf $P_{R|X}^{(t, \, h)}$ for each layer with the iteration-specific parameters $\phi_{ch}^{(t)}$, $\phi_v^{(t)}$, and $\Gamma_v^{(t)}$.
	%
	%
	Note that $\phi_v^{(t)}$, $\phi_{ch}^{(t)}$, and $\tilde P_{S|X}^{(t)}$ can be directly used to determine the threshold set $\Gamma_{e}^{(t)}$ for the bit decision with $\{\Phi_v(l, \mb{s}) : (l, \mb{s}) \in \mc{L} \times \mc{S}^{j}\}$ in (\ref{Pbx_layer}) and $q_m=1$.
	We summarize the iteration-specific design flow of the MIM-LQMS decoder in \textbf{Algorithm \ref{alg:designFlow_layer}}.
	By using the iteration-specific parameters, the MIM-LQMS decoder has the same memory requirement for storing the LUTs as that for the MIM-QMS decoder with the same precision.
	Similar to the MIM-QMS decoder, these LUTs of the iteration-specific parameters are fixed for the decoding process and do not change with the variation of the received SNRs.
	\begin{figure}[!h]
		\centering
		\includegraphics[width=3in]{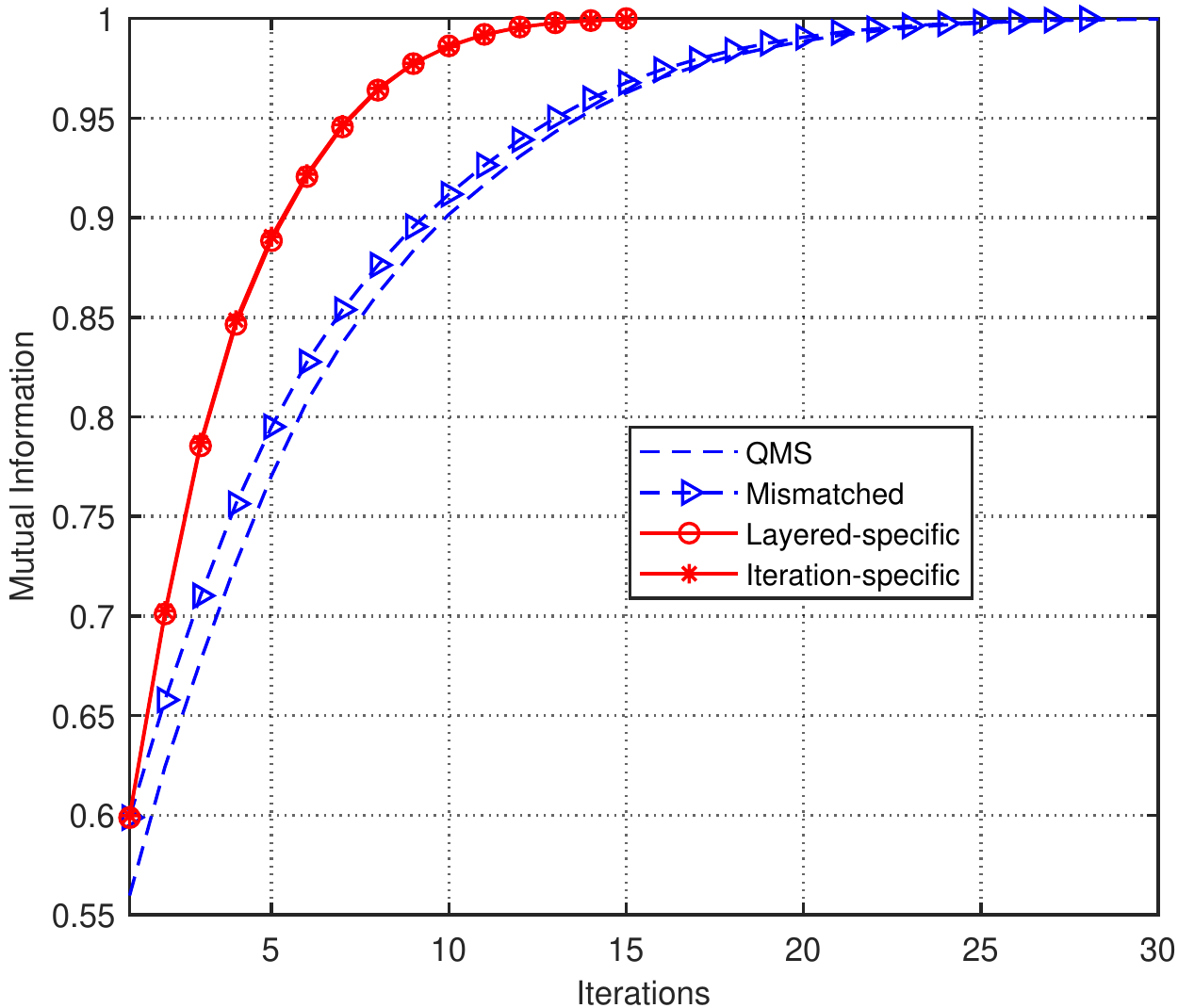}
		\caption{The Comparison of Mutual Information Evolution.} 
		\label{fig: iterMI}
		\vspace{-0.5cm}
	\end{figure}

	In Fig. \ref{fig: iterMI}, we demonstrate the mutual information evolution of the MIM-LQMS decoders with both layer-specific and iteration-specific parameters for the irregular LDPC code in \emph{Example} \ref{802.11n}.
	For comparison, we present the mutual information evolution of the MIM-QMS decoder and the MIM-LQMS decoder with mismatched parameters.
	Note that the simulated SNR for all decoders are selected as ${E_b}/{N_0} = 1.4$ dB.
	For the MIM-LQMS decoder with the layer-specific parameters, we present the average mutual information of $N_b$ layers at each iteration.
	Simulation results show that the mismatched reconstruction functions and threshold sets incur a significant loss of mutual information.
	However, the MIM-LQMS decoders reduce the number of iterations to converge into half compared to the MIM-QMS decoder, which is consistent with the results shown in \cite{Mansour2003cnLayer,Zhang2009cnlayer,Zhang2005vnLayer}.
	More importantly, the MIM-LQMS decoder with iteration-specific parameters has nearly the same convergence speed with neglectable loss of the mutual information compared to the layer-specific design.
	This indicates that the MIM-LQMS decoder with iteration-specific parameters has a high potential for practical implementation.
	\subsection{Simulation Results of MIM-LQMS Decoders}
	\label{section: simulation results}
\begin{figure}[t!]
		\centering
		\includegraphics[width=3in]{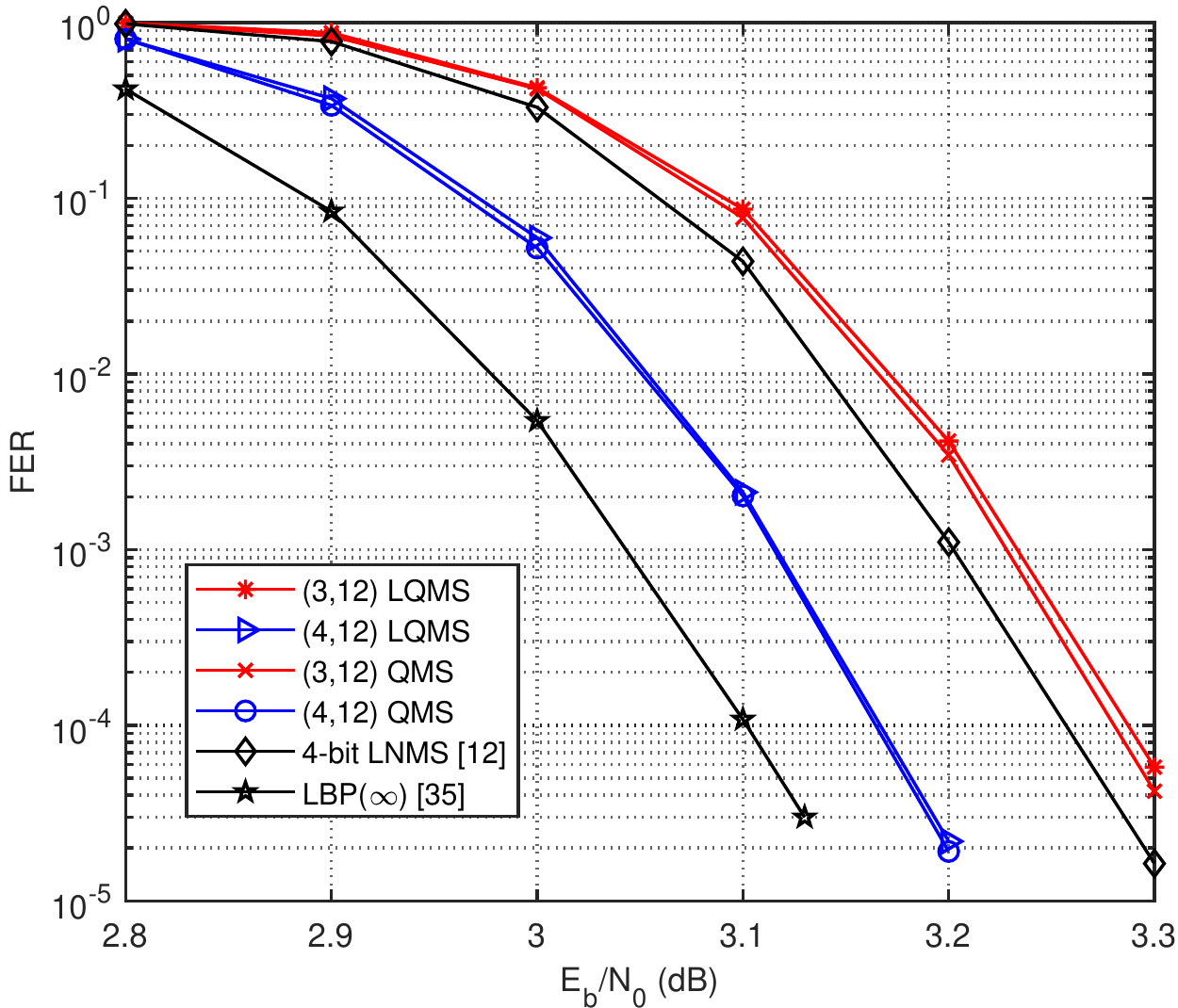}
		\caption{FER of the length-$17664$ IEEE 802.3ca LDPC code \cite{IEEESTD802_3ca} with code rate $0.826$ over the AWGN channels.}
		\label{fig:802_3ca_awgn}
	\end{figure}

\begin{figure}[t!]
		\centering
		\includegraphics[width=3in]{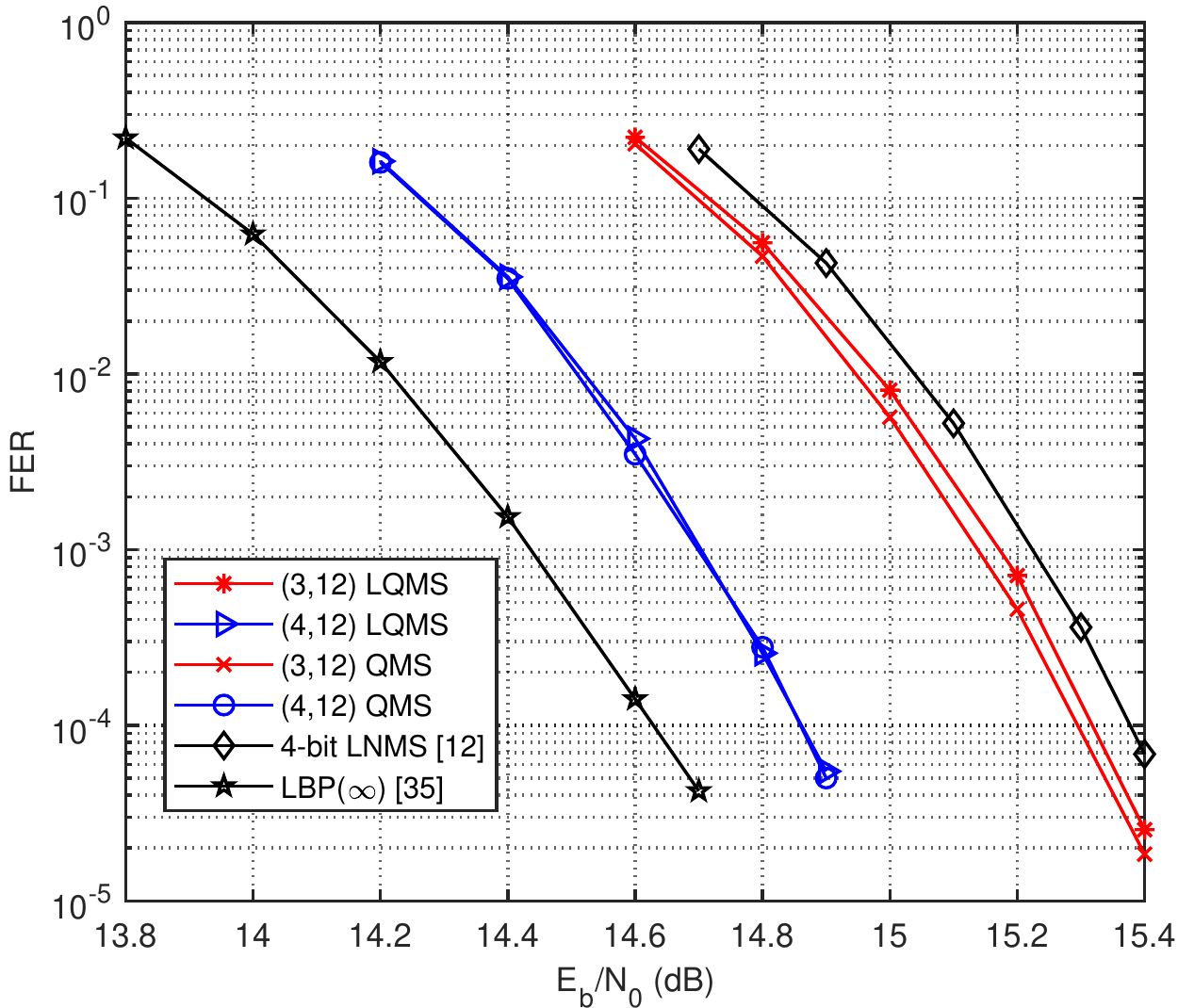}
		\caption{FER of the length-$17664$ IEEE 802.3ca LDPC code \cite{IEEESTD802_3ca} with code rate $0.826$ over the fast fading channels.}
		\label{fig:802_3ca_fading}
	\end{figure}

	\begin{table*}[!h]
		\Large
		\renewcommand{\arraystretch}{1.2}
		\caption{Degree Distributions, $N_b$, and Noise Standard Deviation ${\sigma _d}$ for MIM quantized Decoders}
		\label{table: simulation parameters2}
		\centering
		\resizebox{\textwidth}{!}{
			\begin{tabular}{|c|c|c|c|cc|cc|}
				\hline
				\multirow{2}{*}{} 
				& \multirow{2}{*}{Code rates} & \multirow{2}{*}{Degree distributions ($\rho (x)$, $\theta (x)$)} & \multirow{2}{*}{$N_b$}
				& \multicolumn{2}{c|}{ MIM-LQMS, ${\sigma _d}$} & \multicolumn{2}{c|}{MIM-QMS, ${\sigma _d}$}                 \\ \cline{5-8} 
				& & & & \multicolumn{1}{c|}{($3,12$)}	& ($4,12$) & \multicolumn{1}{c|}{($3,12$)} & ($4,12$) \\ \hline
				\multirow{2}{*}{\emph{Example 2}} & \multirow{2}{*}{$1/2$} 
				& $\rho (x) = {\rm{0}}{\rm{.1039}}{x^3} + {\rm{0}}{\rm{.1948}}{x^4} + {\rm{0}}{\rm{.2338}}{x^5} + {\rm{0}}{\rm{.2078}}{x^{7}} + {\rm{0}}{\rm{.2597}}{x^{9}}$                                  & \multirow{2}{*}{$22$}   & \multicolumn{1}{c|}{\multirow{2}{*}{$0.9821$}} & \multirow{2}{*}{$1.0042$} & \multicolumn{1}{c|}{\multirow{2}{*}{$0.9837$}} & \multirow{2}{*}{$1.0115$} \\
				& & $\theta (x) = {\rm{0}}{\rm{.1039}} + {\rm{0}}{\rm{.0260}}{x} + {\rm{0}}{\rm{.1169}}{x^2} + {\rm{0}}{\rm{.2078}}{x^{3}} + {\rm{0}}{\rm{.1299}}{x^{4}} + {\rm{0}}{\rm{.1818}}{x^{6}} + {\rm{0}}{\rm{.2338}}{x^{8}}$ & &  \multicolumn{1}{c|}{} &  & \multicolumn{1}{c|}{} & \\ \hline
				\multirow{6}{*}{\emph{Example 3}} 
				& \multirow{2}{*}{$1/2$}    
				& $\rho (x) = {\rm{0}}{\rm{.8140}}{x^6} + {\rm{0}}{\rm{.1860}}{x^7}$ & \multirow{6}{*}{$24$}
				& \multicolumn{1}{c|}{\multirow{2}{*}{$0.8362$}} & \multirow{2}{*}{$0.8750$} & \multicolumn{1}{c|}{\multirow{2}{*}{$0.8379$}} & \multirow{2}{*}{$0.8761$} \\
				& 		& $\theta (x) = {\rm{0}}{\rm{.2558}}x + {\rm{0}}{\rm{.3140}}{x^2} + {\rm{0}}{\rm{.0465}}{x^3} + {\rm{0}}{\rm{.3837}}{x^{10}}$ & &\multicolumn{1}{c|}{} & & \multicolumn{1}{c|}{} & \\ \cline{2-3} \cline{5-8} 
				& \multirow{2}{*}{$2/3$}       
				& $\rho (x) = {x^{10}}$  & & \multicolumn{1}{c|}{\multirow{2}{*}{$0.6889$}} & \multirow{2}{*}{$0.7112$} & \multicolumn{1}{c|}{\multirow{2}{*}{$0.6914$}} & \multirow{2}{*}{$0.7112$} \\
				& & $\theta (x) = {\rm{0}}{\rm{.1591}}x + {\rm{0}}{\rm{.4091}}{x^2} + {\rm{0}}{\rm{.1591}}{x^6} + {\rm{0}}{\rm{.2727}}{x^{7}}$ & &\multicolumn{1}{c|}{} & & \multicolumn{1}{c|}{} & \\ \cline{2-3} \cline{5-8} 
				& \multirow{2}{*}{$5/6$}       
				& $\rho (x) = {\rm{0}}{\rm{.7412}}{x^{20}} + {\rm{0}}{\rm{.2588}}{x^{21}}$ & & \multicolumn{1}{c|}{\multirow{2}{*}{$0.5413$}} & \multirow{2}{*}{$0.5502$} & \multicolumn{1}{c|}{\multirow{2}{*}{$0.5420$}} & \multirow{2}{*}{$0.5511$} \\					
				&       & $\theta (x) = {\rm{0}}{\rm{.0706}}x + {\rm{0}}{\rm{.1765}}{x^2} + {\rm{0}}{\rm{.7529}}{x^3}$ & & \multicolumn{1}{c|}{} & &\multicolumn{1}{c|}{} & \\ \hline
				\multirow{2}{*}{\emph{Example 4}} 
				& \multirow{2}{*}{$0.826$}
				& $\rho (x) = {\rm{0}}{\rm{.0800}}{x^{21}} + {\rm{0}}{\rm{.9200}}{x^{22}}$ & \multirow{2}{*}{$69$}   & \multicolumn{1}{c|}{\multirow{2}{*}{$0.5528$}} & \multirow{2}{*}{$0.5599$} & \multicolumn{1}{c|}{\multirow{2}{*}{$0.5529$}} & \multirow{2}{*}{$0.5599$} \\  				
				&       & $\theta (x) = {\rm{0}}{\rm{.5455}}{x^2} + {\rm{0}}{\rm{.3709}}{x^3} + {\rm{0}}{\rm{.0400}}{x^{10}} + {\rm{0}}{\rm{.0436}}{x^{11}}$  & &\multicolumn{1}{c|}{} & & \multicolumn{1}{c|}{} & \\ \hline
			\end{tabular}
		}\vspace{-0.2cm}
	\end{table*}
	In this section, we evaluate the FER performance of the ($3,12$) and ($4,12$)  MIM-LQMS decoders for both the AWGN channels and the fast fading channels via Monte-Carlo simulations.
	We assume the BPSK modulation is adopted and the $j$-th received signal of the fading channel is given by
	\begin{equation}
		{y_j} = h_j \cdot {x_j} + {n_j}, \nonumber
	\end{equation}
	where $h_j$ is the fading coefficient following ${h_j} \sim \mathcal{N}(0,1)$, and $n_j$ is the AWGN with ${n_j} \sim \mathcal{N}(0,{N_0})$.
	Note that $N_0$ is the single-sided noise power spectral density.
	We consider three examples, where the simulated LDPC codes in each example have short, moderate, and long block lengths, respectively.
	More specifically, a length-$560$ 5G LDPC code is considered in \emph{Example} \ref{5g}, which has a code rate of $1/2$ after rate matching \cite{5gChannel1}.
	In \emph{Example} \ref{802_11n2}, the length-$1296$ LDPC codes are selected from the IEEE 802.11n standard \cite{IEEESTD802_11n} with code rates $1/2, 2/3$, and $5/6$, respectively.
	Furthermore, we also consider the length-$17664$ IEEE 802.3ca standard LDPC code \cite{IEEESTD802_3ca} in \emph{Example} \ref{802_3ca}, which has a code rate of $0.826$ without puncturing and shortening. 
	For comparison, we include the FER performance of the floating-point layered BP (LBP) decoder in \cite{Zhang2005vnLayer}, the 4-bit layered normalized min-sum (LNMS) decoder \cite{Jinghu2005reduced}, and the MIM-QMS decoders as references.
	We set the maximum number of iterations $I_{\text{max}}=15$ for all layered decoders, and $I_{\text{max}}=30$ for the MIM-QMS decoders.
	Following the notations in Section \ref{MIM quantized fer}, we summarize the degree distributions and $N_b$ of the LDPC codes and also present ${\sigma _d}$ for the corresponding MIM quantized decoders in TABLE \ref{table: simulation parameters2}.
	(See Appendix B in \cite{kang2020generalized} for details of the constructed LUTs.)
	%
	%
	\begin{figure}[h!]
		\centering
		\includegraphics[width=3in]{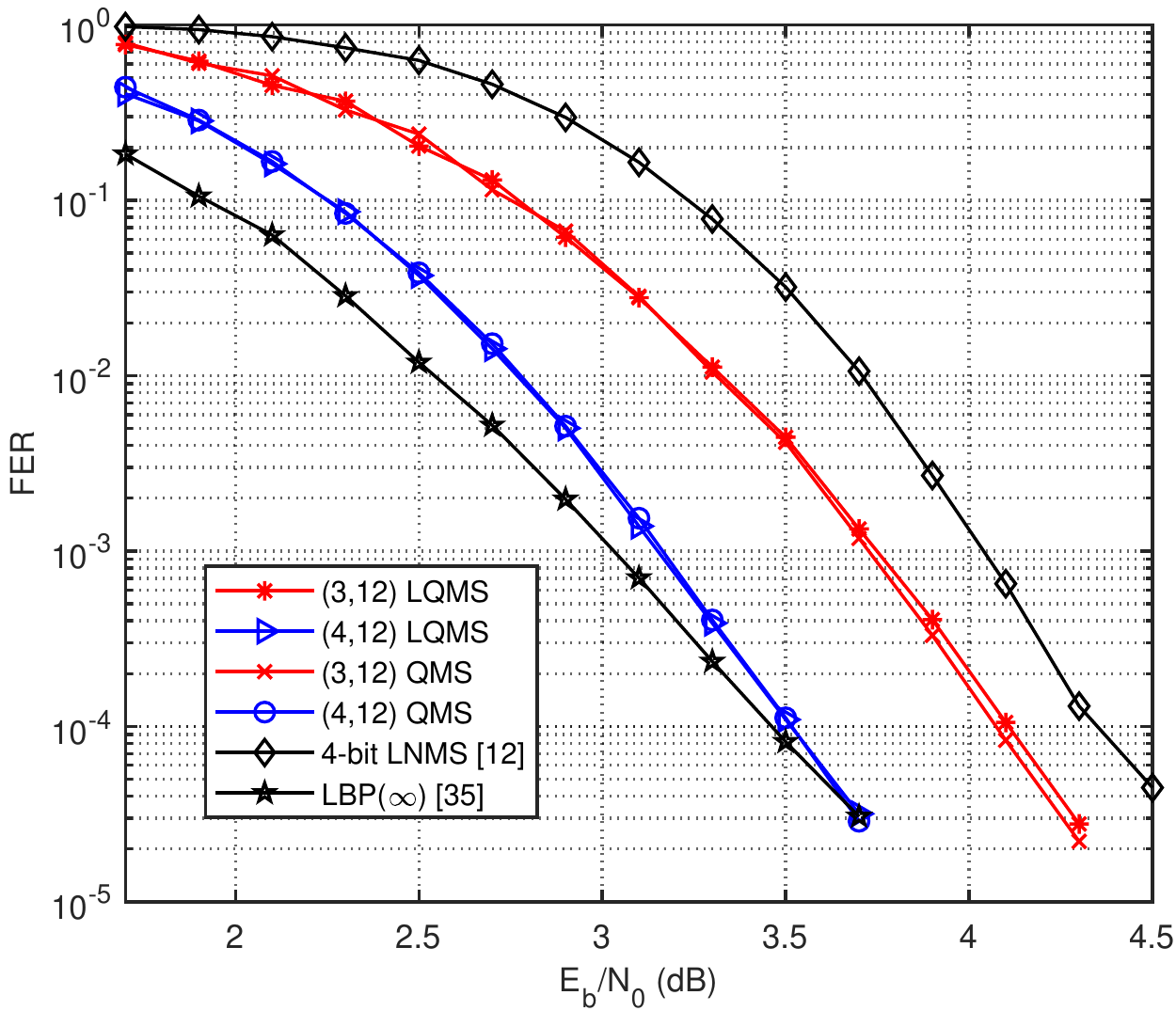}
		\caption{FER of the length-$560$ 5G LDPC code \cite{5gChannel1} with code rate $1/2$ over the AWGN channels.}
		\label{fig:5g_awgn}
		\vspace{-0.4cm}
	\end{figure}
	\begin{figure}[h!]
		\centering
		\includegraphics[width=3in]{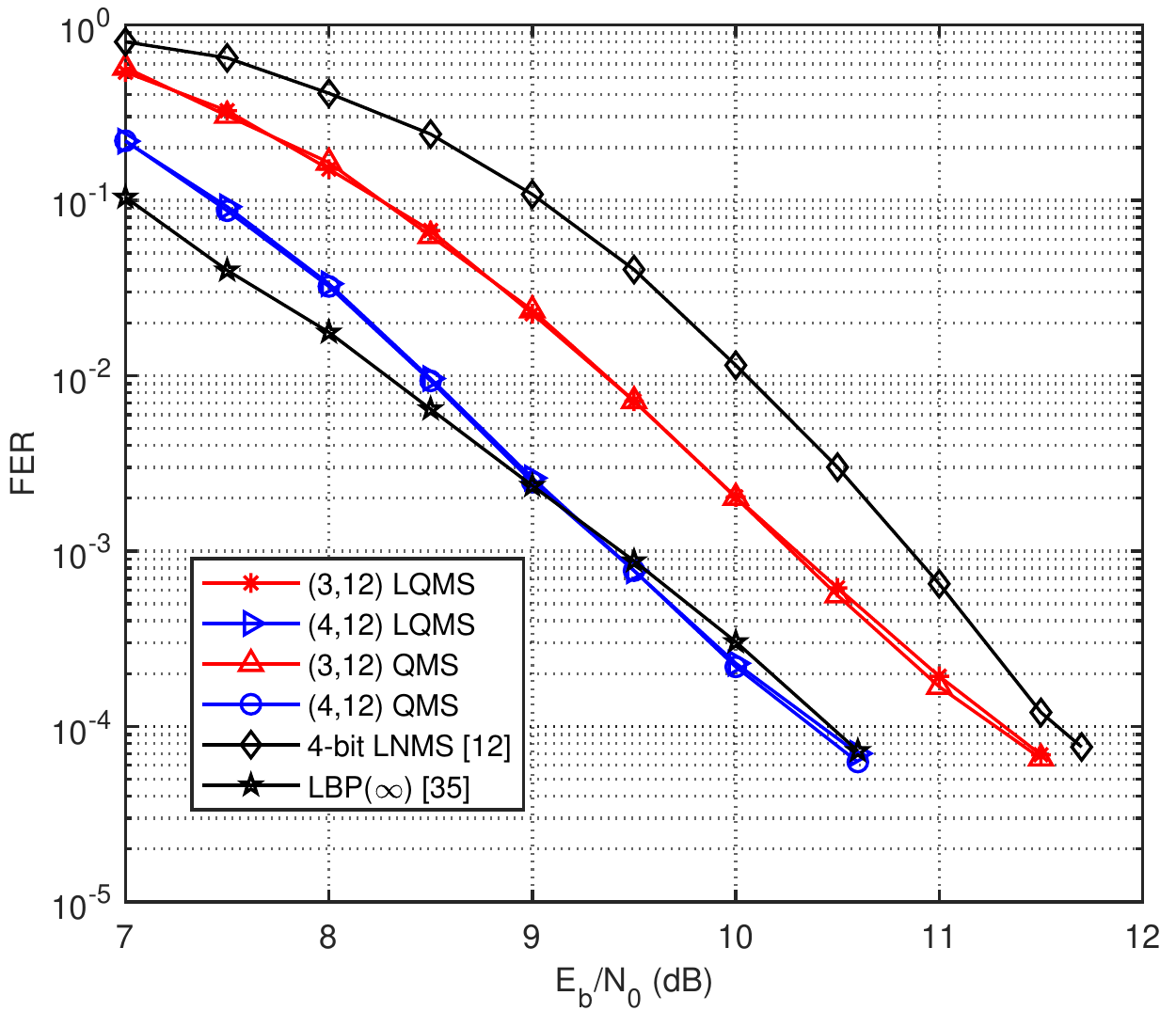}
		\caption{FER of the length-$560$ 5G LDPC code \cite{5gChannel1} with code rate $1/2$ over the fast fading channels.}
		\label{fig:5g_fading}
	\end{figure}
	\begin{example}\label{5g}
		Fig. \ref{fig:5g_awgn} and Fig. \ref{fig:5g_fading} show the FERs of the length-$560$ 5G LDPC code \cite{5gChannel1} over the AWGN channels and the fast fading channels, respectively.
		The scaling factor of the LNMS decoder is optimized as $0.7$.
		We can see that the $(3,12)$ and $(4,12)$ MIM-LQMS decoders have almost identical FER performance compared to the MIM-QMS decoders with the same precision.
		The $(3,12)$ MIM-LQMS decoder outperforms the 4-bit LNMS decoder by about $0.4$ dB and $0.6$ dB at FER $= 10^{-2}$, over the AWGN channels and the fast fading channels, respectively.
		Moreover, the $(4,12)$ MIM-LQMS decoder achieves performance close to the LBP($\infty$) decoder within $0.25$ dB and surpasses the 4-bit LNMS decoder by about $0.9$ dB over the AWGN channels.
		For the fast fading channels, the $(4,12)$ MIM-LQMS decoder achieves a performance gain of about $1.5$ dB at FER $= 10^{-2}$ compared to the 4-bit LNMS decoder, and it approaches the performance of the LBP($\infty$) decoder within $0.1$ dB at FER $= 10^{-4}$.
	\end{example}
	\begin{example}\label{802_11n2}
		Fig. \ref{fig:802_11n_awgn} and Fig. \ref{fig:802_11n_fading} illustrate the FERs of the proposed MIM-LQMS decoders for the length-$1296$ IEEE 802.11n LDPC codes with different code rates over the AWGN channels and the fast fading channels, respectively.
		The scaling factor of the LNMS decoder is set to $0.8$.
		We observe that the $(3,12)$ and $(4,12)$ MIM-LQMS decoders have nearly the same FER performance compared to the MIM-QMS decoders with the same precision.
		More importantly, the performance gaps between the MIM-LQMS decoders and the LBP($\infty$) decoder reduces with the increase of the code rates.
		For the AWGN channels, the FER performance of the $(3,12)$ MIM-LQMS decoder is about $0.4$ dB away from that of the LBP($\infty$) decoder for code rate $1/2$, while it approaches the FER of the LBP($\infty$) decoder within $0.3$ and $0.2$ dB for the code rates of $2/3$ and $5/6$, respectively.
		Moreover, the performance gap between the $(4,12)$ MIM-LQMS decoder and the LBP($\infty$) decoder is less than $0.2$ and $0.1$ dB compared with that of the LBP($\infty$) decoder for code rates $1/2$ and $2/3$, respectively.
		For code rate $5/6$, the $(4,12)$ MIM-LQMS decoder has almost the same FER performance compared to the LBP($\infty$) decoder.
		In addition, compared to the 4-bit LNMS decoder, the $(4,12)$ MIM-LQMS decoder has a performance gain of about $0.4$ dB, $0.3$ dB and $0.2$ dB at FER $=10^{-2}$, for the code rates of $1/2$, $2/3$, and $5/6$, respectively.
		The $(3,12)$ MIM-LQMS decoder can still obtain $0.1$ dB gain at FER $=10^{-2}$ for the code rates of $1/2$ and $2/3$ when compared with the 4-bit LNMS decoder.\vspace{-0.3cm}
		\begin{figure}[t!]
			\centering
			\includegraphics[width=3in]{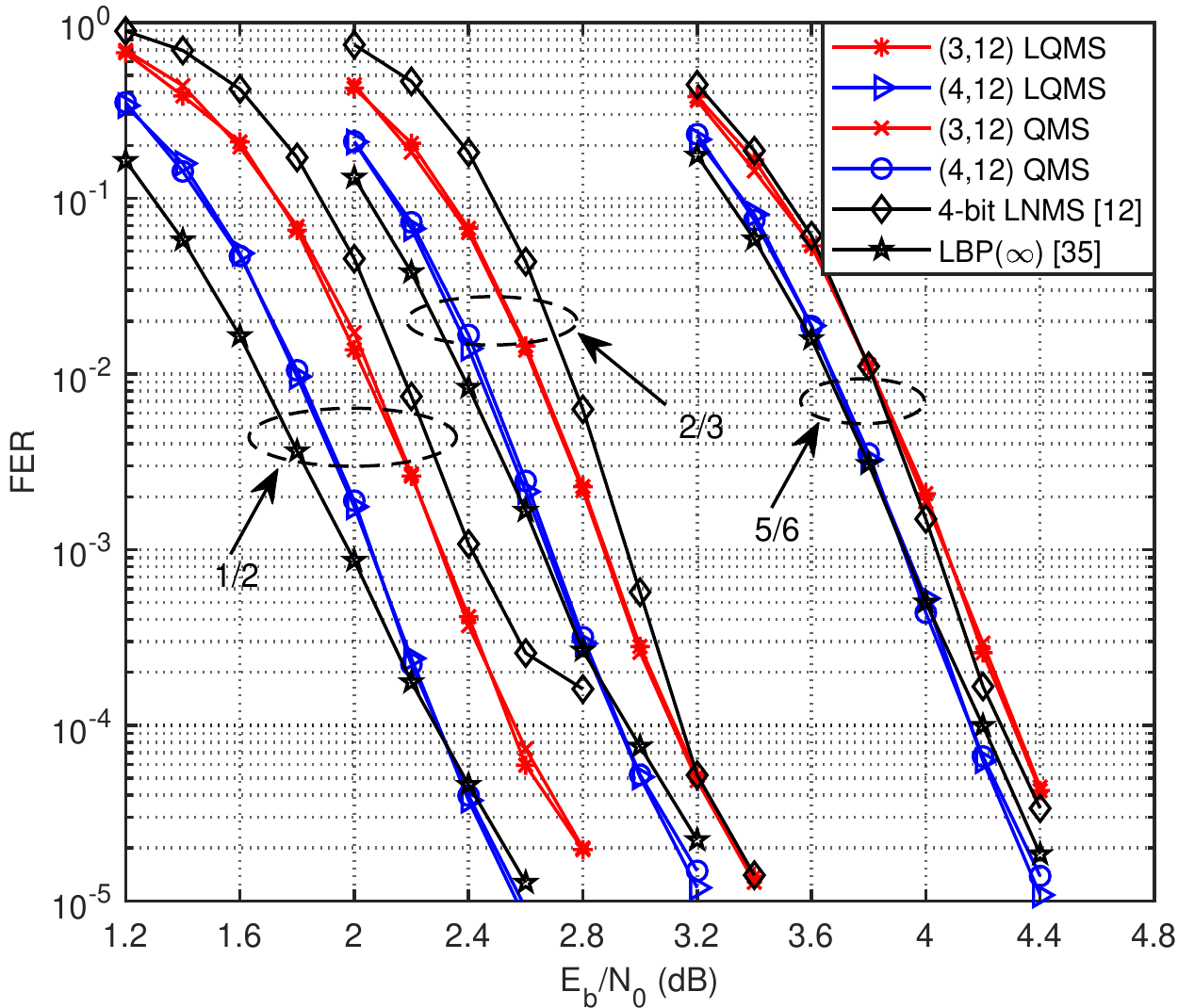}
			\caption{FER of the length-$1296$ IEEE 802.11n LDPC codes \cite{IEEESTD802_11n} with code rates $1/2$, $2/3$, and $5/6$ over the AWGN channels.}
			\label{fig:802_11n_awgn}
		\end{figure}
		\begin{figure}[t!]
			\centering
			\includegraphics[width=3in]{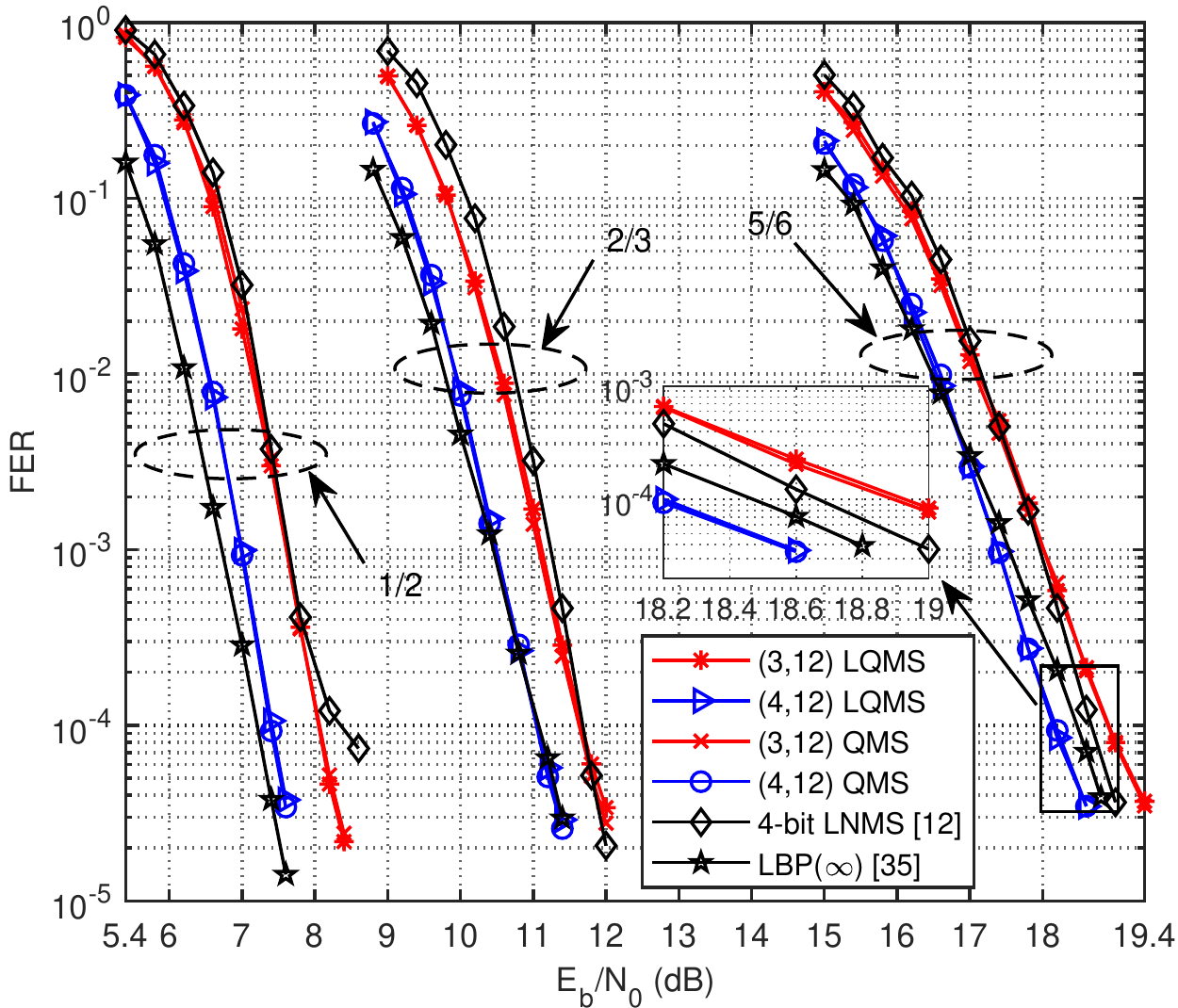}
			\caption{FER of the length-$1296$ IEEE 802.11n LDPC codes \cite{IEEESTD802_11n} with code rates $1/2$, $2/3$, and $5/6$ over the fast fading channels.}
			\label{fig:802_11n_fading}
			\vspace{-0.4cm}
		\end{figure}

		With regard to the fast fading channels, the $(3,12)$ MIM-LQMS decoder approaches the FER performance of the LBP($\infty$) decoder by about $1$ dB, $0.7$ dB, and $0.6$ dB for the code rates of $1/2$, $2/3$ and $5/6$, respectively.
		Furthermore, the $(4,12)$ MIM-LQMS decoder has the FER performance of less than $0.3$ dB and $0.15$ dB away compared to the LBP($\infty$) decoder for code rates $1/2$ and $2/3$, respectively.
		For code rate $5/6$, the $(4,12)$ MIM-LQMS decoder achieves nearly the same FER performance as the LBP($\infty$) decoder at FER $=10^{-2}$, and outperforms the LBP($\infty$) decoder by $0.25$ dB at FER $=10^{-4}$.
		We also observe that the $(4,12)$ MIM-LQMS decoder outperforms the 4-bit LNMS decoder by about $0.6$ dB for all code rates, and even the $(3,12)$ MIM-LQMS decoder can perform slightly better than the 4-bit LNMS decoder for all code rates in the low-to-moderate SNR region.
	\end{example}
	\begin{example}\label{802_3ca}
		In Fig. \ref{fig:802_3ca_awgn} and Fig. \ref{fig:802_3ca_fading}, we compare the FER performance of the proposed MIM-LQMS decoders to that of various decoders over the AWGN channels and the fast fading channels. 
		Note that we optimize the scaling factor of the LNMS decoder as $0.65$.
		It is shown that the $(3,12)$ and $(4,12)$ MIM-LQMS decoders have almost the same FER performance as the $(3,12)$ and $(4,12)$ MIM-QMS decoders, respectively.
		The performance gap between the $(3,12)$ MIM-LQMS decoder and the LBP($\infty$) decoder is about $0.2$ dB over the AWGN channels, and about $0.8$ dB over the fast fading channels.
		Furthermore, the $(4,12)$ MIM-LQMS decoder can approach the FER performance of the LBP($\infty$) decoder within $0.1$ and $0.3$ dB over the AWGN channels and the fast fading channels, respectively.
		Compared to the 4-bit LNMS decoder, the $(3,12)$ MIM-LQMS decoder shows negligible performance loss over the AWGN channel, and it achieves the performance gain of $0.1$ dB for the fast fading channels.
		In addition, the $(4,12)$ MIM-LQMS decoder also performs better than the 4-bit LNMS decoder by about $0.5$ dB over the fast fading channel.
	\end{example}
	\section{Conclusion}
	\label{section: conclusion}
	In this paper, we have proposed a generalized framework of the MIM quantized decoding for the LDPC codes to improve the error performance of the MIM-LUT decoding.
	The proposed decoding method can be designed based on either the BP or the MS algorithm, and implemented by only simple mappings and fixed-point additions.
	Particularly, we have proposed the MIM-DE to construct the LUTs of the reconstruction and quantization mappings for the node updates. 
	For practical concerns, we discussed the computational complexity and the memory requirements of the MIM quantized decoders.
	To speed up the convergence of the decoder for QC-LDPC codes, we developed the LMIM-DE and adopted it to design the LUTs for the MIM-LQMS decoder.
	We also presented an optimization method to reduce the number of LUTs required for the VN update of the MIM-LQMS decoder, which achieves less memory requirement.
	Simulation results show that the MIM quantized decoders with 3-bit and 4-bit precisions outperform the state-of-the-art LUT decoders with the same message precision with respect to the error performance in the low-to-moderate SNR regions and the convergence speed for small decoding iterations.
	Moreover, the 4-bit MIM-LQMS decoder achieves the FER performance of less than $0.3$ dB away from the floating-point LBP decoder in the moderate-to-high SNR regions.
	%
	

	\bibliographystyle{ieeetr}

	\newpage
	\begin{appendices}
		\section{Constructed LUTs for MIM quantized decoders with flooding schedule}
		\subsection{The 3-bit MIM-QBP decoder in Example 1}
		\begin{table}[h!]
			\footnotesize
			\centering
			\setstretch{1.1}
			\caption{Reconstruction Functions $\phi_c$, $\phi_v$, and $\phi_{ch}$}
			\vspace{0.2cm}
\begin{tabular}{l|cccccccc}
				\hline
				\multicolumn{1}{l|}{\multirow{2}{*}{Iteration}} & \multicolumn{8}{c}{$\phi_c(r)$, $r \in \mc{R}$} \\ \cline{2-9} 
				\multicolumn{1}{l|}{}                           & 0  & 1  & 2  & 3   & 4    & 5   & 6   & 7  \\ \hline
				1	& 4  & 26  & 85  & 255  & -255  & -85  & -26  & -4  \\ 
				2	& 3  & 24  & 81  & 244  & -255  & -83  & -24  & -3  \\ 
				3	& 3  & 23  & 83  & 255  & -252  & -80  & -22  & -3  \\ 
				4	& 2  & 21  & 78  & 243  & -255  & -81  & -22  & -2  \\ 
				5	& 2  & 18  & 71  & 237  & -255  & -77  & -19  & -2  \\ 
				6	& 1  & 14  & 60  & 255  & -102  & -34  & -9  & -1  \\
\hline
			\end{tabular}
		\end{table}
		\begin{table}[h!]
			\centering
			\footnotesize
			\setstretch{1.1}
			\begin{tabular}{l|cccccccc}
				\hline
				\multirow{2}{*}{Iteration} & \multicolumn{8}{c}{$\phi_v(s)$, $s \in \mc{S}$}    \\ \cline{2-9} 
				& 0   & 1   & 2  & 3  & 4   & 5   & 6    & 7    \\ \hline
				1	& 62  & 35  & 21  & 8  & 0  & -9  & -28  & -57  \\ 
				2	& 68  & 34  & 11  & 1  & -10  & -25  & -42  & -74  \\ 
				3	& 75  & 38  & 13  & 1  & -11  & -27  & -46  & -81  \\ 
				4	& 85  & 46  & 28  & 10  & -5  & -19  & -45  & -85  \\ 
				5	& 94  & 52  & 32  & 13  & -1  & -15  & -44  & -87  \\ 
				6	& 106  & 70  & 44  & 23  & 0  & -23  & -49  & -93  \\
\hline
			\end{tabular}
		\end{table}		
		\begin{table}[h!]
			\centering
			\footnotesize
			\setstretch{1.1}
			\begin{tabular}{l|cccccccc}
				\hline
				\multirow{2}{*}{Iteration} & \multicolumn{8}{c}{$\phi_{ch}(l)$, $l \in \mc{L}$}    \\ \cline{2-9}
				& 0   & 1  & 2  & 3  & 4   & 5   & 6   & 7    \\ \hline
				1-15                       & 170 & 99 & 54 & 17 & -17 & -54 & -99 & -170 \\
				16	& 136  & 79  & 43  & 14  & -14  & -43  & -79  & -136  \\ 
				17	& 134  & 78  & 42  & 14  & -14  & -42  & -78  & -134  \\ 
				18	& 131  & 76  & 41  & 13  & -13  & -41  & -76  & -131  \\ 
				19	& 131  & 76  & 41  & 13  & -13  & -41  & -76  & -131  \\ 
				20	& 130  & 76  & 41  & 13  & -13  & -41  & -76  & -130  \\
\hline
			\end{tabular}
		\end{table}		
		\newpage
		\begin{table}[htbp]
			\centering
			\footnotesize
			\setstretch{1.1}
			\caption{Threshold sets $\Gamma_c$, $\Gamma_v$, and $\Gamma_{ch}$}
			\begin{tabular}{l|ccccccc}
				\hline
				\multirow{2}{*}{Iteration} & \multicolumn{7}{c}{$\Gamma_c$}                                                          \\ \cline{2-8} 
				& $\gamma_1$ & $\gamma_2$ & $\gamma_3$ & $\gamma_4$ & $\gamma_5$ & $\gamma_6$ & $\gamma_7$ \\ \hline
				1	& 94  & 175  & 279  & 485  & -470  & -252  & -127  \\ 
				2	& 84  & 204  & 420  & -472  & -254  & -159  & -81  \\ 
				3	& 81  & 201  & 420  & -479  & -249  & -152  & -75  \\ 
				4	& 54  & 134  & 232  & 564  & -318  & -164  & -69  \\ 
				5	& 48  & 123  & 222  & 448  & -401  & -182  & -76  \\ 
				6	& 23  & 51  & 97  & 205  & -203  & -91  & -38  \\
\hline
			\end{tabular}
		\end{table}		
		\begin{table}[h!]
			\centering
			\footnotesize
			\setstretch{1.1}
			\begin{tabular}{l|ccccccc}
				\hline
				\multirow{2}{*}{Iteration} & \multicolumn{7}{c}{$\Gamma_v$}                                                          \\ \cline{2-8} 
				& $\gamma_1$ & $\gamma_2$ & $\gamma_3$ & $\gamma_4$ & $\gamma_5$ & $\gamma_6$ & $\gamma_7$ \\ \hline
				1	& 134  & 78  & 38  & 2  & -33  & -74  & -131  \\ 
				2	& 138  & 77  & 35  & -1  & -38  & -81  & -143  \\ 
				3	& 146  & 81  & 38  & 2  & -35  & -78  & -141  \\ 
				4	& 152  & 86  & 40  & 1  & -38  & -84  & -152  \\ 
				5	& 145  & 75  & 28  & -12  & -50  & -95  & -164  \\ 
				6	& 158  & 86  & 38  & -1  & -43  & -89  & -159  \\
\hline
			\end{tabular}
		\end{table}
		\begin{table}[h!]
			\centering
			\small
			\setstretch{1.1}
			\begin{tabular}{|c|c|c|c|c|c|c|}
				\hline
				\multicolumn{7}{|c|}{$\Gamma_{ch}$ (in LLR format)}                                                         \\ \hline
				$\gamma_1$ & $\gamma_2$ & $\gamma_3$ & $\gamma_4$ & $\gamma_5$ & $\gamma_6$ & $\gamma_7$ \\ \hline
				3.36  & 1.94  & 0.90  & 0  & -0.90  & -1.94  & -3.36 \\ \hline
			\end{tabular}
		\end{table}
		\begin{table}[h!]
			\footnotesize
			\setstretch{1.2}
			\centering
			\begin{tabular}{l|c}
				\hline
				\multirow{2}{*}{Iteration} & $\Gamma_e$ \\ \cline{2-2} 
				& $\gamma_1$ \\ \hline
				1	& 2  \\ 
				2	& -1  \\ 
				3	& 2  \\ 
				4	& -1  \\ 
				5	& 0  \\ 
				6	& 1  \\ 
				7	& 2  \\ 
				8	& 0  \\ 
				9	& 0  \\ 
				10	& 1  \\ 
				11	& 0  \\ 
				12	& 1  \\ 
				13	& 1  \\ 
				14	& 0  \\ 
				15	& 1  \\ 
				16	& 1  \\ 
				17	& 0  \\ 
				18	& 1  \\ 
				19	& 1  \\ 
				20	& 1  \\
				21	& 1  \\ 
				22	& 0  \\ 
				23	& -1  \\ 
				24	& 0  \\ 
				25	& 1  \\ 
				26	& 0  \\ 
				27	& 1  \\ 
				28	& 1  \\ 
				29	& 2  \\ 
				30	& 2  \\ 
				31	& 0  \\ 
				32	& 0  \\ 
				33	& 1  \\ 
				34	& 0  \\ 
				35	& 1  \\ 
				36	& 0  \\ 
				37	& 1  \\ 
				38-42	& 0  \\
				43	& 1  \\ 
				44	& 0  \\ 
				45	& 1  \\ 
				46-50	& 0  \\ \hline
			\end{tabular}
		\end{table}
		\clearpage
		\subsection{The 4-bit MIM-QBP decoder in Example 1}
		\begin{table}[!h]
			\centering
			\huge
			\setstretch{1.2}
			\caption{Reconstruction Functions $\phi_c$, $\phi_v$, and $\phi_{ch}$}
			\resizebox{0.5\textwidth}{!}{
			\begin{tabular}{l|cccccccccccccccc}
				\hline
				\multicolumn{1}{l|}{\multirow{2}{*}{Iteration}} & \multicolumn{16}{c}{$\phi_c(r)$, $r \in \mc{R}$} \\ \cline{2-17} 
				\multicolumn{1}{l|}{}  & 0  & 1  & 2  & 3   & 4    & 5   & 6   & 7 & 8 & 9 & 10 & 11 & 12 & 13 & 14 & 15 \\ \hline
				1	& 1  & 5  & 12  & 24  & 43  & 74  & 128  & 255  & -235  & -122  & -72  & -42  & -24  & -12  & -5  & -1  \\ 
				2	& 1  & 5  & 12  & 24  & 43  & 74  & 125  & 239  & -255  & -127  & -75  & -44  & -24  & -11  & -4  & -1  \\ 
				3	& 1  & 4  & 11  & 24  & 45  & 77  & 131  & 255  & -245  & -125  & -73  & -42  & -23  & -10  & -4  & -1  \\ 
				4	& 1  & 4  & 11  & 23  & 42  & 74  & 127  & 255  & -233  & -121  & -71  & -41  & -22  & -10  & -3  & -1  \\ 
				5	& 1  & 1  & 6  & 14  & 31  & 62  & 117  & 255  & -240  & -118  & -64  & -32  & -15  & -6  & -2  & -1  \\ 
				6	& 1  & 1  & 4  & 9  & 18  & 34  & 62  & 116  & -255  & -91  & -46  & -22  & -10  & -4  & -1  & -1  \\
\hline
			\end{tabular}
		}
		\end{table}
		\begin{table}[!h]
			\centering
			\huge
			\setstretch{1.2}
			\resizebox{0.5\textwidth}{!}{
			\begin{tabular}{l|cccccccccccccccc}
				\hline
				\multicolumn{1}{l|}{\multirow{2}{*}{Iteration}} & \multicolumn{16}{c}{$\phi_v(s)$, $s \in \mc{S}$} \\ \cline{2-17} 
				\multicolumn{1}{l|}{}  & 0  & 1  & 2  & 3   & 4    & 5   & 6   & 7 & 8 & 9 & 10 & 11 & 12 & 13 & 14 & 15 \\ \hline
				1	& 63  & 45  & 34  & 25  & 18  & 12  & 7  & 3  & 0  & -4  & -9  & -15  & -22  & -31  & -42  & -60  \\ 
				2	& 77  & 56  & 43  & 32  & 24  & 16  & 10  & 4  & 0  & -5  & -11  & -18  & -28  & -40  & -54  & -76  \\ 
				3	& 86  & 61  & 46  & 35  & 26  & 18  & 11  & 4  & 0  & -5  & -12  & -20  & -30  & -43  & -58  & -83  \\ 
				4	& 89  & 62  & 45  & 32  & 21  & 13  & 5  & 0  & -5  & -12  & -19  & -28  & -36  & -47  & -63  & -89  \\ 
				5	& 102  & 74  & 55  & 40  & 30  & 21  & 14  & 6  & -1  & -7  & -16  & -25  & -39  & -55  & -74  & -102  \\ 
				6	& 111  & 80  & 61  & 47  & 36  & 26  & 16  & 8  & 0  & -10  & -18  & -29  & -42  & -58  & -79  & -111  \\
\hline
			\end{tabular}
		}
		\end{table}
		\begin{table}[!h]
			\centering
			\huge
			\setstretch{1.2}
			\resizebox{0.5\textwidth}{!}{
			\begin{tabular}{l|cccccccccccccccc}
				\hline
				\multicolumn{1}{l|}{\multirow{2}{*}{Iteration}} & \multicolumn{16}{c}{$\phi_{ch}(l)$, $l \in \mc{L}$} \\ \cline{2-17} 
				\multicolumn{1}{l|}{}  & 0  & 1  & 2  & 3   & 4    & 5   & 6   & 7 & 8 & 9 & 10 & 11 & 12 & 13 & 14 & 15 \\ \hline
				1-19	& 169  & 121  & 92  & 70  & 52  & 36  & 21  & 6  & -8  & -22  & -37  & -53  & -71  & -93  & -122  & -170  \\ 
				20	& 169  & 121  & 92  & 70  & 52  & 35  & 20  & 6  & -8  & -22  & -36  & -52  & -71  & -92  & -122  & -169  \\ 
				21	& 165  & 119  & 90  & 68  & 51  & 35  & 20  & 6  & -7  & -21  & -36  & -51  & -69  & -91  & -119  & -166  \\ 
				22	& 159  & 114  & 86  & 66  & 49  & 33  & 19  & 6  & -7  & -20  & -34  & -49  & -66  & -87  & -114  & -159  \\ 
				23	& 163  & 117  & 88  & 67  & 50  & 34  & 20  & 6  & -7  & -21  & -35  & -50  & -68  & -89  & -117  & -163  \\ 
				24	& 159  & 114  & 86  & 66  & 49  & 33  & 19  & 6  & -7  & -20  & -34  & -49  & -66  & -87  & -115  & -160  \\
\hline
			\end{tabular}
		}
		\end{table}
		\begin{table}[!h]
			\centering
			\huge
			\caption{Threshold sets $\Gamma_c$, $\Gamma_v$, and $\Gamma_{ch}$}
			\setstretch{1.2}
			\resizebox{0.5\textwidth}{!}{
			\begin{tabular}{l|ccccccccccccccc}
				\hline
				\multirow{2}{*}{Iteration} & \multicolumn{15}{c}{$\Gamma_c$}   \\ \cline{2-16} 
				& $\gamma_1$   & $\gamma_2$   & $\gamma_3$  & $\gamma_4$  & $\gamma_5$   & $\gamma_6$   & $\gamma_7$    & $\gamma_8$    & $\gamma_9$  & $\gamma_{10}$  & $\gamma_{11}$  & $\gamma_{12}$  & $\gamma_{13}$  & $\gamma_{14}$ & $\gamma_{15}$  \\ \hline	
				1	& 41  & 65  & 91  & 122  & 160  & 209  & 283  & 436  & -376  & -252  & -185  & -139  & -103  & -73  & -46  \\ 
				2	& 28  & 48  & 71  & 99  & 133  & 177  & 245  & 388  & -373  & -236  & -166  & -119  & -81  & -53  & -30  \\ 
				3	& 23  & 43  & 66  & 92  & 126  & 171  & 240  & 384  & -365  & -227  & -158  & -112  & -76  & -48  & -26  \\ 
				4	& 22  & 43  & 69  & 104  & 148  & 215  & 349  & -371  & -229  & -162  & -118  & -87  & -63  & -42  & -23  \\ 
				5	& 18  & 35  & 58  & 88  & 120  & 164  & 235  & 401  & -358  & -217  & -145  & -97  & -60  & -36  & -19  \\ 
				6	& 10  & 19  & 31  & 45  & 64  & 91  & 131  & 215  & -203  & -122  & -83  & -56  & -36  & -21  & -12  \\
\hline
			\end{tabular}
		}
		\end{table}
		\begin{table}[!h]
			\centering
			\huge
			\setstretch{1.2}
			\resizebox{0.5\textwidth}{!}{
			\begin{tabular}{l|ccccccccccccccc}
				\hline
				\multirow{2}{*}{Iteration} & \multicolumn{15}{c}{$\Gamma_v$}   \\ \cline{2-16} 
				& $\gamma_1$   & $\gamma_2$   & $\gamma_3$  & $\gamma_4$  & $\gamma_5$   & $\gamma_6$   & $\gamma_7$    & $\gamma_8$    & $\gamma_9$  & $\gamma_{10}$  & $\gamma_{11}$  & $\gamma_{12}$  & $\gamma_{13}$  & $\gamma_{14}$ & $\gamma_{15}$  \\ \hline
				1	& 146  & 108  & 82  & 62  & 46  & 30  & 16  & 2  & -13  & -28  & -43  & -61  & -81  & -108  & -146  \\ 
				2	& 152  & 109  & 82  & 61  & 43  & 28  & 14  & 0  & -14  & -29  & -45  & -63  & -85  & -113  & -155  \\ 
				3	& 159  & 113  & 85  & 63  & 45  & 29  & 14  & 0  & -14  & -29  & -45  & -63  & -85  & -114  & -159  \\ 
				4	& 191  & 138  & 103  & 76  & 54  & 35  & 17  & -1  & -18  & -35  & -54  & -77  & -104  & -138  & -191  \\ 
				5	& 193  & 139  & 105  & 80  & 59  & 40  & 23  & 6  & -11  & -30  & -50  & -74  & -101  & -136  & -191  \\ 
				6	& 199  & 143  & 107  & 80  & 58  & 38  & 20  & 2  & -14  & -31  & -50  & -74  & -101  & -137  & -194  \\
\hline
			\end{tabular}
		}
		\end{table}
		\begin{table}[h!]
			\centering
			\huge
			\setstretch{1.2}
			\resizebox{0.5\textwidth}{!}{
			\begin{tabular}{|c|c|c|c|c|c|c|c|c|c|c|c|c|c|c|}
				\hline
				\multicolumn{15}{|c|}{$\Gamma_{ch}$ (in LLR format)} \\ \hline
				$\gamma_1$   & $\gamma_2$   & $\gamma_3$  & $\gamma_4$  & $\gamma_5$   & $\gamma_6$   & $\gamma_7$    & $\gamma_8$    & $\gamma_9$  & $\gamma_{10}$  & $\gamma_{11}$  & $\gamma_{12}$  & $\gamma_{13}$  & $\gamma_{14}$ & $\gamma_{15}$  \\ \hline
				4.50  & 3.34  & 2.54  & 1.92  & 1.38  & 0.88  & 0.42  & -0.02  & -0.46  & -0.92  & -1.40  & -1.94  & -2.56  & -3.36  & -4.52   \\ \hline
			\end{tabular}
		}
		\end{table}
		\begin{table}[h!]
				\footnotesize
				\setstretch{1.1}
				\centering
			\begin{tabular}{l|c}
				\hline
				\multirow{2}{*}{Iteration} & $\Gamma_e$ \\ \cline{2-2} 
				& $\gamma_1$ \\ \hline
				1	& 1  \\ 
				2	& 0  \\ 
				3	& 0  \\ 
				4	& -1  \\ 
				5	& 0  \\ 
				6	& 0  \\ 
				7	& 1  \\ 
				8	& 0  \\ 
				9	& 0  \\ 
				10	& 0  \\ 
				11	& 1  \\ 
				12	& 1  \\ 
				13	& 1  \\ 
				14	& 1  \\ 
				15	& 0  \\ 
				16	& 1  \\ 
				17	& 1  \\ 
				18	& 1  \\ 
				19	& 0  \\ 
				20	& 1  \\ 
				21	& 1  \\ 
				22	& 1  \\ 
				23	& 0  \\ 
				24	& 0  \\ 
				25	& 0  \\ 
				26	& 0  \\ 
				27	& 0  \\ 
				28	& 1  \\ 
				29	& 0  \\ 
				30	& 1  \\ 
				31	& 0  \\ 
				32	& 1  \\ 
				33	& 1  \\ 
				34	& 1  \\ 
				35	& 0  \\ 
				36	& 1  \\ 
				37	& 0  \\ 
				38	& 0  \\ 
				39	& 0  \\ 
				40	& 1  \\ 
				41	& 0  \\ 
				42	& 1  \\ 
				43	& 1  \\ 
				44	& 1  \\ 
				45	& 1  \\ 
				46	& 0  \\ 
				47	& 1  \\ 
				48	& 1  \\ 
				49	& 0  \\ 
				50	& 0  \\ \hline
			\end{tabular}
		\end{table}
		\clearpage
		\subsection{The 3-bit MIM-QMS decoder in Example 1}
		\begin{table}[h!]
			\centering
			\footnotesize
			\setstretch{1.2}
			\caption{Reconstruction Functions $\phi_v$ and $\phi_{ch}$}
			\begin{tabular}{l|cccccccc}
				\hline
				\multicolumn{1}{l|}{\multirow{2}{*}{Iteration}} & \multicolumn{8}{c}{$\phi_v(s)$, $s \in \mc{S}$} \\ \cline{2-9} 
				\multicolumn{1}{l|}{}  & 0  & 1  & 2  & 3   & 4    & 5   & 6   & 7  \\ \hline
				1	& 103  & 58  & 26  & 5  & -5  & -26  & -58  & -103  \\ 
				2	& 111  & 65  & 31  & 7  & -7  & -31  & -65  & -111  \\ 
				3	& 125  & 73  & 35  & 8  & -8  & -35  & -73  & -125  \\ 
				4	& 136  & 76  & 37  & 10  & -10  & -37  & -76  & -136  \\ 
				5	& 146  & 83  & 42  & 11  & -11  & -42  & -83  & -146  \\ 
				6	& 155  & 87  & 43  & 11  & -11  & -43  & -87  & -155  \\
\hline
			\end{tabular}
		\end{table}
		\begin{table}[h!]
			\centering
			\footnotesize
			\setstretch{1.2}
			\begin{tabular}{l|cccccccc}
				\hline
				\multicolumn{1}{l|}{\multirow{2}{*}{Iteration}} & \multicolumn{8}{c}{$\phi_{ch}(l)$, $l \in \mc{L}$} \\ \cline{2-9} 
				\multicolumn{1}{l|}{}                           & 0  & 1  & 2  & 3   & 4    & 5   & 6   & 7  \\ \hline
				1-7	& 170  & 99  & 54  & 17  & -17  & -54  & -99  & -170  \\ 
				8	& 156  & 91  & 50  & 16  & -16  & -50  & -91  & -156  \\ 
				9	& 153  & 89  & 49  & 16  & -16  & -49  & -89  & -153  \\ 
				10	& 147  & 85  & 47  & 15  & -15  & -47  & -85  & -147  \\ 
				11	& 143  & 83  & 45  & 15  & -15  & -45  & -83  & -143  \\ 
				12	& 136  & 79  & 43  & 14  & -14  & -43  & -79  & -136  \\
\hline
			\end{tabular}
		\end{table}
		\begin{table}[!h]
			\centering
			\footnotesize
			\setstretch{1.16}
			\caption{Threshold sets $\Gamma_v$ and $\Gamma_{ch}$}
			\begin{tabular}{l|ccccccc}
				\hline
				\multirow{2}{*}{Iteration} & \multicolumn{7}{c}{$\Gamma_v$}                                                          \\ \cline{2-8} 
				& $\gamma_1$ & $\gamma_2$ & $\gamma_3$ & $\gamma_4$ & $\gamma_5$ & $\gamma_6$ & $\gamma_7$ \\ \hline
				1	& 135  & 77  & 37  & 0  & -36  & -76  & -134  \\ 
				2	& 143  & 80  & 37  & 0  & -36  & -79  & -142  \\ 
				3	& 147  & 80  & 37  & 0  & -36  & -79  & -146  \\ 
				4	& 152  & 83  & 39  & 1  & -38  & -82  & -151  \\ 
				5	& 156  & 84  & 39  & 1  & -38  & -83  & -155  \\ 
				6	& 168  & 92  & 44  & 3  & -38  & -90  & -167  \\
\hline
			\end{tabular}
		\end{table}	
		\begin{table}[h!]
			\small
			\centering
			\setstretch{1.2}
			\begin{tabular}{|c|c|c|c|c|c|c|}
				\hline
				\multicolumn{7}{|c|}{$\Gamma_{ch}$ (in LLR format)}                                                         \\ \hline
				$\gamma_1$ & $\gamma_2$ & $\gamma_3$ & $\gamma_4$ & $\gamma_5$ & $\gamma_6$ & $\gamma_7$ \\ \hline
				3.46  & 2.00  & 0.94  & 0  & -0.94  & -2.00  & -3.46  \\ \hline
			\end{tabular}
		\end{table}
		\begin{table}[h!]
			\footnotesize
			\setstretch{1.2}
			\centering
			\begin{tabular}{l|c}
				\hline
				\multirow{2}{*}{Iteration} & $\Gamma_e$ \\ \cline{2-2} 
				& $\gamma_1$ \\ \hline
				1	& 1  \\ 
				2	& 0  \\ 
				3	& 1  \\ 
				4	& 2  \\ 
				5	& 0  \\ 
				6	& -1  \\ 
				7-12	& 1  \\ 
				13	& 0  \\ 
				14	& 1  \\ 
				15	& 1  \\ 
				16	& 1  \\ 
				17	& 0  \\ 
				18	& 0  \\ 
				19	& -1  \\ 
				20	& 1  \\ 
				21	& 0  \\ 
				22	& 1  \\ 
				23	& 0  \\ 
				24	& 1  \\ 
				25	& 1  \\ 
				26	& 1  \\ 
				27	& 0  \\ 
				28-34	& 1  \\ 
				35	& 0  \\ 
				36	& 0  \\ 
				37	& 1  \\ 
				38	& 1  \\ 
				39	& 1  \\ 
				40	& 3  \\ 
				41	& 1  \\ 
				42	& 0  \\ 
				43	& 0  \\ 
				44	& 1  \\ 
				45	& 2  \\ 
				46	& 1  \\ 
				47	& 1  \\ 
				48	& 2  \\ 
				49	& 1  \\ 
				50	& 1  \\  \hline
			\end{tabular}
		\end{table}
		\clearpage
		\subsection{The 4-bit MIM-QMS decoder in Example 1}
		\begin{table}[!h]
			\centering
			\huge
			\setstretch{1.2}
			\caption{Reconstruction Functions $\phi_v$ and $\phi_{ch}$}
			\resizebox{0.5\textwidth}{!}{
			\begin{tabular}{l|cccccccccccccccc}
				\hline
				\multicolumn{1}{l|}{\multirow{2}{*}{Iteration}} & \multicolumn{16}{c}{$\phi_v(s)$, $s \in \mc{S}$} \\ \cline{2-17} 
				\multicolumn{1}{l|}{}  & 0  & 1  & 2  & 3   & 4    & 5   & 6   & 7 & 8 & 9 & 10 & 11 & 12 & 13 & 14 & 15 \\ \hline
				1	& 113  & 81  & 58  & 41  & 27  & 15  & 7  & 2  & -2  & -7  & -15  & -27  & -41  & -58  & -81  & -113  \\ 
				2	& 118  & 85  & 63  & 44  & 29  & 18  & 9  & 2  & -2  & -9  & -18  & -29  & -44  & -63  & -85  & -118  \\ 
				3	& 128  & 93  & 68  & 49  & 34  & 21  & 10  & 3  & -3  & -10  & -21  & -34  & -49  & -68  & -93  & -128  \\ 
				4	& 145  & 107  & 79  & 57  & 39  & 24  & 11  & 3  & -3  & -11  & -24  & -39  & -57  & -79  & -107  & -145  \\ 
				5	& 153  & 110  & 81  & 58  & 39  & 24  & 13  & 3  & -3  & -13  & -24  & -39  & -58  & -81  & -110  & -153  \\ 
				6	& 170  & 123  & 89  & 65  & 45  & 28  & 15  & 4  & -4  & -15  & -28  & -45  & -65  & -89  & -123  & -170  \\
\hline
			\end{tabular}
		}
		\end{table}
		\begin{table}[!h]
			\centering
			\huge
			\setstretch{1.2}
			\resizebox{0.5\textwidth}{!}{
			\begin{tabular}{l|cccccccccccccccc}
				\hline
				\multicolumn{1}{l|}{\multirow{2}{*}{Iteration}} & \multicolumn{16}{c}{$\phi_{ch}(l)$, $l \in \mc{L}$} \\ \cline{2-17} 
				\multicolumn{1}{l|}{}  & 0  & 1  & 2  & 3   & 4    & 5   & 6   & 7 & 8 & 9 & 10 & 11 & 12 & 13 & 14 & 15 \\ \hline
				1-5	& 170  & 122  & 93  & 71  & 52  & 36  & 22  & 8  & -6  & -21  & -36  & -52  & -70  & -92  & -121  & -169  \\ 
				6	& 166  & 119  & 91  & 69  & 51  & 36  & 21  & 7  & -6  & -20  & -35  & -51  & -68  & -90  & -118  & -165  \\ 
				7	& 170  & 122  & 93  & 71  & 52  & 36  & 22  & 8  & -6  & -21  & -36  & -52  & -70  & -92  & -121  & -169  \\ 
				8	& 152  & 109  & 83  & 63  & 47  & 33  & 19  & 7  & -6  & -18  & -32  & -46  & -63  & -82  & -108  & -151  \\ 
				9	& 152  & 109  & 83  & 63  & 47  & 33  & 19  & 7  & -6  & -18  & -32  & -46  & -63  & -83  & -108  & -151  \\ 
				10	& 142  & 102  & 78  & 59  & 44  & 30  & 18  & 6  & -5  & -17  & -30  & -43  & -59  & -77  & -101  & -141  \\
\hline
			\end{tabular}
		}
		\end{table}
		\begin{table}[!h]
			\centering
			\huge
			\caption{Threshold sets $\Gamma_v$ and $\Gamma_{ch}$}
			\setstretch{1.2}
			\resizebox{0.5\textwidth}{!}{
			\begin{tabular}{l|ccccccccccccccc}
				\hline
				\multirow{2}{*}{Iteration} & \multicolumn{15}{c}{$\Gamma_v$}   \\ \cline{2-16} 
				& $\gamma_1$   & $\gamma_2$   & $\gamma_3$  & $\gamma_4$  & $\gamma_5$   & $\gamma_6$   & $\gamma_7$    & $\gamma_8$    & $\gamma_9$  & $\gamma_{10}$  & $\gamma_{11}$  & $\gamma_{12}$  & $\gamma_{13}$  & $\gamma_{14}$ & $\gamma_{15}$  \\ \hline
				1	& 146  & 108  & 82  & 61  & 43  & 27  & 13  & -1  & -16  & -30  & -45  & -62  & -81  & -106  & -145  \\ 
				2	& 153  & 112  & 86  & 65  & 47  & 32  & 16  & 1  & -14  & -30  & -46  & -64  & -84  & -111  & -151  \\ 
				3	& 160  & 116  & 87  & 65  & 46  & 27  & 10  & -6  & -21  & -37  & -55  & -76  & -101  & -134  & -182  \\ 
				4	& 163  & 116  & 86  & 63  & 44  & 27  & 12  & -5  & -21  & -37  & -55  & -76  & -102  & -136  & -186  \\ 
				5	& 190  & 138  & 103  & 76  & 54  & 35  & 17  & 0  & -18  & -36  & -55  & -77  & -102  & -136  & -188  \\ 
				6	& 179  & 127  & 93  & 69  & 48  & 31  & 15  & 1  & -14  & -29  & -46  & -66  & -91  & -125  & -177  \\
\hline
			\end{tabular}
		}
		\end{table}
		\begin{table}[h!]
			\centering
			\huge
			\setstretch{1.2}
			\resizebox{0.5\textwidth}{!}{
			\begin{tabular}{|c|c|c|c|c|c|c|c|c|c|c|c|c|c|c|}
				\hline
				\multicolumn{15}{|c|}{$\Gamma_{ch}$ (in LLR format)} \\ \hline
				$\gamma_1$   & $\gamma_2$   & $\gamma_3$  & $\gamma_4$  & $\gamma_5$   & $\gamma_6$   & $\gamma_7$    & $\gamma_8$    & $\gamma_9$  & $\gamma_{10}$  & $\gamma_{11}$  & $\gamma_{12}$  & $\gamma_{13}$  & $\gamma_{14}$ & $\gamma_{15}$  \\ \hline
				4.54  & 3.38  & 2.58  & 1.94  & 1.40  & 0.92  & 0.46  & 0.02  & -0.42  & -0.88  & -1.38  & -1.92  & -2.5600  & -3.34  & -4.50 \\ \hline
			\end{tabular}
		}
		\end{table}
		\begin{table}[h!]
			\footnotesize
			\setstretch{1.1}
			\centering
			\begin{tabular}{l|c}
				\hline
				\multirow{2}{*}{Iteration} & $\Gamma_e$ \\ \cline{2-2} 
				& $\gamma_1$ \\ \hline
				1	& 1  \\ 
				2	& 0  \\ 
				3	& 0  \\ 
				4	& 1  \\ 
				5-9	& 0  \\ 
				10-13	& 1  \\ 
				14	& 0  \\ 
				15	& 0  \\ 
				16-19	& 1  \\ 
				20	& 0  \\ 
				21	& 0  \\ 
				22	& 1  \\ 
				23	& 0  \\ 
				24	& 1  \\ 
				25	& 1  \\ 
				26	& 0  \\ 
				27	& 0  \\ 
				28	& 0  \\ 
				29	& 1  \\ 
				30	& 1  \\ 
				31	& 1  \\ 
				32	& 0  \\ 
				33	& 1  \\ 
				34	& 1  \\ 
				35	& 1  \\ 
				36-40	& 0  \\ 
				
				41	& 1  \\ 
				42	& 1  \\ 
				43	& 0  \\ 
				44	& 1  \\ 
				45	& 0  \\ 
				46	& 0  \\ 
				47	& 1  \\ 
				48	& 0  \\ 
				49	& 1  \\ 
				50	& 0  \\  \hline
			\end{tabular}
		\end{table}
		\clearpage			
		\subsection{The 3-bit MIM-QMS decoder in Example 2}
		\begin{table}[h!]
			\centering
			\footnotesize
			\setstretch{1.2}
			\caption{Reconstruction Functions $\phi_v$ and $\phi_{ch}$ for the length-$560$ 5G LDPC code with code rate $1/2$}
			\begin{tabular}{l|cccccccc}
				\hline
				\multicolumn{1}{l|}{\multirow{2}{*}{Iteration}} & \multicolumn{8}{c}{$\phi_v(s)$, $s \in \mc{S}$} \\ \cline{2-9} 
				\multicolumn{1}{l|}{}                           & 0  & 1  & 2  & 3   & 4    & 5   & 6   & 7  \\ \hline
				1	& 138  & 74  & 29  & 5  & -5  & -29  & -74  & -138  \\ 
				2	& 144  & 79  & 33  & 6  & -6  & -33  & -79  & -144  \\ 
				3	& 153  & 84  & 35  & 7  & -7  & -35  & -84  & -153  \\ 
				4	& 160  & 86  & 37  & 7  & -7  & -37  & -86  & -160  \\ 
				5	& 167  & 90  & 39  & 8  & -8  & -39  & -90  & -167  \\ 
				6	& 169  & 91  & 40  & 9  & -9  & -40  & -91  & -169  \\
\hline
			\end{tabular}
		\end{table}
		\begin{table}[h!]
			\centering
			\footnotesize
			\setstretch{1.2}
			\begin{tabular}{l|cccccccc}
				\hline
				\multicolumn{1}{l|}{\multirow{2}{*}{Iteration}} & \multicolumn{8}{c}{$\phi_{ch}(l)$, $l \in \mc{L}$} \\ \cline{2-9} 
				\multicolumn{1}{l|}{}                           & 0  & 1  & 2  & 3   & 4    & 5   & 6   & 7  \\ \hline
				1-13	& 204  & 120  & 66  & 21  & -21  & -66  & -120  & -204  \\ 
				14	& 185  & 109  & 60  & 19  & -19  & -60  & -109  & -185  \\ 
				15	& 157  & 92  & 51  & 16  & -16  & -51  & -92  & -157  \\ 
				16	& 152  & 89  & 49  & 16  & -16  & -49  & -89  & -152  \\ 
				17	& 146  & 86  & 47  & 15  & -15  & -47  & -86  & -146  \\ 
				18	& 142  & 83  & 46  & 15  & -15  & -46  & -83  & -142  \\ 
				19	& 137  & 81  & 44  & 14  & -14  & -44  & -81  & -137  \\ 
				20	& 131  & 77  & 42  & 14  & -14  & -42  & -77  & -131  \\ 
				21	& 126  & 74  & 41  & 13  & -13  & -41  & -74  & -126  \\ 
				22	& 121  & 71  & 39  & 13  & -13  & -39  & -71  & -121  \\ 
				23	& 117  & 69  & 38  & 12  & -12  & -38  & -69  & -117  \\ 
				24	& 115  & 68  & 37  & 12  & -12  & -37  & -68  & -115  \\ 
				25	& 113  & 66  & 36  & 12  & -12  & -36  & -66  & -113  \\ 
				26	& 111  & 65  & 36  & 12  & -12  & -36  & -65  & -111  \\ 
				27	& 110  & 65  & 35  & 11  & -11  & -35  & -65  & -110  \\ 
				28	& 109  & 64  & 35  & 11  & -11  & -35  & -64  & -109  \\ 
				29	& 109  & 64  & 35  & 11  & -11  & -35  & -64  & -109  \\ 
				30	& 108  & 64  & 35  & 11  & -11  & -35  & -64  & -108  \\  \hline
			\end{tabular}
		\end{table}
		\begin{table}[!h]
			\centering
			\footnotesize
			\setstretch{1.2}
			\caption{Threshold sets $\Gamma_v$ and $\Gamma_{ch}$ for the length-$560$ 5G LDPC code with code rate $1/2$}
			\begin{tabular}{l|ccccccc}
				\hline
				\multirow{2}{*}{Iteration} & \multicolumn{7}{c}{$\Gamma_v$}                                                          \\ \cline{2-8} 
				& $\gamma_1$ & $\gamma_2$ & $\gamma_3$ & $\gamma_4$ & $\gamma_5$ & $\gamma_6$ & $\gamma_7$ \\ \hline
				1	& 162  & 93  & 45  & 1  & -44  & -92  & -161  \\ 
				2	& 169  & 96  & 44  & 0  & -43  & -95  & -168  \\ 
				3	& 173  & 97  & 45  & 1  & -44  & -96  & -172  \\ 
				4	& 177  & 98  & 46  & 0  & -45  & -97  & -176  \\ 
				5	& 178  & 99  & 46  & 1  & -45  & -98  & -177  \\ 
				6	& 179  & 99  & 47  & 1  & -46  & -98  & -178  \\
\hline
			\end{tabular}
		\end{table}
		\begin{table}[h!]
			\small
			\centering
			\setstretch{1.2}
			\begin{tabular}{|c|c|c|c|c|c|c|}
				\hline
				\multicolumn{7}{|c|}{$\Gamma_{ch}$ (in LLR format)}                                                         \\ \hline
				$\gamma_1$ & $\gamma_2$ & $\gamma_3$ & $\gamma_4$ & $\gamma_5$ & $\gamma_6$ & $\gamma_7$ \\ \hline
				3.06  & 1.78  & 0.84  & 0  & -0.84  & -1.78  & -3.06  \\ \hline
			\end{tabular}
		\end{table}
		\begin{table}[h!]
			\small
			\setstretch{1.1}
			\centering
			\begin{tabular}{l|c}
				\hline
				\multirow{2}{*}{Iteration} & $\Gamma_e$ \\ \cline{2-2} 
				& $\gamma_1$ \\ \hline
				1	& 1  \\ 
				2	& -1  \\ 
				3	& 1  \\ 
				4	& -2  \\ 
				5	& 1  \\ 
				6	& 0  \\ 
				7	& 0  \\ 
				8	& 1  \\ 
				9	& 0  \\ 
				10	& 1  \\ 
				11	& 0  \\ 
				12	& 0  \\ 
				13	& 1  \\ 
				14	& -1  \\ 
				15	& 1  \\ 
				16	& -1  \\ 
				17	& 4  \\ 
				18	& 3  \\ 
				19-22	& 0  \\ 
				23	& 1  \\ 
				24	& 0  \\ 
				25	& 1  \\ 
				26	& 0  \\ 
				27	& 1  \\ 
				28	& 0  \\ 
				29	& 1  \\ 
				30	& 0  \\ \hline
			\end{tabular}
		\end{table}
		\clearpage
		\subsection{The 4-bit MIM-QMS decoder in Example 2}
		\begin{table}[!h]
			\centering
			\huge
			\setstretch{1.2}
			\caption{Reconstruction Functions $\phi_v$ and $\phi_{ch}$ for the length-$560$ 5G LDPC code with code rate $1/2$}
			\resizebox{0.5\textwidth}{!}{
			\begin{tabular}{l|cccccccccccccccc}
				\hline
				\multicolumn{1}{l|}{\multirow{2}{*}{Iteration}} & \multicolumn{16}{c}{$\phi_v(s)$, $s \in \mc{S}$} \\ \cline{2-17} 
				\multicolumn{1}{l|}{}  & 0  & 1  & 2  & 3   & 4    & 5   & 6   & 7 & 8 & 9 & 10 & 11 & 12 & 13 & 14 & 15 \\ \hline
			1	& 154  & 111  & 79  & 54  & 33  & 18  & 8  & 2  & -2  & -8  & -18  & -33  & -54  & -79  & -111  & -154  \\ 
			2	& 161  & 115  & 82  & 56  & 36  & 20  & 9  & 2  & -2  & -9  & -20  & -36  & -56  & -82  & -115  & -161  \\ 
			3	& 169  & 120  & 84  & 58  & 37  & 21  & 10  & 2  & -2  & -10  & -21  & -37  & -58  & -84  & -120  & -169  \\ 
			4	& 175  & 123  & 86  & 60  & 39  & 22  & 10  & 3  & -3  & -10  & -22  & -39  & -60  & -86  & -123  & -175  \\ 
			5	& 180  & 125  & 88  & 61  & 40  & 23  & 11  & 3  & -3  & -11  & -23  & -40  & -61  & -88  & -125  & -180  \\ 
			6	& 197  & 141  & 101  & 72  & 48  & 30  & 15  & 4  & -4  & -15  & -30  & -48  & -72  & -101  & -140  & -196  \\
\hline
			\end{tabular}
		}
		\end{table}
		\begin{table}[!h]
			\centering
			\huge
			\setstretch{1.2}
			\resizebox{0.5\textwidth}{!}{
			\begin{tabular}{l|cccccccccccccccc}
				\hline
				\multicolumn{1}{l|}{\multirow{2}{*}{Iteration}} & \multicolumn{16}{c}{$\phi_{ch}(l)$, $l \in \mc{L}$} \\ \cline{2-17} 
				\multicolumn{1}{l|}{}  & 0  & 1  & 2  & 3   & 4    & 5   & 6   & 7 & 8 & 9 & 10 & 11 & 12 & 13 & 14 & 15 \\ \hline
				1-7	& 204  & 148  & 113  & 86  & 64  & 44  & 26  & 9  & -8  & -26  & -44  & -64  & -86  & -113  & -148  & -204  \\ 
				8	& 182  & 132  & 101  & 77  & 57  & 39  & 24  & 8  & -7  & -23  & -39  & -57  & -77  & -101  & -132  & -182  \\ 
				9	& 179  & 130  & 99  & 76  & 56  & 39  & 23  & 8  & -7  & -22  & -39  & -56  & -76  & -99  & -130  & -179  \\ 
				10	& 178  & 129  & 99  & 75  & 56  & 39  & 23  & 8  & -7  & -22  & -38  & -56  & -75  & -99  & -129  & -178  \\ 
				11	& 174  & 126  & 96  & 74  & 54  & 38  & 22  & 8  & -7  & -22  & -37  & -54  & -74  & -96  & -126  & -174  \\ 
				12	& 171  & 124  & 94  & 72  & 53  & 37  & 22  & 8  & -7  & -21  & -37  & -53  & -72  & -94  & -124  & -171  \\
\hline
			\end{tabular}
		}
		\end{table}
		\begin{table}[!h]
			\centering
			\huge
			\caption{Threshold sets $\Gamma_v$ and $\Gamma_{ch}$ for the length-$560$ 5G LDPC code with code rate $1/2$}
			\setstretch{1.2}
			\resizebox{0.5\textwidth}{!}{
			\begin{tabular}{l|ccccccccccccccc}
				\hline
				\multirow{2}{*}{Iteration} & \multicolumn{15}{c}{$\Gamma_v$}   \\ \cline{2-16} 
				& $\gamma_1$   & $\gamma_2$   & $\gamma_3$  & $\gamma_4$  & $\gamma_5$   & $\gamma_6$   & $\gamma_7$    & $\gamma_8$    & $\gamma_9$  & $\gamma_{10}$  & $\gamma_{11}$  & $\gamma_{12}$  & $\gamma_{13}$  & $\gamma_{14}$ & $\gamma_{15}$  \\ \hline
			1	& 179  & 132  & 101  & 76  & 55  & 36  & 18  & 1  & -17  & -35  & -54  & -75  & -100  & -131  & -178  \\ 
			2	& 183  & 134  & 101  & 76  & 55  & 36  & 18  & 0  & -17  & -35  & -54  & -76  & -101  & -133  & -183  \\ 
			3	& 187  & 135  & 102  & 77  & 55  & 36  & 18  & 1  & -17  & -35  & -54  & -76  & -102  & -135  & -187  \\ 
			4	& 191  & 137  & 103  & 77  & 55  & 36  & 18  & 1  & -17  & -35  & -55  & -76  & -102  & -136  & -190  \\ 
			5	& 229  & 170  & 130  & 99  & 74  & 51  & 29  & 7  & -14  & -34  & -54  & -76  & -103  & -138  & -194  \\ 
			6	& 229  & 170  & 130  & 100  & 74  & 51  & 29  & 6  & -15  & -34  & -54  & -76  & -102  & -137  & -194  \\
\hline
			\end{tabular}
		}
		\end{table}
		\begin{table}[h!]
			\centering
			\huge
			\setstretch{1.2}
			\resizebox{0.5\textwidth}{!}{
			\begin{tabular}{|c|c|c|c|c|c|c|c|c|c|c|c|c|c|c|}
				\hline
				\multicolumn{15}{|c|}{$\Gamma_{ch}$ (in LLR format)} \\ \hline
				$\gamma_1$   & $\gamma_2$   & $\gamma_3$  & $\gamma_4$  & $\gamma_5$   & $\gamma_6$   & $\gamma_7$    & $\gamma_8$    & $\gamma_9$  & $\gamma_{10}$  & $\gamma_{11}$  & $\gamma_{12}$  & $\gamma_{13}$  & $\gamma_{14}$ & $\gamma_{15}$  \\ \hline
				4.12  & 3.08  & 2.36  & 1.78  & 1.28  & 0.84  & 0.42  & 0.02  & -0.40  & -0.82  & -1.28  & -1.78  & -2.36  & -3.08  & -4.12  \\ \hline
			\end{tabular}
		}
		\end{table}
		\begin{table}[h!]
			\small
			\setstretch{1.1}
			\centering
			\begin{tabular}{l|c}
				\hline
				\multirow{2}{*}{Iteration} & $\Gamma_e$ \\ \cline{2-2} 
				& $\gamma_1$ \\ \hline
				1	& 1  \\ 
				2	& 0  \\ 
				3	& 1  \\ 
				4-7	& 0  \\ 
				8	& 1  \\ 
				9	& 1  \\ 
				10	& 0  \\ 
				11	& 0  \\ 
				12-18	& 1  \\ 
				19-23	& 0  \\ 
				24-30	& 1  \\ 
				\hline
			\end{tabular}
		\end{table}
		\clearpage			
		\subsection{The 3-bit MIM-QMS decoders in Example 3}
		\begin{table}[h!]
			\centering
			\footnotesize
			\setstretch{1.1}
			\caption{Reconstruction Functions $\phi_v$ and $\phi_{ch}$ for the length-$1296$ IEEE 802.11n LDPC code with code rate $1/2$}
			\begin{tabular}{l|cccccccc}
				\hline
				\multicolumn{1}{l|}{\multirow{2}{*}{Iteration}} & \multicolumn{8}{c}{$\phi_v(s)$, $s \in \mc{S}$} \\ \cline{2-9} 
				\multicolumn{1}{l|}{}                           & 0  & 1  & 2  & 3   & 4    & 5   & 6   & 7  \\ \hline
			1	& 103  & 58  & 27  & 6  & -6  & -27  & -58  & -103  \\ 
			2	& 113  & 66  & 31  & 7  & -7  & -31  & -66  & -113  \\ 
			3	& 127  & 74  & 36  & 9  & -9  & -36  & -74  & -127  \\ 
			4	& 140  & 82  & 41  & 11  & -11  & -41  & -82  & -140  \\ 
			5	& 152  & 86  & 42  & 11  & -11  & -42  & -86  & -152  \\ 
			6	& 168  & 98  & 48  & 12  & -12  & -48  & -98  & -168  \\
\hline
			\end{tabular}
		\end{table}
		\begin{table}[h!]
			\centering
			\footnotesize
			\setstretch{1.1}
			\begin{tabular}{l|cccccccc}
				\hline
				\multicolumn{1}{l|}{\multirow{2}{*}{Iteration}} & \multicolumn{8}{c}{$\phi_{ch}(l)$, $l \in \mc{L}$} \\ \cline{2-9} 
				\multicolumn{1}{l|}{}                           & 0  & 1  & 2  & 3   & 4    & 5   & 6   & 7  \\ \hline
				1-6	& 170  & 99  & 54  & 17  & -17  & -54  & -99  & -170  \\ 
				7	& 152  & 88  & 48  & 15  & -15  & -48  & -88  & -152  \\ 
				8	& 151  & 88  & 48  & 15  & -15  & -48  & -88  & -151  \\ 
				9	& 145  & 84  & 46  & 15  & -15  & -46  & -84  & -145  \\ 
				10	& 144  & 83  & 45  & 14  & -14  & -45  & -83  & -144  \\ 
				11	& 139  & 81  & 44  & 14  & -14  & -44  & -81  & -139  \\ 
				12	& 131  & 76  & 41  & 13  & -13  & -41  & -76  & -131  \\ 
				13	& 129  & 75  & 41  & 13  & -13  & -41  & -75  & -129  \\ 
				14	& 125  & 72  & 39  & 13  & -13  & -39  & -72  & -125  \\ 
				15	& 122  & 71  & 38  & 12  & -12  & -38  & -71  & -122  \\ 
				16	& 122  & 71  & 38  & 12  & -12  & -38  & -71  & -122  \\ 
				17	& 120  & 70  & 38  & 12  & -12  & -38  & -70  & -120  \\ 
				18	& 118  & 69  & 37  & 12  & -12  & -37  & -69  & -118  \\ 
				19	& 116  & 67  & 37  & 12  & -12  & -37  & -67  & -116  \\ 
				20	& 114  & 66  & 36  & 11  & -11  & -36  & -66  & -114  \\ 
				21	& 111  & 65  & 35  & 11  & -11  & -35  & -65  & -111  \\ 
				22	& 109  & 63  & 34  & 11  & -11  & -34  & -63  & -109  \\ 
				23	& 107  & 62  & 34  & 11  & -11  & -34  & -62  & -107  \\ 
				24	& 104  & 60  & 33  & 10  & -10  & -33  & -60  & -104  \\ 
				25	& 102  & 59  & 32  & 10  & -10  & -32  & -59  & -102  \\ 
				26	& 101  & 59  & 32  & 10  & -10  & -32  & -59  & -101  \\ 
				27	& 97  & 56  & 31  & 10  & -10  & -31  & -56  & -97  \\ 
				28	& 97  & 56  & 30  & 10  & -10  & -30  & -56  & -97  \\ 
				29	& 95  & 55  & 30  & 10  & -10  & -30  & -55  & -95  \\ 
				30	& 94  & 54  & 30  & 9  & -9  & -30  & -54  & -94  \\  \hline
			\end{tabular}
		\end{table}
		\begin{table}[!h]
			\centering
			\footnotesize
			\setstretch{1.1}
			\caption{Threshold sets $\Gamma_v$ and $\Gamma_{ch}$ for the length-$1296$ IEEE 802.11n LDPC code with code rate $1/2$}
			\begin{tabular}{l|ccccccc}
				\hline
				\multirow{2}{*}{Iteration} & \multicolumn{7}{c}{$\Gamma_v$}                                                          \\ \cline{2-8} 
				& $\gamma_1$ & $\gamma_2$ & $\gamma_3$ & $\gamma_4$ & $\gamma_5$ & $\gamma_6$ & $\gamma_7$ \\ \hline
				1	& 136  & 78  & 36  & 0  & -35  & -77  & -135  \\ 
				2	& 142  & 79  & 38  & 1  & -37  & -78  & -141  \\ 
				3	& 150  & 83  & 38  & 0  & -38  & -82  & -149  \\ 
				4	& 155  & 84  & 39  & 0  & -38  & -83  & -154  \\ 
				5	& 166  & 90  & 40  & 2  & -38  & -89  & -165  \\ 
				6	& 207  & 126  & 71  & 23  & -25  & -81  & -166  \\
\hline
			\end{tabular}
		\end{table}
		\begin{table}[h!]
			\small
			\centering
			\setstretch{1.2}
			\begin{tabular}{|c|c|c|c|c|c|c|}
				\hline
				\multicolumn{7}{|c|}{$\Gamma_{ch}$ (in LLR  format)}                                                         \\ \hline
				$\gamma_1$ & $\gamma_2$ & $\gamma_3$ & $\gamma_4$ & $\gamma_5$ & $\gamma_6$ & $\gamma_7$ \\ \hline
				3.50  & 2.02  & 0.94  & 0  & -0.94  & -2.02  & -3.50  \\ \hline
			\end{tabular}
		\end{table}
		\begin{table}[h!]
			\footnotesize
			\setstretch{1.1}
			\centering
			\begin{tabular}{l|c}
				\hline
				\multirow{2}{*}{Iteration} & $\Gamma_e$ \\ \cline{2-2} 
				& $\gamma_1$ \\ \hline
				1	& 2  \\ 
				2	& 1  \\ 
				3	& 1  \\ 
				4	& 0  \\ 
				5	& 0  \\ 
				6	& 1  \\ 
				7	& 2  \\ 
				8-12	& 1  \\
				13	& 0  \\ 
				14	& 0  \\ 
				15	& 1  \\ 
				16	& -1  \\ 
				17	& 0  \\ 
				18	& 1  \\ 
				19	& 1  \\ 
				20	& 1  \\ 
				21	& 0  \\ 
				22	& 1  \\ 
				23	& 0  \\ 
				24	& 2  \\ 
				25-29	& 1  \\
				30	& 2  \\  \hline
			\end{tabular}
		\end{table}
		\begin{table}[h!]
			\centering
			\footnotesize
			\setstretch{1.2}
			\caption{Reconstruction Functions $\phi_v$ and $\phi_{ch}$ for the length-$1296$ IEEE 802.11n LDPC code with code rate $2/3$}
			\begin{tabular}{l|cccccccc}
				\hline
				\multicolumn{1}{l|}{\multirow{2}{*}{Iteration}} & \multicolumn{8}{c}{$\phi_v(s)$, $s \in \mc{S}$} \\ \cline{2-9} 
				\multicolumn{1}{l|}{}                           & 0  & 1  & 2  & 3   & 4    & 5   & 6   & 7  \\ \hline
				1	& 132  & 74  & 34  & 7  & -7  & -34  & -74  & -132  \\ 
				2	& 137  & 81  & 39  & 9  & -9  & -39  & -81  & -137  \\ 
				3	& 139  & 83  & 41  & 10  & -10  & -41  & -83  & -139  \\ 
				4	& 145  & 86  & 43  & 11  & -11  & -43  & -86  & -145  \\ 
				5	& 160  & 96  & 48  & 12  & -12  & -48  & -96  & -160  \\ 
				6	& 170  & 103  & 53  & 14  & -14  & -53  & -103  & -170  \\
\hline
			\end{tabular}
		\end{table}
		\begin{table}[h!]
			\centering
			\footnotesize
			\setstretch{1.2}
			\begin{tabular}{l|cccccccc}
				\hline
				\multicolumn{1}{l|}{\multirow{2}{*}{Iteration}} & \multicolumn{8}{c}{$\phi_{ch}(l)$, $l \in \mc{L}$} \\ \cline{2-9} 
				\multicolumn{1}{l|}{}                           & 0  & 1  & 2  & 3   & 4    & 5   & 6   & 7  \\ \hline
				1-15	& 227  & 129  & 69  & 22  & -22  & -69  & -129  & -227  \\
				16	& 216  & 123  & 66  & 21  & -21  & -66  & -123  & -216  \\ 
				17	& 218  & 123  & 67  & 21  & -21  & -67  & -123  & -218  \\ 
				18	& 210  & 119  & 64  & 20  & -20  & -64  & -119  & -210  \\ 
				19	& 209  & 118  & 64  & 20  & -20  & -64  & -118  & -209  \\ 
				20	& 197  & 112  & 60  & 19  & -19  & -60  & -112  & -197  \\ 
				21	& 187  & 106  & 57  & 18  & -18  & -57  & -106  & -187  \\ 
				22	& 183  & 104  & 56  & 18  & -18  & -56  & -104  & -183  \\ 
				23	& 178  & 101  & 54  & 17  & -17  & -54  & -101  & -178  \\ 
				24	& 167  & 95  & 51  & 16  & -16  & -51  & -95  & -167  \\ 
				25	& 158  & 89  & 48  & 15  & -15  & -48  & -89  & -158  \\ 
				26	& 150  & 85  & 46  & 15  & -15  & -46  & -85  & -150  \\ 
				27	& 143  & 81  & 44  & 14  & -14  & -44  & -81  & -143  \\ 
				28	& 136  & 77  & 41  & 13  & -13  & -41  & -77  & -136  \\ 
				29	& 131  & 74  & 40  & 13  & -13  & -40  & -74  & -131  \\ 
				30	& 126  & 71  & 38  & 12  & -12  & -38  & -71  & -126  \\ 
				\hline
			\end{tabular}
		\end{table}
		\begin{table}[!h]
			\centering
			\footnotesize
			\setstretch{1.2}
			\caption{Threshold sets $\Gamma_v$ and $\Gamma_{ch}$ for the length-$1296$ IEEE 802.11n LDPC code with code rate $2/3$}
			\begin{tabular}{l|ccccccc}
				\hline
				\multirow{2}{*}{Iteration} & \multicolumn{7}{c}{$\Gamma_v$}                                                          \\ \cline{2-8} 
				& $\gamma_1$ & $\gamma_2$ & $\gamma_3$ & $\gamma_4$ & $\gamma_5$ & $\gamma_6$ & $\gamma_7$ \\ \hline
				1	& 173  & 99  & 46  & 1  & -45  & -98  & -172  \\ 
				2	& 172  & 98  & 45  & 0  & -45  & -97  & -171  \\ 
				3	& 175  & 98  & 46  & 0  & -45  & -97  & -174  \\ 
				4	& 186  & 105  & 46  & -2  & -51  & -104  & -185  \\ 
				5	& 194  & 110  & 51  & 1  & -50  & -109  & -193  \\ 
				6	& 192  & 105  & 49  & 2  & -48  & -104  & -191  \\
\hline
			\end{tabular}
		\end{table}
		\begin{table}[h!]
			\small
			\centering
			\setstretch{1.2}
			\begin{tabular}{|c|c|c|c|c|c|c|}
				\hline
				\multicolumn{7}{|c|}{$\Gamma_{ch}$ (in LLR format)}                                                         \\ \hline
				$\gamma_1$ & $\gamma_2$ & $\gamma_3$ & $\gamma_4$ & $\gamma_5$ & $\gamma_6$ & $\gamma_7$ \\ \hline
				4.08  & 2.32  & 1.08  & 0  & -1.08  & -2.32  & -4.08  \\ \hline
			\end{tabular}
		\end{table}
		\begin{table}[h!]
			\small
			\setstretch{1.2}
			\centering
			\begin{tabular}{l|c}
				\hline
				\multirow{2}{*}{Iteration} & $\Gamma_e$ \\ \cline{2-2} 
				& $\gamma_1$ \\ \hline
				1	& 0  \\ 
				2	& 0  \\ 
				3	& 1  \\ 
				4	& 3  \\ 
				5	& 1  \\ 
				6	& 2  \\ 
				7	& 1  \\ 
				8	& 1  \\ 
				9	& 0  \\ 
				10	& 0  \\ 
				11	& 1  \\ 
				12	& 1  \\ 
				13	& 0  \\ 
				14	& 1  \\ 
				15	& 2  \\ 
				16	& 0  \\ 
				17	& 1  \\ 
				18	& 1  \\ 
				19	& 0  \\ 
				20	& 0  \\ 
				21	& 1  \\ 
				22	& 1  \\ 
				23	& 0  \\ 
				24	& 1  \\ 
				25	& 1  \\ 
				26	& 0  \\ 
				27	& 1  \\ 
				28	& 3  \\ 
				29	& 3  \\ 
				30	& 0  \\ \hline
			\end{tabular}
		\end{table}
		\begin{table}[h!]
			\centering
			\footnotesize
			\setstretch{1.2}
			\caption{Reconstruction Functions $\phi_v$ and $\phi_{ch}$ for the length-$1296$ IEEE 802.11n LDPC code with code rate $5/6$}
			\begin{tabular}{l|cccccccc}
				\hline
				\multicolumn{1}{l|}{\multirow{2}{*}{Iteration}} & \multicolumn{8}{c}{$\phi_v(s)$, $s \in \mc{S}$} \\ \cline{2-9} 
				\multicolumn{1}{l|}{}                           & 0  & 1  & 2  & 3   & 4    & 5   & 6   & 7  \\ \hline
				1	& 229  & 129  & 62  & 13  & -13  & -62  & -129  & -229  \\ 
				2	& 227  & 138  & 68  & 16  & -16  & -68  & -138  & -227  \\ 
				3	& 235  & 144  & 74  & 18  & -18  & -74  & -144  & -235  \\ 
				4	& 244  & 150  & 77  & 19  & -19  & -77  & -150  & -244  \\ 
				5	& 248  & 154  & 80  & 20  & -20  & -80  & -154  & -248  \\ 
				6	& 252  & 157  & 80  & 20  & -20  & -80  & -157  & -252  \\
\hline
			\end{tabular}
		\end{table}
		\begin{table}[h!]
			\centering
			\footnotesize
			\setstretch{1.2}
			\begin{tabular}{l|cccccccc}
				\hline
				\multicolumn{1}{l|}{\multirow{2}{*}{Iteration}} & \multicolumn{8}{c}{$\phi_{ch}(l)$, $l \in \mc{L}$} \\ \cline{2-9} 
				\multicolumn{1}{l|}{}                           & 0  & 1  & 2  & 3   & 4    & 5   & 6   & 7  \\ \hline
				1-24	& 409  & 222  & 117  & 37  & -37  & -117  & -222  & -409  \\ 
				25	& 372  & 202  & 107  & 34  & -34  & -107  & -202  & -372  \\ 
				26	& 329  & 178  & 94  & 30  & -30  & -94  & -178  & -329  \\ 
				27	& 290  & 157  & 83  & 26  & -26  & -83  & -157  & -290  \\ 
				28	& 248  & 135  & 71  & 22  & -22  & -71  & -135  & -248  \\ 
				29	& 225  & 122  & 65  & 20  & -20  & -65  & -122  & -225  \\ 
				30	& 205  & 111  & 59  & 19  & -19  & -59  & -111  & -205  \\ 
				\hline
			\end{tabular}
		\end{table}
		\begin{table}[!h]
			\centering
			\footnotesize
			\setstretch{1.2}
			\caption{Threshold sets $\Gamma_v$ and $\Gamma_{ch}$ for the length-$1296$ IEEE 802.11n LDPC code with code rate $5/6$}
			\begin{tabular}{l|ccccccc}
				\hline
				\multirow{2}{*}{Iteration} & \multicolumn{7}{c}{$\Gamma_v$}                                                          \\ \cline{2-8} 
				& $\gamma_1$ & $\gamma_2$ & $\gamma_3$ & $\gamma_4$ & $\gamma_5$ & $\gamma_6$ & $\gamma_7$ \\ \hline
				1	& 295  & 165  & 76  & 1  & -75  & -164  & -293  \\ 
				2	& 300  & 166  & 78  & 3  & -75  & -165  & -297  \\ 
				3	& 308  & 177  & 84  & 7  & -71  & -163  & -301  \\ 
				4	& 312  & 169  & 74  & -5  & -82  & -173  & -305  \\ 
				5	& 309  & 169  & 73  & -3  & -81  & -171  & -309  \\ 
				6	& 312  & 175  & 83  & 10  & -72  & -169  & -310  \\
\hline
			\end{tabular}
		\end{table}
		\begin{table}[h!]
			\small
			\centering
			\setstretch{1.2}
			\begin{tabular}{|c|c|c|c|c|c|c|}
				\hline
				\multicolumn{7}{|c|}{$\Gamma_{ch}$ (in LLR format)}                                                         \\ \hline
				$\gamma_1$ & $\gamma_2$ & $\gamma_3$ & $\gamma_4$ & $\gamma_5$ & $\gamma_6$ & $\gamma_7$ \\ \hline
				4.88  & 2.72  & 1.24  & 0  & -1.24  & -2.72  & -4.88  \\  \hline
			\end{tabular} 
		\end{table}
		\begin{table}[h!]
			\small
			\setstretch{1.1}
			\centering
			\begin{tabular}{l|c}
				\hline
				\multirow{2}{*}{Iteration} & $\Gamma_e$ \\ \cline{2-2} 
				& $\gamma_1$ \\ \hline
				1-5	& 1  \\ 
				6	& 2  \\ 
				7	& 1  \\ 
				8	& 1  \\ 
				9	& 0  \\ 
				10	& 1  \\ 
				11	& 1  \\ 
				12	& 0  \\ 
				13	& 0  \\ 
				14	& 2  \\ 
				15	& 0  \\ 
				16	& 1  \\ 
				17	& 1  \\ 
				18	& 1  \\ 
				19	& 3  \\ 
				20	& 1  \\ 
				21	& 1  \\ 
				22	& 0  \\ 
				23	& 0  \\ 
				24	& 0  \\ 
				25	& 1  \\ 
				26	& 1  \\ 
				27	& 0  \\ 
				28	& 2  \\ 
				29	& 3  \\ 
				30	& 7  \\ 
				\hline
			\end{tabular}
		\end{table}		 
		\clearpage
		\subsection{The 4-bit MIM-QMS decoders in Example 3}
		\begin{table}[!h]
			\centering
			\huge
			\setstretch{1.1}
			\caption{Reconstruction Functions $\phi_v$ and $\phi_{ch}$ for the length-$1296$ IEEE 802.11n LDPC code with code rate $1/2$}
			\resizebox{0.5\textwidth}{!}{
			\begin{tabular}{l|cccccccccccccccc}
				\hline
				\multicolumn{1}{l|}{\multirow{2}{*}{Iteration}} & \multicolumn{16}{c}{$\phi_v(s)$, $s \in \mc{S}$} \\ \cline{2-17} 
				\multicolumn{1}{l|}{}  & 0  & 1  & 2  & 3   & 4    & 5   & 6   & 7 & 8 & 9 & 10 & 11 & 12 & 13 & 14 & 15 \\ \hline
				1	& 114  & 82  & 60  & 42  & 28  & 16  & 8  & 2  & -2  & -8  & -16  & -28  & -42  & -60  & -82  & -114  \\ 
				2	& 121  & 88  & 65  & 46  & 31  & 19  & 10  & 2  & -2  & -10  & -19  & -31  & -46  & -65  & -88  & -121  \\ 
				3	& 134  & 98  & 72  & 52  & 36  & 22  & 11  & 3  & -3  & -11  & -22  & -36  & -52  & -72  & -98  & -134  \\ 
				4	& 154  & 112  & 82  & 60  & 42  & 26  & 13  & 3  & -3  & -13  & -26  & -42  & -60  & -82  & -112  & -154  \\ 
				5	& 170  & 122  & 88  & 62  & 42  & 25  & 13  & 4  & -4  & -13  & -25  & -42  & -62  & -88  & -122  & -170  \\ 
				6	& 170  & 120  & 88  & 63  & 44  & 27  & 14  & 4  & -4  & -14  & -27  & -44  & -63  & -88  & -120  & -170  \\
\hline
			\end{tabular}
		}
		\end{table}
		\begin{table}[!h]
			\centering
			\huge
			\setstretch{1.1}
			\resizebox{0.5\textwidth}{!}{
			\begin{tabular}{l|cccccccccccccccc}
				\hline
				\multicolumn{1}{l|}{\multirow{2}{*}{Iteration}} & \multicolumn{16}{c}{$\phi_{ch}(l)$, $l \in \mc{L}$} \\ \cline{2-17} 
				\multicolumn{1}{l|}{}  & 0  & 1  & 2  & 3   & 4    & 5   & 6   & 7 & 8 & 9 & 10 & 11 & 12 & 13 & 14 & 15 \\ \hline
				1-4	& 170  & 122  & 93  & 71  & 52  & 36  & 21  & 7  & -7  & -21  & -36  & -52  & -71  & -93  & -122  & -170  \\ 
				5	& 161  & 115  & 88  & 67  & 50  & 34  & 20  & 7  & -7  & -20  & -34  & -50  & -67  & -88  & -115  & -161  \\ 
				6	& 153  & 110  & 83  & 64  & 47  & 33  & 19  & 6  & -6  & -19  & -33  & -47  & -64  & -83  & -110  & -153  \\ 
				7	& 152  & 108  & 83  & 63  & 47  & 32  & 19  & 6  & -6  & -19  & -32  & -47  & -63  & -83  & -108  & -152  \\ 
				8	& 138  & 98  & 75  & 57  & 42  & 29  & 17  & 6  & -6  & -17  & -29  & -42  & -57  & -75  & -98  & -138  \\ 
				9	& 134  & 96  & 73  & 56  & 41  & 29  & 17  & 6  & -6  & -17  & -29  & -41  & -56  & -73  & -96  & -134  \\
\hline
			\end{tabular}
		}
		\end{table}
		\begin{table}[!h]
			\centering
			\huge
			\caption{Threshold sets $\Gamma_v$ and $\Gamma_{ch}$ for the length-$1296$ IEEE 802.11n LDPC code with code rate $1/2$}
			\setstretch{1.2}
			\resizebox{0.5\textwidth}{!}{
			\begin{tabular}{l|ccccccccccccccc}
				\hline
				\multirow{2}{*}{Iteration} & \multicolumn{15}{c}{$\Gamma_v$}   \\ \cline{2-16} 
				& $\gamma_1$   & $\gamma_2$   & $\gamma_3$  & $\gamma_4$  & $\gamma_5$   & $\gamma_6$   & $\gamma_7$    & $\gamma_8$    & $\gamma_9$  & $\gamma_{10}$  & $\gamma_{11}$  & $\gamma_{12}$  & $\gamma_{13}$  & $\gamma_{14}$ & $\gamma_{15}$  \\ \hline
				1	& 148  & 109  & 84  & 64  & 47  & 32  & 17  & 2  & -12  & -26  & -41  & -59  & -81  & -107  & -147  \\ 
				2	& 157  & 114  & 87  & 66  & 47  & 30  & 16  & 1  & -14  & -29  & -46  & -65  & -86  & -113  & -156  \\ 
				3	& 186  & 136  & 103  & 78  & 58  & 40  & 24  & 8  & -8  & -26  & -44  & -64  & -87  & -116  & -163  \\ 
				4	& 191  & 137  & 101  & 73  & 51  & 32  & 16  & 1  & -15  & -31  & -50  & -72  & -100  & -136  & -190  \\ 
				5	& 185  & 132  & 98  & 73  & 52  & 32  & 16  & 1  & -15  & -31  & -51  & -72  & -97  & -131  & -184  \\ 
				6	& 172  & 118  & 86  & 63  & 44  & 28  & 14  & 1  & -13  & -27  & -43  & -62  & -85  & -117  & -171  \\
\hline
			\end{tabular}
		}
		\end{table}
		\begin{table}[h!]
			\centering
			\huge
			\setstretch{1.2}
			\resizebox{0.5\textwidth}{!}{
			\begin{tabular}{|c|c|c|c|c|c|c|c|c|c|c|c|c|c|c|}
				\hline
				\multicolumn{15}{|c|}{$\Gamma_{ch}$ (in LLR format)} \\ \hline
				$\gamma_1$   & $\gamma_2$   & $\gamma_3$  & $\gamma_4$  & $\gamma_5$   & $\gamma_6$   & $\gamma_7$    & $\gamma_8$    & $\gamma_9$  & $\gamma_{10}$  & $\gamma_{11}$  & $\gamma_{12}$  & $\gamma_{13}$  & $\gamma_{14}$ & $\gamma_{15}$  \\ \hline
				4.62  & 3.42  & 2.62  & 1.98  & 1.42  & 0.92  & 0.46  & 0 & -0.46  & -0.92  & -1.42  & -1.98  & -2.62  & -3.42  & -4.62 \\ \hline
			\end{tabular}
		}
		\end{table}
		\begin{table}[h!]
			\small
			\setstretch{1.1}
			\centering
			\begin{tabular}{l|c}
				\hline
				\multirow{2}{*}{Iteration} & $\Gamma_e$ \\ \cline{2-2} 
				& $\gamma_1$ \\ \hline
				1	& 1  \\ 
				2	& 1  \\ 
				3	& 0  \\ 
				4	& 1  \\ 
				5	& 1  \\ 
				6	& 0  \\ 
				7	& 1  \\ 
				8	& 0  \\ 
				9-17	& 1  \\ 
				
				18	& 0  \\ 
				19	& 1  \\ 
				20	& 0  \\ 
				21	& 1  \\ 
				22	& 0  \\ 
				23	& 0  \\ 
				24-27	& 1  \\ 
				
				28	& 0  \\ 
				29	& 0  \\ 
				30	& 1  \\ \hline
			\end{tabular}
		\end{table}
		\begin{table}[!h]
			\centering
			\huge
			\setstretch{1.2}
			\caption{Reconstruction Functions $\phi_v$ and $\phi_{ch}$ for the length-$1296$ IEEE 802.11n LDPC code with code rate $2/3$}
			\resizebox{0.5\textwidth}{!}{
			\begin{tabular}{l|cccccccccccccccc}
				\hline
				\multicolumn{1}{l|}{\multirow{2}{*}{Iteration}} & \multicolumn{16}{c}{$\phi_v(s)$, $s \in \mc{S}$} \\ \cline{2-17} 
				\multicolumn{1}{l|}{}  & 0  & 1  & 2  & 3   & 4    & 5   & 6   & 7 & 8 & 9 & 10 & 11 & 12 & 13 & 14 & 15 \\ \hline
				1	& 148  & 105  & 77  & 54  & 36  & 21  & 10  & 2  & -2  & -10  & -21  & -36  & -54  & -77  & -105  & -148  \\ 
				2	& 149  & 108  & 81  & 58  & 39  & 24  & 12  & 3  & -3  & -12  & -24  & -39  & -58  & -81  & -108  & -149  \\ 
				3	& 156  & 114  & 86  & 63  & 43  & 27  & 13  & 3  & -3  & -13  & -27  & -43  & -63  & -86  & -114  & -156  \\ 
				4	& 160  & 117  & 89  & 65  & 45  & 28  & 14  & 4  & -4  & -14  & -28  & -45  & -65  & -89  & -117  & -160  \\ 
				5	& 168  & 123  & 92  & 67  & 46  & 28  & 14  & 3  & -3  & -14  & -28  & -46  & -67  & -92  & -123  & -168  \\ 
				6	& 174  & 128  & 95  & 68  & 47  & 28  & 14  & 4  & -4  & -14  & -28  & -47  & -68  & -95  & -128  & -174  \\
\hline
			\end{tabular}
		}
		\end{table}
		\begin{table}[!h]
			\centering
			\huge
			\setstretch{1.2}
			\resizebox{0.5\textwidth}{!}{
			\begin{tabular}{l|cccccccccccccccc}
				\hline
				\multicolumn{1}{l|}{\multirow{2}{*}{Iteration}} & \multicolumn{16}{c}{$\phi_{ch}(l)$, $l \in \mc{L}$} \\ \cline{2-17} 
				\multicolumn{1}{l|}{}  & 0  & 1  & 2  & 3   & 4    & 5   & 6   & 7 & 8 & 9 & 10 & 11 & 12 & 13 & 14 & 15 \\ \hline
				1-17	& 227  & 159  & 120  & 91  & 67  & 46  & 27  & 9  & -9  & -27  & -46  & -67  & -91  & -120  & -159  & -227  \\ 
				18	& 226  & 159  & 120  & 91  & 67  & 46  & 27  & 9  & -9  & -27  & -46  & -67  & -91  & -120  & -159  & -226  \\ 
				19	& 221  & 155  & 117  & 89  & 65  & 45  & 26  & 9  & -9  & -26  & -45  & -65  & -89  & -117  & -155  & -221  \\ 
				20	& 217  & 152  & 115  & 87  & 64  & 44  & 26  & 9  & -9  & -26  & -44  & -64  & -87  & -115  & -152  & -217  \\ 
				21	& 205  & 144  & 109  & 82  & 61  & 42  & 25  & 8  & -8  & -25  & -42  & -61  & -82  & -109  & -144  & -205  \\ 
				22	& 203  & 142  & 107  & 81  & 60  & 41  & 24  & 8  & -8  & -24  & -41  & -60  & -81  & -107  & -142  & -203  \\ 
				23	& 197  & 138  & 104  & 79  & 58  & 40  & 24  & 8  & -8  & -24  & -40  & -58  & -79  & -104  & -138  & -197  \\ 
				24	& 194  & 137  & 103  & 78  & 58  & 40  & 23  & 8  & -8  & -23  & -40  & -58  & -78  & -103  & -137  & -194  \\ 
				25	& 187  & 131  & 99  & 75  & 55  & 38  & 22  & 7  & -7  & -22  & -38  & -55  & -75  & -99  & -131  & -187  \\ 
				26	& 178  & 125  & 94  & 71  & 53  & 36  & 21  & 7  & -7  & -21  & -36  & -53  & -71  & -94  & -125  & -178  \\ 
				27	& 176  & 124  & 93  & 71  & 52  & 36  & 21  & 7  & -7  & -21  & -36  & -52  & -71  & -93  & -124  & -176  \\ 
				28	& 170  & 119  & 90  & 68  & 50  & 35  & 20  & 7  & -7  & -20  & -35  & -50  & -68  & -90  & -119  & -170  \\ 
				29	& 161  & 113  & 85  & 65  & 48  & 33  & 19  & 6  & -6  & -19  & -33  & -48  & -65  & -85  & -113  & -161  \\ 
				30	& 155  & 109  & 82  & 62  & 46  & 32  & 19  & 6  & -6  & -19  & -32  & -46  & -62  & -82  & -109  & -155  \\ \hline
			\end{tabular}
		}
		\end{table}
		\begin{table}[!h]
			\centering
			\huge
			\caption{Threshold sets $\Gamma_v$ and $\Gamma_{ch}$ for the length-$1296$ IEEE 802.11n LDPC code with code rate $2/3$}
			\setstretch{1.2}
			\resizebox{0.5\textwidth}{!}{
			\begin{tabular}{l|ccccccccccccccc}
				\hline
				\multirow{2}{*}{Iteration} & \multicolumn{15}{c}{$\Gamma_v$}   \\ \cline{2-16} 
				& $\gamma_1$   & $\gamma_2$   & $\gamma_3$  & $\gamma_4$  & $\gamma_5$   & $\gamma_6$   & $\gamma_7$    & $\gamma_8$    & $\gamma_9$  & $\gamma_{10}$  & $\gamma_{11}$  & $\gamma_{12}$  & $\gamma_{13}$  & $\gamma_{14}$ & $\gamma_{15}$  \\ \hline
				1	& 188  & 138  & 105  & 79  & 57  & 38  & 19  & 1  & -17  & -35  & -55  & -77  & -104  & -137  & -187  \\ 
				2	& 193  & 141  & 108  & 81  & 58  & 37  & 18  & 0  & -20  & -39  & -60  & -83  & -109  & -142  & -193  \\ 
				3	& 195  & 142  & 109  & 82  & 59  & 38  & 19  & 0  & -18  & -37  & -58  & -81  & -108  & -141  & -194  \\ 
				4	& 202  & 147  & 111  & 83  & 59  & 37  & 18  & 1  & -17  & -36  & -58  & -82  & -110  & -146  & -201  \\ 
				5	& 207  & 150  & 112  & 83  & 59  & 39  & 21  & 3  & -15  & -34  & -57  & -82  & -111  & -149  & -206  \\ 
				6	& 225  & 168  & 129  & 98  & 73  & 51  & 31  & 11  & -10  & -32  & -54  & -79  & -108  & -146  & -204  \\
\hline
			\end{tabular}
		}
		\end{table}
		\begin{table}[h!]
			\centering
			\huge
			\setstretch{1.2}
			\resizebox{0.5\textwidth}{!}{
			\begin{tabular}{|c|c|c|c|c|c|c|c|c|c|c|c|c|c|c|}
				\hline
				\multicolumn{15}{|c|}{$\Gamma_{ch}$ (in LLR format)} \\ \hline
				$\gamma_1$   & $\gamma_2$   & $\gamma_3$  & $\gamma_4$  & $\gamma_5$   & $\gamma_6$   & $\gamma_7$    & $\gamma_8$    & $\gamma_9$  & $\gamma_{10}$  & $\gamma_{11}$  & $\gamma_{12}$  & $\gamma_{13}$  & $\gamma_{14}$ & $\gamma_{15}$  \\ \hline
				5.44  & 4  & 3.04  & 2.28  & 1.64  & 1.06  & 0.52  & 0  & -0.52  & -1.06  & -1.64  & -2.28  & -3.04  & -4.00  & -5.44 \\ \hline
			\end{tabular}
		}
		\end{table}
		\begin{table}[h!]
			\small
			\setstretch{1.1}
			\centering
			\begin{tabular}{l|c}
				\hline
				\multirow{2}{*}{Iteration} & $\Gamma_e$ \\ \cline{2-2} 
				& $\gamma_1$ \\ \hline
				1	& 0  \\ 
				2	& 0  \\ 
				3	& 1  \\ 
				4-8	& 0  \\ 
				
				9-15	& 1  \\ 
			
				16	& 0  \\ 
				17	& 1  \\ 
				18	& 1  \\ 
				19	& 1  \\ 
				20	& 0  \\ 
				21	& 1  \\ 
				22	& 0  \\ 
				23	& 1  \\ 
				24	& 0  \\ 
				25	& 1  \\ 
				26	& 0  \\ 
				27	& 1  \\ 
				28	& 1  \\ 
				29	& 0  \\ 
				30	& 0  \\ \hline
			\end{tabular}
		\end{table}
		\begin{table}[!h]
			\centering
			\huge
			\setstretch{1.2}
			\caption{Reconstruction Functions $\phi_v$ and $\phi_{ch}$ for the length-$1296$ IEEE 802.11n LDPC code with code rate $5/6$}
			\resizebox{0.5\textwidth}{!}{
			%
		}
		\end{table}
		\begin{table}[!h]
			\centering
			\huge
			\caption{Threshold sets $\Gamma_v$ and $\Gamma_{ch}$ for the length-$1296$ IEEE 802.11n LDPC code with code rate $5/6$}
			\setstretch{1.2}
			\resizebox{0.5\textwidth}{!}{
			%
		\end{table}
		\clearpage			
		\subsection{The 3-bit MIM-QMS decoder in Example 4}
		\begin{table}[h!]
			\centering
			\footnotesize
			\setstretch{1.2}
			\caption{Reconstruction Functions $\phi_v$ and $\phi_{ch}$ for the length-$17664$ IEEE 802.3ca LDPC code with code rate $0.826$}
			%
		\end{table}
		\begin{table}[!h]
			\centering
			\footnotesize
			\setstretch{1.2}
			\caption{Threshold sets $\Gamma_v$ and $\Gamma_{ch}$ for the length-$17664$ IEEE 802.3ca LDPC code with code rate $0.826$}
			%
		\end{table}
		\clearpage
		\subsection{The 4-bit MIM-QMS decoder in Example 4}
		\begin{table}[!h]
			\centering
			\huge
			\setstretch{1.2}
			\caption{Reconstruction Functions $\phi_v$ and $\phi_{ch}$ for the length-$17664$ IEEE 802.3ca LDPC code with code rate $0.826$}
			\resizebox{0.5\textwidth}{!}{
			%
		}
		\end{table}
		\begin{table}[!h]
			\centering
			\huge
			\caption{Threshold sets $\Gamma_v$ and $\Gamma_{ch}$ for the length-$17664$ IEEE 802.3ca LDPC code with code rate $0.826$}
			\setstretch{1.2}
			\resizebox{0.5\textwidth}{!}{
			%
		\end{table}
	\clearpage
	\section{Constructed LUTs for MIM-LQMS decoders}
	\subsection{The 3-bit MIM-LQMS decoder in Example 2}
	\begin{table}[h!]
		\centering
		\footnotesize
		\setstretch{1.2}
		\caption{Reconstruction Functions $\phi_v$ and $\phi_{ch}$ for the length-$560$ 5G LDPC code with code rate $1/2$}
		%
	\end{table}
	\begin{table}[!h]
		\centering
		\footnotesize
		\setstretch{1.2}
		\caption{Threshold sets $\Gamma_v$ and $\Gamma_{ch}$ for the length-$560$ 5G LDPC code with code rate $1/2$}
		%
	\end{table}
	\clearpage
	\subsection{The 4-bit MIM-LQMS decoder in Example 2}
	\begin{table}[!h]
		\centering
		\huge
		\setstretch{1.2}
		\caption{Reconstruction Functions $\phi_v$ and $\phi_{ch}$ for the length-$560$ 5G LDPC code with code rate $1/2$}
		\resizebox{0.5\textwidth}{!}{
		%
	}
	\end{table}
	\begin{table}[!h]
		\centering
		\huge
		\caption{Threshold sets $\Gamma_v$ and $\Gamma_{ch}$ for the length-$560$ 5G LDPC code with code rate $1/2$}
		\setstretch{1.2}
		\resizebox{0.5\textwidth}{!}{
		%
	\end{table}
	\clearpage
	\subsection{The 3-bit MIM-LQMS decoders in Example 3}
	\begin{table}[h!]
		\centering
		\footnotesize
		\setstretch{1.2}
		\caption{Reconstruction Functions $\phi_v$ and $\phi_{ch}$ for the length-$1296$ IEEE 802.11n LDPC code with code rate $1/2$}
		%
	\end{table}
	\clearpage	
	\begin{table}[h!]
		\centering
		\footnotesize
		\setstretch{1.2}
		\caption{Reconstruction Functions $\phi_v$ and $\phi_{ch}$ for the length-$1296$ IEEE 802.11n LDPC code with code rate $2/3$}
		%
	\end{table}
	\clearpage
	\begin{table}[h!]
		\centering
		\footnotesize
		\setstretch{1.2}
		\caption{Reconstruction Functions $\phi_v$ and $\phi_{ch}$ for the length-$1296$ IEEE 802.11n LDPC code with code rate $5/6$}
		%
	\end{table}		 
	\clearpage
	\subsection{The 4-bit MIM-LQMS decoders in Example 3}
	\begin{table}[!h]
		\centering
		\huge
		\setstretch{1.2}
		\caption{Reconstruction Functions $\phi_v$ and $\phi_{ch}$ for the length-$1296$ IEEE 802.11n LDPC code with code rate $1/2$}
		\resizebox{0.5\textwidth}{!}{
		%
	\end{table}
	\begin{table}[!h]
		\centering
		\huge
		\setstretch{1.2}
		\caption{Reconstruction Functions $\phi_v$ and $\phi_{ch}$ for the length-$1296$ IEEE 802.11n LDPC code with code rate $2/3$}
		\resizebox{0.5\textwidth}{!}{
		%
	\end{table}
	\begin{table}[!h]
		\centering
		\huge
		\setstretch{1.2}
		\caption{Reconstruction Functions $\phi_v$ and $\phi_{ch}$ for the length-$1296$ IEEE 802.11n LDPC code with code rate $5/6$}
		\resizebox{0.5\textwidth}{!}{
		%
	\end{table}
	\clearpage
	\subsection{The 3-bit MIM-LQMS decoder in Example 4}
	\begin{table}[h!]
		\footnotesize
		\setstretch{1.2}
		\caption{Reconstruction Functions $\phi_v$ and $\phi_{ch}$ for the length-$17664$ IEEE 802.3ca LDPC code with code rate $0.826$}
		%
	\end{table}
	\begin{table}[!h]
		\centering
		\footnotesize
		\setstretch{1.2}
		\caption{Threshold sets $\Gamma_v$ and $\Gamma_{ch}$ for the length-$17664$ IEEE 802.3ca LDPC code with code rate $0.826$}
		%
	\end{table}
	\clearpage
	\subsection{The 4-bit MIM-LQMS decoder in Example 4}
	\begin{table}[!h]
		\centering
		\huge
		\setstretch{1.2}
		\caption{Reconstruction Functions $\phi_v$ and $\phi_{ch}$ for the length-$17664$ IEEE 802.3ca LDPC code with code rate $0.826$}
		\resizebox{0.5\textwidth}{!}{
		%
	}
	\end{table}
	\begin{table}[!h]
		\centering
		\huge
		\caption{Threshold sets $\Gamma_v$ and $\Gamma_{ch}$ for the length-$17664$ IEEE 802.3ca LDPC code with code rate $0.826$}
		\setstretch{1.2}
		\resizebox{0.5\textwidth}{!}{
		%
	\end{table}
	\end{appendices}

\end{document}